\documentclass[aps,prc,twocolumn,showpacs,floatfix,superscriptaddress,amsmath,amssymb,nofootinbib,longbibliography]{revtex4-1}
   
\usepackage{graphicx}
\graphicspath{{./../notebooks/}}
\usepackage{hyperref}
\usepackage{color}
\usepackage{amsmath}
\usepackage{dcolumn}
\usepackage{multirow}
\usepackage{bm}
\usepackage{verbatim}

\newcommand{\be}{\begin{equation}}
\newcommand{\ee}{  \end{equation}}
\newcommand{\ba}{\begin{eqnarray}}
\newcommand{\ea}{  \end{eqnarray}}
\newcommand{\bas}{\begin{eqnarray*}}
\newcommand{\eas}{  \end{eqnarray*}}

\begin{document}

\title{Low-energy bound states, resonances, and scattering of light ions} 

\author{Benjamin K. Luna} 
\affiliation{Department of Physics, Tennessee Technological
   University, Cookeville, Tennessee 38505, USA}
\affiliation{Joint Institute for Nuclear Physics and Applications, Oak
  Ridge National Laboratory, Oak Ridge, Tennessee 37831, USA}

\author{T. Papenbrock} 
\affiliation{Department of Physics and Astronomy, University of
  Tennessee, Knoxville, Tennessee 37996, USA}
\affiliation{Physics Division, Oak Ridge National Laboratory, Oak
  Ridge, Tennessee 37831, USA}

\begin{abstract}
  We describe bound states, resonances and elastic scattering of light
  ions using a $\delta$-shell potential. Focusing on low-energy data
  such as energies of bound states and resonances, charge radii,
  asymptotic normalization coefficients, effective-range parameters,
  and phase shifts, we adjust the two parameters of the potential to
  some of these observables and make predictions for the nuclear
  systems $d+\alpha$, $\mbox{$^3$He}+\alpha$, $\mbox{$^3$He}+\alpha$,
  $\alpha+\alpha$, and $p+\mbox{$^{16}$O}$. We identify relevant
  momentum scales for Coulomb halo nuclei and propose how to apply
  systematic corrections to the potentials. This allows us to quantify
  statistical and systematic uncertainties. We present a constructive
  criticism of Coulomb halo effective field theory and compute the
  unknown charge radius of $^{17}$F.
\end{abstract}


\maketitle

%
\section{Introduction}
Low-energy reactions between light ions fuel stars and are relevant to
stellar nucleosynthesis~\cite{adelberger2011}. Because of the Coulomb
barrier, fusion cross sections decrease exponentially with decreasing
kinetic energy of the reactants, and this makes it difficult to
measure them in laboratories. For the extrapolation of data to low
energies, and a quantitative understanding of the reactions one thus
has to turn to theoretical calculations.

Theoretical approaches can roughly be divided into two kinds, taking
either the ions as degrees of freedom or starting from individual
nucleons. The former approach includes a variety of
models~\cite{buck1975,kulik2003,typel2005,grassi2017}, effective range
expansions~\cite{hamilton1973,mur1985,mur1993,sparenberg2010,yarmukhamedov2011,suarez2017,blokhintsev2018},
and effective field theories
(EFTs)~\cite{higa2008,ryberg2014,zhang2014,higa2018,capel2018,zhang2018};
the microscopic approach ranges from simpler models~\cite{neff2011} to
ab initio
computations~\cite{nollett2001,quaglioni2008,hupin2015,doheteraly2016}. Unfortunately,
there are still significant uncertainties~\cite{adelberger2011}, and
data tables for relevant quantities such as asymptotic normalization
coefficients (ANCs) or astrophysical $S$ factors
may~\cite{dubovichenko2017} or may
not~\cite{descouvemont2004,huang2010} contain theoretical
uncertainties.

There are various tools available for computing theoretical
uncertainties~\cite{dobaczewski2014,furnstahl2014c}.  Systematic
errors are accessible within EFTs (because of a power
counting)~\cite{schindler2009,furnstahl2014c,furnstahl2015,coelloperez2015b,carlsson2016}
but much harder to quantify for models. Nevertheless, all models are
constrained by data with errors, and the propagation of the latter to
computed observables, or the employment of a set of models provides us
with means to uncertainty estimates~\cite{dobaczewski2014}.

In this work, we revisit low-energy bound states, resonances, and
scattering within simple two-parameter models, using ions as the
relevant degrees of freedom. In an attempt to estimate uncertainties,
we quantify the sensitivity of the computed results to the input
data. We also propose systematic improvements of the simple models.
This allows us to estimate model uncertainties. As we will see, this
approach yields accurate results when compared to data. One of the key
results is the prediction for the unknown charge radius of
$^{17}$F. We contrast our approach to Coulomb halo EFT (which is not
accurate at leading order for $^8$Be~\cite{higa2008} and
$^{17}$F~\cite{ryberg2014}) and present a constructive criticism based
on a finite range and a modified derivative expansion. 

This paper is organized as follows. In Section~\ref{theory} we present
arguments in support of finite-range interactions, review key formulas
for the $\delta$-shell potential, and discuss systematic improvements.
Section~\ref{results} shows the results for a number of interesting
light-ion systems. We conclude with a summary in
Sect.~\ref{summary}. Several details are relegated to the
Appendix~\ref{appendix}.

\section{Theoretical background}
\label{theory}

\subsection{Energy scales and estimates for observables}
\label{eft}
While effective range expansions~\cite{mur1985,mur1993,sparenberg2010}
established relations between low-energy observables, we still lack
simple expressions that give estimates for such observables when only
basic properties such as energies and radii of the involved ions are
available. In applications of EFTs to low-energy ion scattering one
makes assumptions about the relevant momentum scales to propose a
power counting~\cite{higa2008,ryberg2014,zhang2014,capel2018}. This
makes it important to understand the relevant scales. As it turns out,
the presence of Coulomb interactions modifies expectations from
neutron-halo EFT or pion-less EFT significantly. To see this, we
explore how a finite-range potential differs from a zero-range
potential.

The range of the strong nuclear force is close to the sum of the
(charge) radii $D$ of two interacting particles. This is true for
both, the nucleon-nucleon interaction and for the strong force between
ions considered in this work. It is in this sense that the nuclear
interaction is short ranged. This implies that the two-body wave
function essentially acquires its ``free'' asymptotic form for
inter-particle distances $r\gtrsim D$.

The relevant asymptotic properties of a low-energy bound-state wave
function are its binding momentum and ANC. In the absence of the
Coulomb interaction, the $s$-wave ANC $C_0$ is related to the
bound-state momentum $\gamma$ for weakly bound states via
$C_0^2\approx 2\gamma$. Similarly , the $s$-wave scattering length
$a_0$ fulfills $a_0\approx 1/\gamma$. This allows one -- at leading
order -- to work with zero-range potentials. We note that the
effective range scales as $r_0\sim{\cal O}(D)$. Finite-range
effects of the potential enter at next-to-leading order. Pion-less EFT
and neutron-halo EFT are based on these
insights~\cite{bedaque2002,bertulani2002,hammer2017}.

Let us now contrast this to the case when the Coulomb potential
\be
\label{VC}
V_C(r) = {\hbar^2 k_c\over mr} 
\ee
is added.  Here, $m$ is the reduced mass and $k_c$ is the Coulomb
momentum (or inverse Bohr radius)
\be
k_c \equiv {Z_1Z_2\alpha m\over \hbar} .
\ee
It is given in terms of the fine structure constant $\alpha\approx
1/137$ and the charge numbers $Z_1$ and $Z_2$ of the two ions. As we
will see, this new momentum scale significantly modifies the discussion
of low-energy observables.

We consider a weakly bound state with energy $-\hbar^2\gamma^2/(2m)$
and bound-state momentum $\gamma$, and assume $\gamma\ll k_c$; for
resonances we consider a low-energy resonance with energy
$\hbar^2\kappa^2/(2m)$ and momentum $\kappa$, and also assume $\kappa
\ll k_c$. In what follows, we will simply refer to these momenta as
$k$, setting $k=i\gamma$ for bound states and $k=\kappa$ for
resonances. The Sommerfeld parameter is
\be
\eta \equiv {k_c\over k} .
\ee
For radial distances $r$ approximately exceeding the sum $D$ of the
charge radii of the two ions, the strong interaction potential
vanishes, and the Hamiltonian consists of the kinetic energy and the
Coulomb potential. Thus, for $r\gtrsim D$, the wave functions are
combinations of Coulomb wave functions. For small momenta $|k|\ll
k_c$, the Coulomb wave functions can be expanded in a series of
modified Bessel functions, where coefficients fall off as inverse
powers of $\eta$, while the modified Bessel functions have arguments
$2\sqrt{2k_cr}$ (see the Appendix for details). Thus, low-energy
observables (such as ANCs, radii, scattering lengths, and effective
ranges) become series of functions of $2\sqrt{2k_c D}$, with
coefficients that fall off as inverse powers of $|\eta|$. Let us
contrast the case $D\to 0$ of zero-range interactions and the case
$2\sqrt{2k_c D}\gg 1$. Estimates for several low-energy $s$-wave
observables are given in Table~\ref{tabnaive}, based on calculations
with a $\delta$-shell potential~\cite{mur1993} (with details of the
calculation presented in the Appendix).

\begin{table}[tb]
  \centering
  \begin{tabular}{c|c|c}\hline
    Observable & $2\sqrt{2k_c D}\gg 1$  & $D\to 0$ \\\hline
    $a_0$  & $-(\pi\kappa^2 D)^{-1} e^{4\sqrt{2k_cD}}$  &  $-{6k_c\over \kappa^2}$\\
    $r_0$  & $(3 k_c)^{-1}$  &  ${\cal O}(D)$\\
    $C_0$  & $(\pi D)^{-1/2}\Gamma(1+k_c/\gamma)e^{2\sqrt{2k_cD}}$ & $\sqrt{6k_c} \Gamma(1+k_c/\gamma)$ \\
    ${\Gamma\over E}$ & $4{k_c\over\kappa^2 D} e^{4\sqrt{2k_cD}} e^{-2\pi{k_c\over \kappa}}$ & $24 \pi {k_c^2\over \kappa^2}  e^{-2\pi{k_c\over \kappa}}$ \\
    $\langle r^2\rangle$ & $D^2$ & ${\cal O}(k_c^{-2})$\\\hline
  \end{tabular}
  \caption{Simple estimates for low-energy observables of a two-ion
    system with a bound-state momentum $\gamma$ or a resonance
    momentum $\kappa$, in presence of a Coulomb potential with the
    Coulomb momentum $k_c$, and a $\delta$-shell potential with the
    range $D$, in the limit $\kappa, \gamma \ll k_c$. Here $a_0$,
    $r_0$, $C_0$, and $\Gamma/E$ are the $s$-wave scattering length,
    effective range, and ANC, respectively. The resonance energy is
    $E=\hbar^2\kappa^2/(2m)$, and the corresponding width is denoted
    as $\Gamma$, not to be confused with the Gamma function
    $\Gamma(1+k_c/\gamma)$. The inter-ion distance is $\langle
    r^2\rangle$.}
    \label{tabnaive}
\end{table}

We see that the scattering length $a_0$, the squared ANC $C_0^2$, and
the resonance width are exponentially enhanced by a factor
$e^{4\sqrt{2k_cD}}$ when $k_cD\gg 1$ compared to the case $D\to 0$. We
also see that the inter-ion distance squared $\langle r^2\rangle$ is
not large, though we considered the limit of vanishing bound-state
momentum. However, this distance becomes very small in the zero-range
limit. It is clear that a zero-range potential is not compatible with
nuclei: As ions have finite charge radii they must be separated by a
distance that is similar to the sum of their charge radii in order to
retain their identities. An EFT that employs a contact at leading
order fails short of this requirement.  These arguments confirm the
need to include finite-range potentials or a finite effective range at
leading order~\cite{higa2008,papenbrock2019,schmickler2019}.

On the first view, the quantities displayed in the second column of
Table~\ref{tabnaive} appear to be model dependent for
$2\sqrt{2k_cD}\gg 1$ (as they depend on the parameter $D$). However,
in the considered limit, the inter-ion distance fulfills $\langle
r^2\rangle = D^2$, and this links observable quantities to each other.

The inter-ion distance is related to the charge radius. Let the ions
(labeled by $i=1,2$) have masses $m_i$ and charge radii squared
$\langle r_i^2\rangle$. Then, the charge radius squared of the bound
state is~\cite{buck1977}
\ba
\label{rCharge}
\langle R_c^2\rangle = {Z_1 \langle r_1^2\rangle + Z_2\langle r_2^2\rangle \over Z_1 + Z_2} 
+ \frac{(Z_1 m_2^2 + Z_2 m_1^2)\langle r^2\rangle}{(Z_1 + Z_2)(m_1 + m_2)^2} .
\ea
Here, the first term account for the finite charge radii of the ions,
and the second term is the contribution of the ions (taken as point
charges) in the center-of-mass system. The derivation of
Eq.~(\ref{rCharge}) is elementary and this expression is well
known~\cite{buck1975,buck1977}; for a recent EFT discussion of
contributions to charge radii in halo nuclei we refer the reader to
Ref.~\cite{ryberg2019}.  We note that the consistency of any two-ion
model (or EFT) requires that the distance between the two ions is
larger than the sum of their individual charge radii. As we will see
below, our results are largely consistent with the assumption of
separated ions.

We also note that cluster systems consisting of an even-even
and an odd-mass nucleus have magnetic moments (in units of nuclear magnetons)
\be
\label{magmom}
\mu = \mu_{\rm odd} +{Z\over A} l .
\ee
Here, $\mu_{\rm odd}$ is the magnetic moment of the odd-mass
constituent, $l$ is the orbital angular momentum, and $Z$ and $A$ are
the charge and mass number, respectively, of the compound system. Here
we assumed that the magnetic moment due to the spin $S$ of the
odd-mass ion and the magnetic moment due to the orbital angular
momentum $l$ add up. This is the case for states with total spin
$j=l+S$.

We note that (for $2\sqrt{2k_cD}\gg 1$) the effective range in
Table~\ref{tabnaive} does not depend on $D$, and that it decreases
with increasing Coulomb momentum. Its value, $r_0=1/(3k_c)$, is that
of a Coulomb system with a zero-energy bound state (see Appendix for
details), and $1/(3k_c)$ is also at the causality limit imposed by the
Wigner bound~\cite{mur1993,konig2013}. We can define the nontrivial
regime of strong Coulomb interactions by the model-independent
relation $3k_cr_0\approx 1$.  For the $\delta$-shell potential, is
interesting to compute corrections that are due to a finite value of
$2k_cD$. This yields~\cite{mur1993} (see the Appendix for details)
\be
\label{effrange}
r_0 - {1\over 3k_c} = -\pi D e^{-4\sqrt{2k_cD}} .
\ee
This equation expresses model-dependent quantities on its right-hand
side in terms of observables.  Combining it with the expression for
the scattering length in Table~\ref{tabnaive} yields the
model-independent relation
\be
\label{scattrelation}
\kappa^{-2} = a_0\left(r_0-{1\over 3k_c}\right) .
\ee
This formula was derived (for bound states) by \textcite{sparenberg2010}
and very recently rederived by \textcite{schmickler2019}.

Other notable relations that can be obtained from Table~\ref{tabnaive}
are
\ba
\label{new}
a_0 &\approx& -(4\pi k_c)^{-1} {\Gamma\over E}e^{2\pi{k_c\over \kappa}} ,
\ea
relating the scattering length to resonance properties, and 
\ba
   C_0^2 &\approx& \gamma^2 a_0 \left[\Gamma(1+k_c/\gamma)\right]^2 ,
\ea
relating the ANC to the bound-state energy and the scattering length
(after replacing $\kappa$ by $\gamma$). This last expression agrees
with the result in Refs.~\cite{sparenberg2010,konig2013}. It seems to
us that Eq.~(\ref{new}) was not yet known. These model-independent
expressions are valuable. They relate quantities that are often
unknown or hard to measure (such as the ANC or the effective range
parameters) to others that are better known (such as energies or
widths).

We believe the expressions in Table~\ref{tabnaive} are also useful,
because they allow us to estimate these hard-to-measure
quantities. Table~\ref{tabsystems} lists relevant parameters for
two-ion systems of interest. Of the considered systems, only the last
two approximately fulfill both $|\eta|\gg 1$ and $2\sqrt{2k_cD}\gg 1$.
Thus, for theses systems, finite-range models will yield significantly
different values than zero-range models.  Applying the simple
expressions of Table~\ref{tabnaive} and the estimates for $D$ from
Table~\ref{tabsystems} to $\alpha-\alpha$ scattering yields a very
large scattering length of about $a_0\approx -2482$~fm, an effective
range $r_0\approx 1.2$~fm, and a resonance width of $\Gamma\approx
7.5$~eV. These values are reasonably close to actual values. For the
weakly bound $J^\pi=1/2^+$ state of the $p+\mbox{$^{16}$O}$ system, we
note that the simple estimate from Table~\ref{tabnaive} yields an ANC
of about $C_0\approx 80$~fm$^{-1/2}$, close to the empirical
estimates~\cite{gagliardi1999,artemov2009,huang2010,yarmukhamedov2011}.
Thus, we gained an understanding of the scales involved in Coulomb
halo nuclei.

\begin{table}[tb]
  \centering
  \begin{tabular}{c|c|c|c|c|c}\hline
    System                  & $J^\pi$  & $\gamma$ or $\kappa$~(fm$^{-1}$) & $k_c$~(fm$^{-1}$) & $D$~(fm) & $2\sqrt{2k_cD}$ \\\hline
    $d+\alpha$              & $1^+$    & 0.31 & 0.09 & 3.82 & 1.68 \\
    $\mbox{$^3$H}+\alpha$   & $3/2^-$  & 0.45 & 0.12 & 3.43 & 1.80 \\
    $\mbox{$^3$He}+\alpha$  & $3/2^-$  & 0.36 & 0.24 & 3.64 & 2.63 \\
    $p+\mbox{$^7$Be}$       & $1/2^-$  & 0.08 & 0.12 & 3.52 & 1.85 \\
    $\alpha+\alpha$         & $0^+$    & 0.09 & 0.28 & 3.35 & 2.72 \\
    $p+\mbox{$^{16}$O}$     & $1/2^+$  & 0.07 & 0.26 & 3.58 & 2.73 \\\hline
  \end{tabular}
  \caption{Bound-state momentum $\gamma$ (or momentum $\kappa$ of the
    resonant state), Coulomb momentum $k_c$, and sum of charge radii
    $D$ for two-ion systems in the state with
    spin/parity $J^\pi$. The dimensionless quantity $2\sqrt{2k_cD}$ is
    also shown. }
  \label{tabsystems}
  \end{table}

Table~\ref{tabsystems} shows that $2\sqrt{2k_cD}\gtrsim 1$ for
essentially all Coulomb halo nuclei of interest. As a consequence,
$r_0-1/(3k_c)$ is very small for $s$ waves, and this makes scattering
lengths, resonance widths, and ANCs large. We note that these are
natural properties of Coulomb-halo nuclei. In contrast, the
smallness of $r_0-1/(3k_c)$ is viewed as a fine tuning in Coulomb halo
EFT~\cite{higa2008,ryberg2014,higa2018}.

In what follows, we will exploit a separation of scales between the
low momentum scale we are interested in and a higher-lying breakdown
scale. The breakdown momentum $\Lambda_b$ is set by the smaller of an
empirical and a theoretical breakdown scale. The empirical breakdown
scale is set by the energy of excited states of the two clusters or of
the resulting nucleus; however, only states with relevant quantum
numbers count. In $^8$Be, for instance, the ground state has
spin/parity $J^\pi=0^+$, and the empirical breakdown scale is set by
first excited $0^+$ state at about 20~MeV (and not by the energy of
the lowest $2^+$ state at 3~MeV). There is also a theoretical
breakdown scale. The strong interaction potential has a range that is
of the size of the sum $D$ of the charge radii of the clusters
involved. Thus, at momenta $\pi/D$, the details of our model are fully
resolved. As we cannot expect that the $\delta$-shell model would be
accurate at such a high momentum, it sets the theoretical breakdown
scale. In other words: this is the momentum where different models
with a physical range $D$ will differ significantly from each other.

The phenomena we seek to describe are simple because of the empirical
scale separation. Scattering phase shifts at low energies are
typically either close to zero or close to $\pi$. Only in presence of
a narrow resonance do phase shifts vary rapidly in a small energy
region of the size of the resonance width. Thus, away from the
resonance energy, the asymptotic wave function consists mostly of the
regular Coulomb wave function, which is exponentially small under the
Coulomb barrier. This implies that the wave function cannot resolve
any details of a finite-range potential as long as the classical
turning point is larger than the range $D$ of our potential. The
corresponding ``model'' momentum $\Lambda_m$ fulfills
\ba
\Lambda_m \equiv \sqrt{2k_c/D} .
\ea
Thus, for momenta below $\Lambda_m$, it will be hard to distinguish
between different finite-range models that have been adjusted to
low-energy data. In this sense, one deals with universal and
model-independent phenomena.  For momenta $k$ with $\Lambda_m\lesssim
k\lesssim \pi/D$ differences between models start to show up and
eventually become fully resolved. Some models might accurately
describe data even for momenta beyond $\Lambda_m$; we would view such
models as fortuitous but useful picks. The systematic improvements
presented in the previous Subsection can be used to estimate what a
different model would yield; we refer to resulting uncertainties as
``systematic uncertainties'' in what follows. In EFT parlance, the
momentum regime below $\Lambda_m$ would be that where
``leading-order'' results are expected to be accurate and
precise. Higher-order corrections should become visible beyond that
scale.

In this work, we employ simple finite-range models for the nuclear
potential that essentially exhibit two parameters (a range and a
strength). Most calculations will be done with the $\delta$-shell
potential, but for $^8$Be we also employ a simple square well or the
Breit model~\cite{breit1948}, a hard-core potential plus a boundary
condition. As we will see, at sufficiently low energies, and when
adjusted to low-energy data, such simple models will describe data
accurately and precisely. We will also propose how to make systematic
improvements to these models.

\subsection{$\delta$-shell potential}

The $\delta$-shell potential plus the Coulomb interaction is well
understood and can be solved analytically~\cite{kok1982,mur1985,mur1993}. In
this Subsection, we briefly summarize some of the relevant
results. The Hamiltonian is
\be
\label{ham}
H = H_0 + V . 
\ee
The strong interaction potential is $V$, and the ``free'' Hamiltonian
$H_0$ consists of the kinetic energy and the Coulomb interaction
\be
H_0 = -{\hbar^2\over 2m}\Delta + V_C(r).
\ee
Here, $m$ denotes the reduced mass of the two-ion system and $V_C$ is
the Coulomb potential~(\ref{VC}).  The $\delta$-shell potential is
parameterized as
\be
\label{deltashell}
V(r) = {\hbar^2\lambda_0\over m}\delta(r-R) .
\ee
Here, $\lambda_0$ and $R$ denote the strength and the physical range of
the potential, respectively. We work in the center-of-mass system and
employ spherical coordinates. The radial wave function $\psi_l(r) =
u_l(r)/r$ must be continuous at $r=R$, and its derivative $u_l'\equiv
{du_l\over dr}$ fulfills
\be
\label{match}
u_l'(R^+)- u_l'(R^-) = \lambda_0 u_l(R) .
\ee
The radii $R^+$ and $R^-$ are infinitesimal larger and smaller than
$R$, respectively.

\subsubsection{Bound states}

For bound states with energy $E=-{\hbar^2\gamma^2\over 2m}$ we make
the ansatz
\ba
\label{u_bound}
u_l(r) = \left\{
\begin{array}{ll}
  N \frac{H_l^+\left({k_c\over i\gamma}, i\gamma R\right)}{F_l\left({k_c\over i\gamma}, i\gamma R\right)} F_l\left({k_c\over i\gamma}, i\gamma r\right),  & r < R\\
  N H_l^+\left({k_c\over i\gamma}, i\gamma r\right), & r > R .
\end{array}
\right.
\ea
Here, we employed the Coulomb wave functions $F_l$ and $H_l^+$.  As we
employ the Coulomb wave functions at imaginary arguments, some care
must be taken in their numerical implementation; we followed
\textcite{gaspard2018} and present details in the Appendix. In
Eq.~(\ref{u_bound}), the constant $N$ ensures the proper normalization
\be
\label{norm}
\int\limits_0^\infty dr \left|u_l(r)\right|^2 = 1
\ee
of the wave function. Because of the particular ansatz of the wave
function for $r>R$, the ANC is
\be
\label{ANC}
C_l = N\frac{W_{-k_c/\gamma,l+1/2}(2\gamma R)}{H_l^+\left({k_c\over i\gamma}, i\gamma R\right)} .
\ee

The matching condition~(\ref{match}) yields 
\ba
\label{lambda_bound}
{\gamma\over \lambda_0} = i F_l\left({k_c\over i\gamma}, i\gamma R\right)H_l^+\left({k_c\over i\gamma}, i\gamma R\right) .
\ea
The inter-ion distance squared 
\be
\label{iid}
\langle r^2\rangle = \int\limits_0^\infty dr r^2 \left|u_l(r)\right|^2
\ee
enters the computation of the charge radius~(\ref{rCharge}).

\subsubsection{Scattering}

For positive energies $E={\hbar^2k^2\over 2m}$ we make the ansatz
\ba
u_l(r) = \left\{
\begin{array}{ll}
  B F_l\left({k_c\over k}, kr\right),  & r < R\\
  F_l\left({k_c\over k}, kr\right)\cos\delta +G_l\left({k_c\over k}, kr\right)\sin\delta, & r > R .
\end{array}
\right.\nonumber
\ea
Here, $G_l$ is the irregular Coulomb wave function, $\delta$ denotes
the phase shift, and we employed the shorthand
\be
B\equiv \frac{F_l\left({k_c\over k}, kR\right)\cos\delta +G_l\left({k_c\over k}, kR\right)\sin\delta}{F_l\left({k_c\over k}, kR\right)} .
\ee
The matching condition~(\ref{match}) yields 
\be
\label{lambda_scatt}
{k\over\lambda_0} = -F^2_l\left({k_c\over k}, kR\right)\cot\delta -F_l\left({k_c\over k}, kR\right)G_l\left({k_c\over k}, kR\right)
\ee
Given the phase shifts, one can use this equation to adjust
$\lambda_0$. Alternatively, for fixed parameters $(\lambda_0, R)$ this
equation can be solved for the phase shifts. This yields
\ba
\label{phase}
\cot\delta = - \frac{{k\over\lambda_0} + F_l\left({k_c\over k}, kR\right)G_l\left({k_c\over k}, kR\right)}{F^2_l\left({k_c\over k}, kR\right)} .
\ea

The $\delta$-shell potential can at most exhibit one bound state. It is
interesting to identify the critical strength $\lambda_*$ at which the
bound state enters. To do so, we start from Eq.~(\ref{lambda_scatt}),
and consider a resonance by setting $\delta=\pi/2$. In order to take
the limit $k\to 0$, we employ asymptotic approximations of the Coulomb
wave functions (see Appendix for details). This yields
\ba
\label{lambda*}
\lambda_*^{-1} = - 2R I_1\left(2\sqrt{2k_cR}\right) K_1\left(2\sqrt{2k_cR}\right) .
\ea
Here, $I_1$ and $K_1$ are modified Bessel functions. 

The effective range-expansion for the $\delta$-shell
potential is~\cite{kok1982,mur1985}
\ba
\label{ere}
a_l^{-1} &=&\frac{2k_c^{2l+1}}{\left(l! I_{2l+1}\right)^2}\left({1\over \lambda_0 R}+2I_{2l+1}K_{2l+1}\right)\nonumber\\
r_l &=& -\frac{2k_c^{2l-1}}{3 \left(l!I_{2l+1}\right)^2}
\bigg[ 2{k_c\over \lambda_0} {lI_{2l+3}+\sqrt{2k_cR}I_{2l+2}\over I_{2l+1}}\nonumber\\
  &&+2l(l+1)(l+2)I_{2l+1}K_{2l+1}\nonumber\\
  &&-{1\over 2} \left(I_{2l+1}\right)^2 -l(l+1) -k_cR\bigg] .
\ea
Here, we used the shorthands
\be
I_l \equiv I_l\left(2\sqrt{2k_cR}\right)
\ee
for the modified Bessel functions.

\subsubsection{Resonances}
As $\lambda_0$ is decreased from $0$ at fixed $R$, the potential becomes
increasingly more attractive. Just before the critical
strength~(\ref{lambda*}) is reached, the phase shift exhibits a quick
rise through $\pi/2$ at a low momentum $\kappa$.  This is reminiscent
of a resonance, and we can indeed model this physical phenomenon. To
do so, we set $\delta=\pi/2$ in Eq.~(\ref{lambda_scatt}) and find
\be
\label{lambda_res}
      {\kappa\over\lambda_0} = -F_l\left({k_c\over \kappa}, \kappa R\right)
      G_l\left({k_c\over \kappa}, \kappa R\right) . 
\ee
This relates the parameters of our potential to the resonance momentum
$\kappa$. The resonance energy is $E=\hbar^2\kappa^2/(2\mu)$. To
compute the resonance width $\Gamma$, we use the
relation~\cite{wigner1955}
\be
{d\delta\over dE} = {2\over \Gamma }.
\ee
We denote the momentum derivative of a function $f$ as ${df\over
  dk}\equiv \dot{f}$, take the derivative with respect to momentum of
Eq.~(\ref{lambda_scatt}), and set $\delta=\pi/2$. This yields
\ba
\label{wid}
\lambda_0^{-1} = \left(F_l\right)^2 \dot{\delta} - \dot{F}_l G_l -F_l\dot{G}_l . 
\ea
Here and in what follows we suppress the arguments
$(k_c/\kappa,\kappa R)$ of the Coulomb wave functions. Combining
Eqs.~(\ref{lambda_res}) and (\ref{wid}), and using $\dot{\delta}=
4E/(\kappa \Gamma)$ yields an expression for the width that depends on $R$
alone
\be
\label{width}
{E\over \Gamma} = \frac{\kappa\left(\dot{F}_lG_l+F_l\dot{G}_l\right)-F_lG_l}{4(F_l)^2} .
\ee
Given the width and the resonance energy, one can solve
Eq.~(\ref{width}) for the parameter $R$; substitution of the result
into Eq.~(\ref{lambda_res}) then yields the parameter $\lambda_0$.

It is now interesting to combine the result~(\ref{lambda_res}) with
Eq.~(\ref{lambda_scatt}) to compute the phase shift. We find
\ba
\cot\delta =
\frac{{k\over\kappa}F_l\left({k_c\over \kappa}, \kappa R\right)G_l\left({k_c\over \kappa}, \kappa R\right)-F_l\left({k_c\over k}, k R\right)G_l\left({k_c\over k}, k R\right)}{F_l^2\left({k_c\over k}, k R\right)} . \nonumber\\
\ea

\subsection{Systematic improvements}

Let us discuss systematic improvements. Consider the operator
\ba
\label{W}
W_n \equiv {1\over 2}\left(H_0\right)^n \delta(r-R^+) + {1\over 2}\delta(r-R^+)\left(H_0\right)^n . 
\ea
Here, $R^+$ denotes a point that is larger than $R$ by an arbitrarily
small amount, and $n$ is a non-negative integer~\footnote{We could
  envision also more ``democratic'' ways to write powers of $H_0$ left
  and right from the $\delta$ function, but this is not important at
  this stage.}. Consider the Hamiltonian ($n\ge 1$)
\ba
\tilde{H}_n = H_0 + V +g_n W_n , 
\ea
where $g_n$ denotes a low-energy constant. We write down the
Schr\"odinger equation for the Hamiltonian $\tilde{H}_n$ acting on the
eigenfunction of $H_0$ with eigenvalue $E$ and integrate over the
neighborhood of the singularities at $r=R$. This yields
\ba
0 &=& \int\limits_{R^-}^{R^+} dr \tilde{H}_n u_l(r) \nonumber\\
&=&-{\hbar^2\over 2m}\left[u'_l(R^+) - u'_l(R^-) - \lambda_0 u_l(R)\right] + g_n E^n u_l(R^+) \nonumber\\
\ea
Comparison with Eq.~(\ref{match}) shows that the matching condition becomes
\be
u'_l(R^+) - u'_l(R^-) = \tilde\lambda u_l(R) , 
\ee
where we introduced the energy-dependent coupling constant 
\ba
\tilde{\lambda} &\equiv& \lambda_0 + {2m \over \hbar^2}g_n E^n .
\ea
One might prefer to convert energy dependence into a momentum dependence. We employ 
the shorthand
\be
g_n = \left({2m\over \hbar^2}\right)^{n-1}\tilde{g}_n , 
\ee
and $E=\hbar^2 k^2/(2m)$, noting that $k$ can be real (for positive
energies) or purely imaginary $k=i\gamma$ for bound states. Then, the
momentum-dependent coupling constant is
\be
\label{tlamk}
\tilde{\lambda}(k) = \lambda_0 + \tilde{g}_n k^{2n} .
\ee
We remind ourselves that this is only correct if the Hamiltonian acts
on eigenstates of $H_0$. Let us discuss the power counting. The
breakdown momentum is $\Lambda_b$. By definition, the leading-order
Hamiltonian~(\ref{ham}) and the perturbation~(\ref{W}) have the same
energy $\hbar^2\Lambda_b^2/(2m)$ at the breakdown scale. Equating the
respective energies yields the scaling
\be
\tilde{g}_n \sim \frac{R}{\Lambda_b^{2(n-1)}} . 
\ee
Thus, for ``natural'' coefficients of that size, the
momentum-dependent coupling constant~(\ref{tlamk}) is a small
correction at low momenta, and contributions systematically decrease
with increasing $n$. We propose that $W_2$ is the
next-to-leading-order correction to the leading-order Hamiltonian
$H$. The rationale is as follows: The two parameters of our theory
allow us to fit, for instance, the scattering length and the effective
range. Then, a quartic correction at next-to-leading order should
affect the shape parameters in the effective range expansion.

We note that the same result could have been obtained from
perturbation theory. We also note that the same systematic corrections
apply to the Breit model or the square-well potential. The reason is
that also for these models the eigenstates of $H=H_0+V$ are wave
functions of the ``free'' Hamiltonian $H_0$ for $r>R$.  Thus, the
expectation value of $g_nW_n$ in a state with energy
$E=\hbar^2k^2/(2m)$ is $(k^2/\Lambda_b^2)^{n-1} E C_l^2 R$. The power
counting is clearly exhibited, and a systematically improvable
Hamiltonian is (terms are ordered in terms of decreasing importance)
\ba
\label{power}
H &=& H_0 + {\hbar^2\lambda_0\over m}W_0 + g_2W_2 + g_3W_3 +\ldots\nonumber\\
&=& H_0 + {\hbar^2\lambda(k)\over m}\delta(r-R)
\ea
In the first line, we have replaced the $\delta$-shell
potential~(\ref{deltashell}) by $W_0$. In the second line we reminded
ourselves that this corresponds to introducing a momentum-dependent
coupling constant
\be
\lambda(k) = \lambda_0 + \tilde{g}_2 k^4 +\tilde{g}_3 k^6 +\ldots
\ee
when acting on eigenstates of $H=H_0+V$.  In what follows, we will
simply denote the coupling constant as $\lambda$, suppressing its
momentum dependence. In practical applications, we will use $\lambda =
\lambda(0)=\lambda_0$, and employ the missing correction at
next-to-leading order to estimate systematic uncertainties.

On the one hand, the proposed way to include corrections to the
$\delta$-shell Hamiltonian~(\ref{ham}) exhibits a power counting and
thereby follows central ideas from EFT. On the other hand, the
approach is not simply a derivative expansion of the unknown strong
interaction, because $H_0$ contains the Coulomb potential. This is
important, because the contributions from the potential and the
kinetic energy are large when the Sommerfeld parameter is large; only
the combination of kinetic and potential energy yields a small total
energy. To see this, we note that the expectation value of the
``Coulombic'' term $\delta(r-R^+){\hbar^2k_c/(mr)}$ for a state with
energy $E$ is $C_l^2 \hbar^2k_c/(mR)$. As this expectation value can
be very large (compared to $C_l^2 E$), the contribution of a
derivative contact such as $\delta(r-R^+)\hbar^2\Delta/(2m)$ must be
large in size, too, when compared to $C_l^2 E$. This analysis suggests
that systematic improvements to Coulomb systems should be based on a
Coulomb-corrected derivative expansion such as Eq.~(\ref{power}),
rather than on a purely derivative expansion as done in Coulomb halo
EFT.

To further illuminate this point, we consider the Coulomb wave
functions $F_0(k_c/k,kr)$ and $G_0(k_c/k,kr)$ for the case of low
momentum (i.e. for $k\to 0$) and large Coulomb momentum (i.e. for
$k_cr\gg 1$). Then (details are presented in the Appendix)
\ba
{d\over dr} F_0(k_c/k,kr) &\approx& +4k_c F_0(k_c/k,kr) , \nonumber\\
{d\over dr} G_0(k_c/k,kr) &\approx& -4k_c G_0(k_c/k,kr) .
\ea
Thus, the derivative of the Coulomb wave function (even with a small
momentum $k\ll k_c$) yields the large Coulomb momentum $k_c$. This
casts some doubts on using a derivative expansion when the Coulomb
momentum is large compared to the momentum scale of interest.

\section{Results}
\label{results}
In this Section we present our results for various systems of
interest. Our emphasis is on uncertainty estimates and a comparison
with results from Coulomb halo EFT. The prediction of the $^{17}$F
charge radius is subject to confrontation with
data~\cite{garciaruiz2016c}. For completeness, we display the
parameters of the $\delta$-shell potential in Table~\ref{tab2}.  We
note that the values of $D$ (i.e. the sum of the ions' charge radii)
in Table~\ref{tabsystems} are smaller than the values of $R$ displayed
in Table~\ref{tab2} (except for the $3/2^-$ ground state of
$^7$Li). Thus, the strong interaction is peripheral in the cluster
model we employ. In what follows we will employ energies of bound and
resonant states. These are all taken from National Nuclear Data
Center~\cite{nndc}.

\begin{table}[tb]
  \centering
  \begin{tabular}{|c|c|c|d|d|}\hline
    Nucleus & $J^\pi$ & $l$ & \lambda_0~\mbox{(fm$^{-1}$)} & R~\mbox{(fm)}\\\hline
    $^6$Li  & $1^+$   & 0 & -0.89 & 3.84 \\
    $^7$Li  & $3/2^-$ & 1 & -1.45 & 3.14 \\
    $^7$Li  & $1/2^-$ & 1 & -1.31 & 3.50 \\
    $^7$Be  & $3/2^-$ & 1 & -1.42 & 3.22 \\
    $^7$Be  & $1/2^-$ & 1 & -1.25 & 3.75 \\
    $^{17}$F& $5/2^+$ & 2 & -1.63 & 3.60  \\
    $^{17}$F& $1/2^+$ & 0 & -0.79 & 3.85 \\
    $^8$Be  & $0^+$   & 0 & -0.81 & 3.54 \\\hline
  \end{tabular}
  \caption{Potential parameters ($\lambda_0, R$) of the $\delta$-shell
    potential that reproduce the central values for the nuclei
    described in this paper.}
    \label{tab2}
\end{table}

\subsection{$^8$Be as $\alpha+\alpha$ resonance}

The nucleus $^8$Be is not bound, but rather a $J^\pi=0^+$ resonant
state at an energy $E\approx 92$~keV and a width of $\Gamma\approx
6$~eV above the $\alpha+\alpha$ threshold. The next known $0^+$ state
is at 20.2~MeV of excitation. We note that this energy is equal to the
energy of the first $0^+$ state of the $\alpha$ particle to three
significant digits.  Assuming there are indeed no other $0^+$ states,
20~MeV sets the empirical breakdown energy for any cluster model or
EFT that describes $^8$Be in terms of ``elementary'' $\alpha$
particles. The corresponding breakdown momentum is $1.4$~fm$^{-1}$.

However, $\alpha$ particles have a finite size, and the sum of the two
charge radii of the $\alpha$ particles is $D\approx 3.3$~fm.  The
Coulomb momentum is $k_c\approx 0.28$~fm$^{-1}$.  At a momentum
$\pi/D\approx 0.95$~fm$^{-1}$, the details of any Hamiltonian with a
physical range $D$ can be resolved.  The corresponding breakdown
energy in the center-of-mass system is $E_b\approx 9.4$~MeV. This
energy is lower than the empirical breakdown scale and therefor sets
the breakdown scale. We note that it is not precluded to construct a
model that describes data accurately even at the breakdown
scale. However, that would seem to be fortuitous, as a generic
finite-range model that is adjusted to low-energy data is expected to
not be accurate at such energies.  We expect model dependencies to
become visible above the momentum $\Lambda_m\approx
0.4$~fm$^{-1}$. This corresponds to a center-of-mass energy of about
$E_m\approx 1.7$~MeV.

To summarize the arguments: Virtually any model with a physical range
of size $D$ that is adjusted to low-energy data is expected to
describe data accurately up to about $E_m=1.7$~MeV. At higher
energies, model dependencies start getting resolved, and a de-facto
breakdown of models with a range of size $D$ occurs at about
$E_b=9.4$~MeV. The model dependencies of the $\delta$-shell potential can
be estimated by employing the momentum dependent coupling
$\lambda(k)=\lambda_0 + g R k^4/\Lambda_b^2$. Here, $g$ is a number of
order one.

\begin{figure}[tb]
  \centering
  \includegraphics[width=\columnwidth]{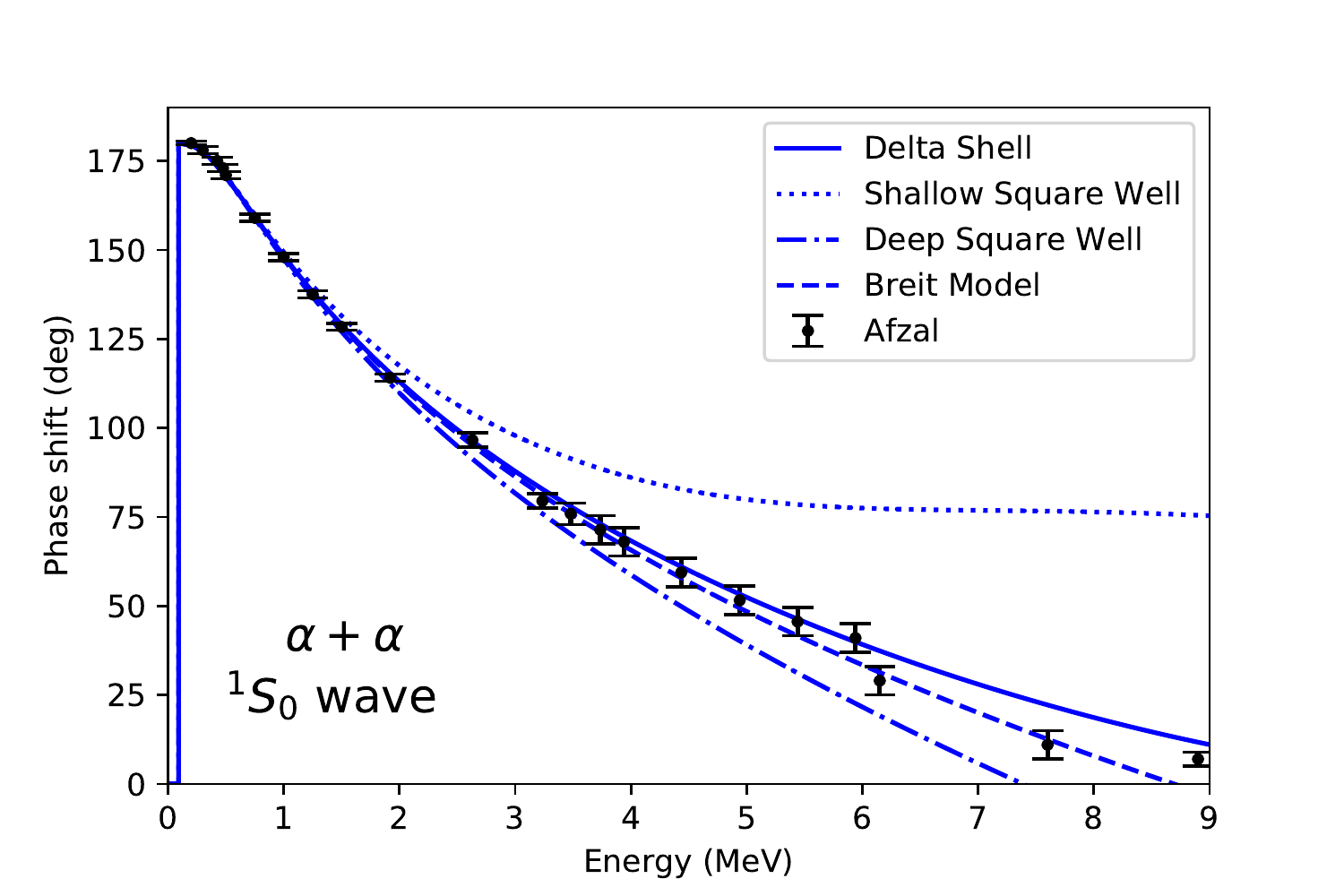}
  \caption{(Color online) Phase shifts of $\alpha-\alpha$ scattering
    in the $s$ wave, as a function of the energy in the laboratory
    frame computed with a shallow square-well potential (dotted line),
    the $\delta$-shell (solid line), the Breit model (dashed line) and
    a deep square well (dashed-dotted line). Data taken from
    Refs.~\cite{heydenburg1956,afzal1969}.}
  \label{8BeModels}
\end{figure}

Figure~\ref{8BeModels} shows $s$-wave phase shifts for $\alpha-\alpha$
scattering computed from different models, and compares them to
data. The two-parameter models have been adjusted to the resonance
energy and its width. All models are practically indistinguishable
below $E_m\approx 1.7$~MeV and differ significantly at the breakdown
energy $E_b\approx 9.4$~MeV.

\begin{figure}[tb]
  \centering
  \includegraphics[width=\columnwidth]{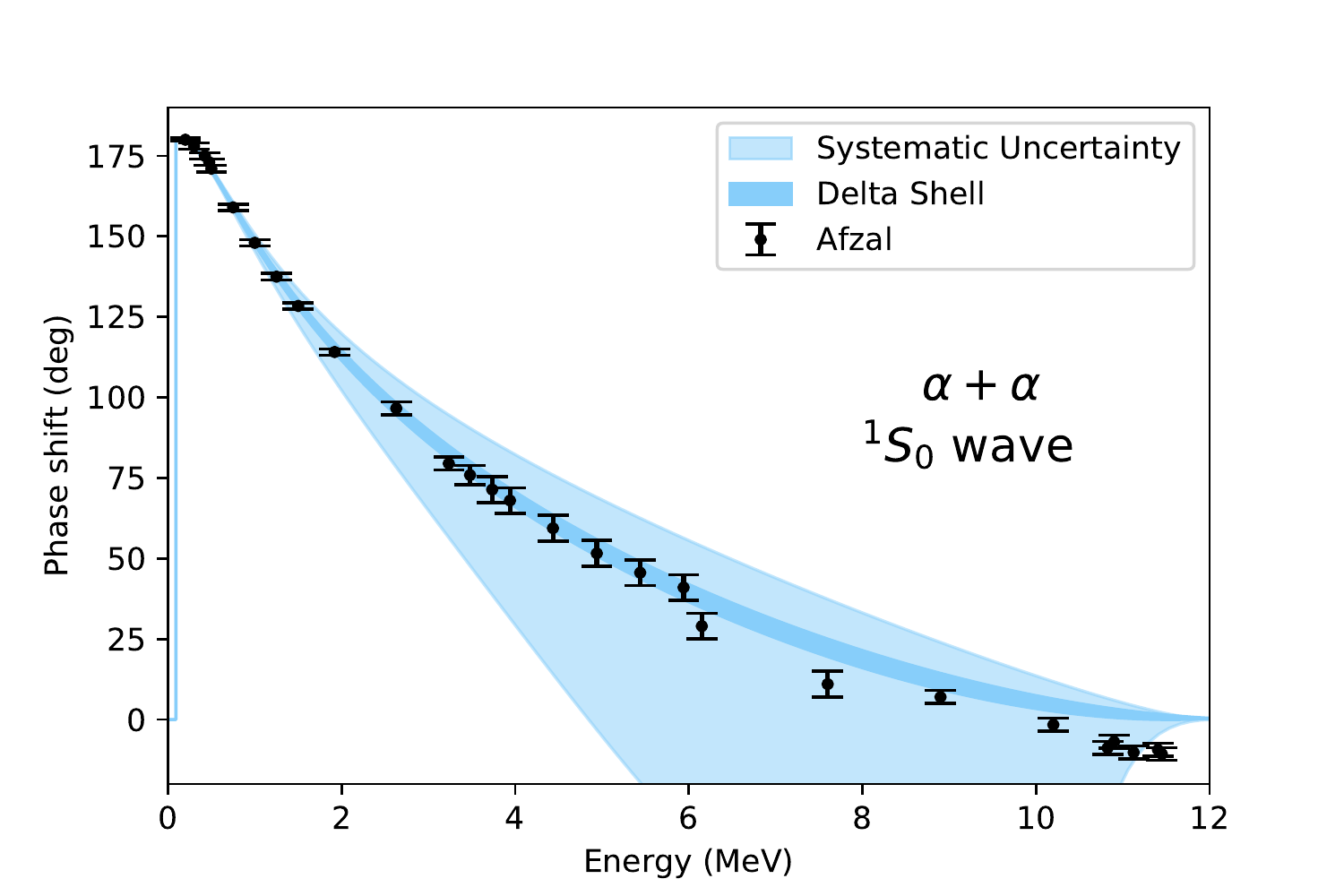}
  \caption{(Color online) Phase shifts of $\alpha-\alpha$ scattering
    in the $s$ wave, as a function of the energy in the center-of-mass
    frame. The dark band shows the uncertainty from the resonance
    width. The light band shows the theoretical uncertainty
    estimate. Data taken from Refs.~\cite{heydenburg1956,afzal1969}.}
  \label{8Be}
\end{figure}

We adjust the parameters of the $\delta$-shell potential to the
resonance energy and width and computed the resulting phase shifts.
Figure~\ref{8Be} shows the phase shifts predicted from our
approach. The central line is obtained from adjusting to the resonance
energy and the central value of its width. Varying the resonance width
$\Gamma=5.57\pm 0.25$~eV within its uncertainty produces the dark
band. The systematic uncertainty estimate, i.e. the range that
different models would explore, is shown as a light band.  Its extent
is generated by employing $\lambda(k)=\lambda_0 \pm R
k^4/\Lambda_b^2$.  We see that the prediction of the $\delta$-shell
potential agrees well with data, even for energies beyond
$E_m=1.7$~MeV. This model happens to be accurate.

We computed the scattering length and effective range and obtained
$a_0=-2020\pm 100$~fm and $r_0=1.106\pm 0.005$~fm, respectively. The
uncertainties stem from the uncertainty in the resonance width. Let us
compare this with effective range parameters from the literature.
Overall, there is a consensus on the effective range, which is close
to the estimate $1/(3k_c)=1.21$~fm shown in Table~\ref{tabnaive}. The
scattering length, of course, is sensitive to the precise difference
$r_0-1/(3k_c)$ [see the approximation~(\ref{scattrelation})], and it
is probably only known to about 5 to 10\%. The effective range
expansions by \textcite{rasche1967}, \textcite{higa2008}, and
\textcite{kamouni2007} found $a_0=1650\pm 150$~fm, $a_0=1920\pm
90$~fm, and $a_0=2390$~fm, respectively. The potential models by
\textcite{kulik2003} yielded $a_0=2030\pm 100$~fm. Ab initio
computations have not yet reached the precision to extract very large
scattering lengths precisely~\cite{elhatisari2015}.

\textcite{kulik2003} uses simple models for the computation of phase
shifts and effective range parameters. The two-parameter models are
(i) the $\delta$-shell potential, and (ii) the Breit
model~\cite{breit1948}, i.e. a hard-core potential where the wave
function's logarithmic derivative at the hard core is set. These
models are adjusted to the resonance energy and to phase shifts, and
they virtually agree with each other for energies in the
center-of-mass system up to 2~MeV. They agree with data over an even
wider range. Interestingly, these models yield an accurate description
of the resonance width when adjusted to phase
shifts. \textcite{kamouni2007} use the resonating group method and
R-matrix theory to extract an effective range expansion. This approach
adjusts about two parameters in each partial wave.

\begin{figure}[tb]
  \centering
  \includegraphics[width=\columnwidth]{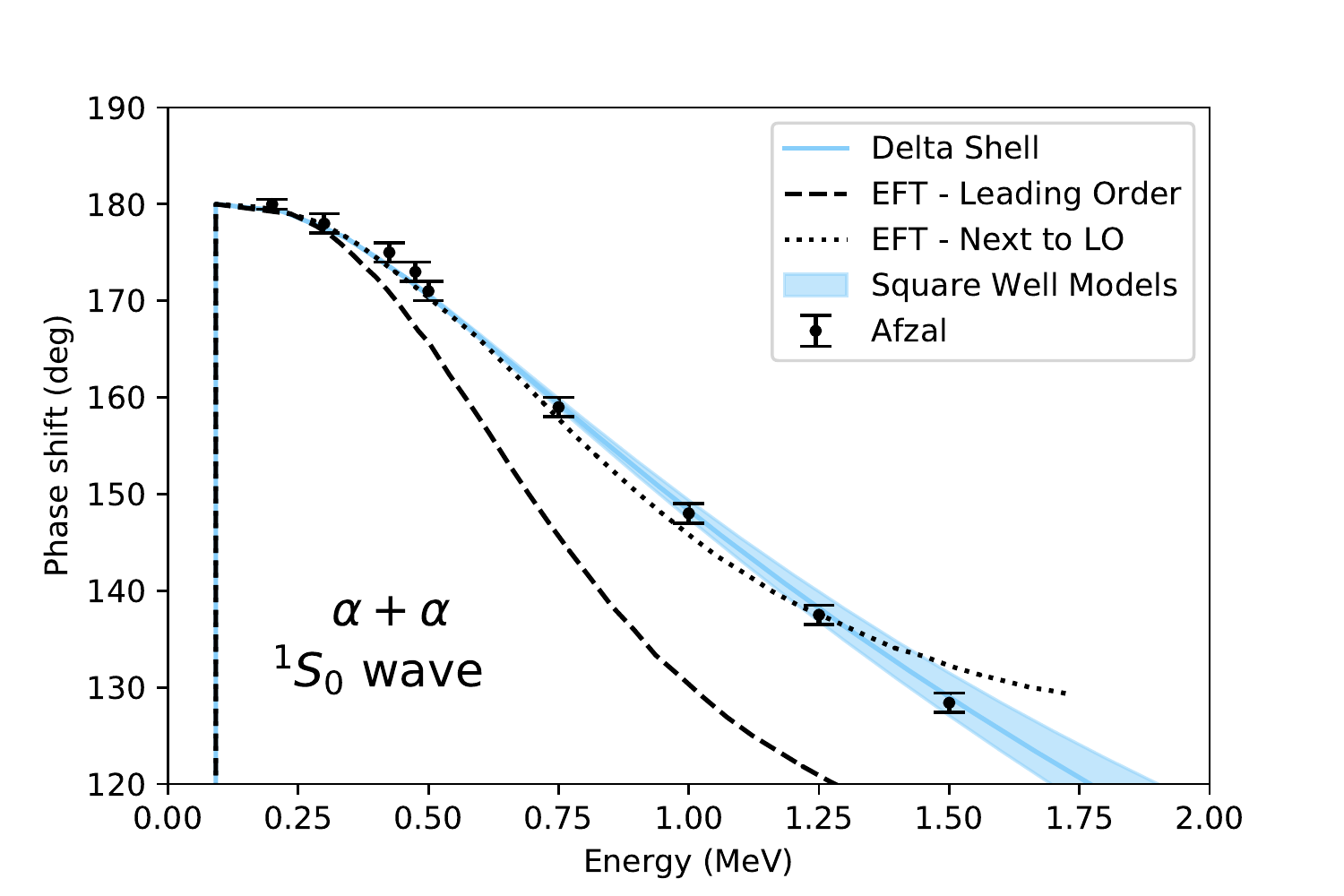}
  \caption{(Color online) Phase shifts of $\alpha+\alpha$ scattering
    in the $s$ wave computed from finite-range models and in leading
    order (LO) and next-to-leading order (NLO) Coulomb halo
    EFT~\cite{higa2008}, as a function of the energy in the laboratory
    frame. Data taken from Refs.~\cite{heydenburg1956,afzal1969}.}
  \label{8BeEFT}
\end{figure}

Let us contrast our approach to the halo EFT work by
\textcite{higa2008}. That approach is based on a dimer formulation
with contact interactions. At leading order (LO), a fit to the
resonance energy and width yields phase shifts that agree with data
only up to 0.3~MeV in the center-of-mass frame. At
next-to-leading-order (NLO), three parameters are adjusted to the
resonance energy, its width, and phase shifts. The resulting phase
shifts clearly deviate from data above 0.7~MeV of center-of-mass
energy. Figure~\ref{8BeEFT} compares the EFT results at LO and NLO to
models. The EFT results are not accurate. This is somewhat surprising,
because the effective-range expansion by the same authors yielded
phase shifts that agree with data.

\subsection{$^{17}$F as $^{16}\mbox{O}+p$}
The $^{17}$F nucleus plays a role in nucleosynthesis. Its
$J^\pi=5/2^+$ ground state and its first excited $1/2^+$ states are
bound by about 0.6 and 0.1~MeV, respectively. These energies are small
compared to 6~MeV, the energy it takes to excite the doubly-magic
nucleus $^{16}$O, and we can thus approximate $^{17}$F as a
$^{16}\mbox{O}+p$ system at sufficiently low energies. The next
excited states in $^{17}$F with quantum numbers $5/2^+$ and $1/2^+$
are separated by 6.7 and 6.5 MeV, respectively, from the corresponding
bound states. Thus, the empirical breakdown energy is about
$E_b\approx 6$~MeV. The sum of the charge radii of the proton and
$^{16}$O is about $D\approx 3.6$~fm. This sets the theoretical
breakdown momentum to $\pi/D\approx 0.88$~fm$^{-1}$, corresponding to
an energy of about 17~MeV. Thus, the breakdown scale is set by the
empirical breakdown energy. The Coulomb momentum is $k_c\approx
0.26$~fm$^{-1}$.  Thus, potentials with a physical range $D$ are
expected to exhibit model dependencies above about
$\Lambda_m=(2k_c/D)^{1/2}\approx 0.27$~fm$^{-1}$, corresponding to an
energy $E_m\approx 1.6$~MeV.

\begin{figure}[tb]
  \centering
  \includegraphics[width=\columnwidth]{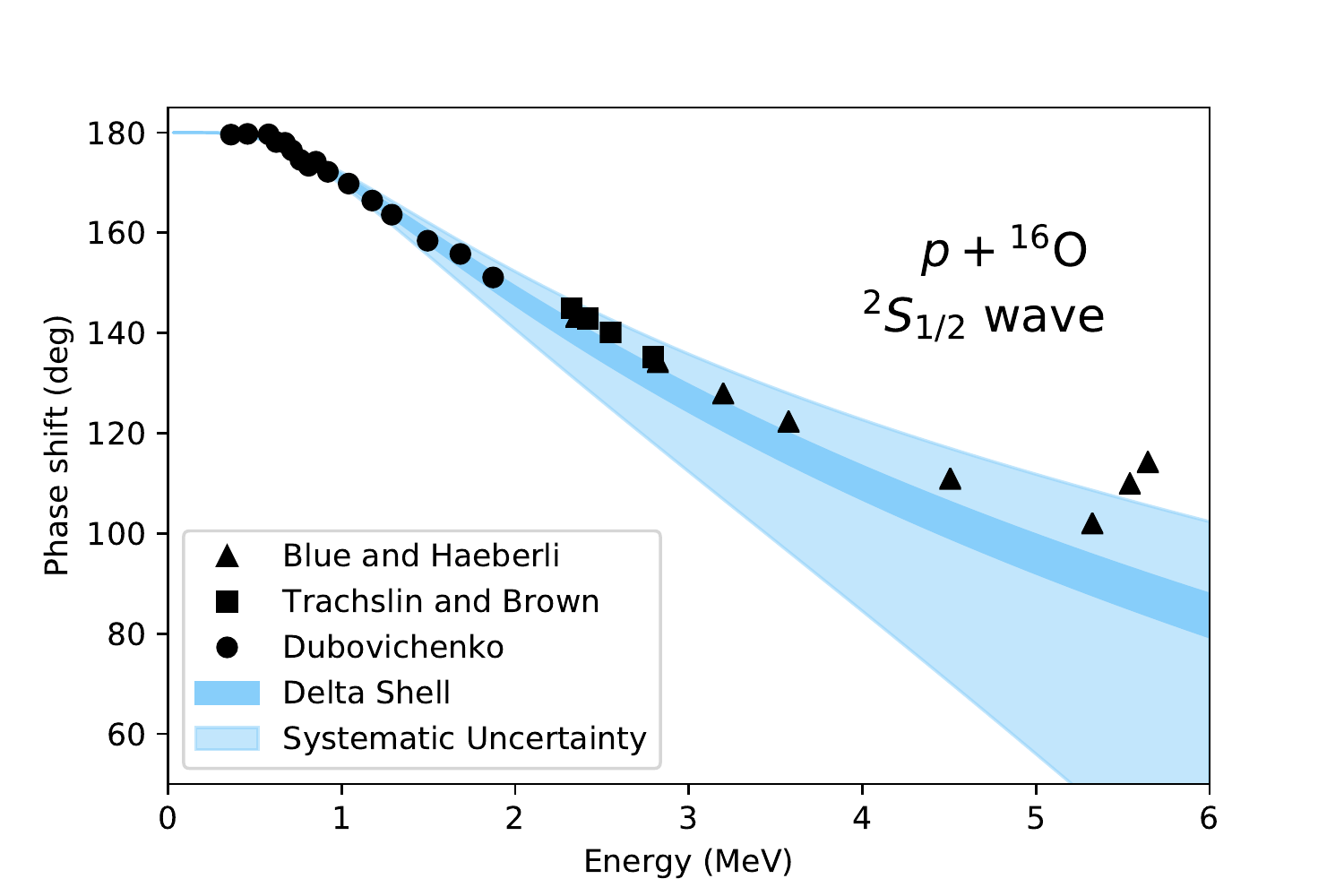}
  \caption{(Color online) Phase shifts of $p+^{16}\mbox{O}$ scattering in the
    $^2S_{1/2}$ partial wave, as a function of the energy in the
    center-of-mass frame. Data taken from Ref.~\cite{blue1965,trachslin1967,dubovichenko2017b}.}
  \label{17F*}
\end{figure}

Let us consider the excited $J^\pi=1/2^+$ halo
state~\cite{morlock1997}. We adjust the model parameters to the
binding energy and the $^2S_{1/2}$ phase shift data from
Ref.~\cite{dubovichenko2017b}. The results are shown in
Fig.~\ref{17F*}. We then predict the ANC to be $C_0=78.9\pm
4.2$~fm$^{-1/2}$, and the charge radius of the excited state is
$R^*_c= 3.096 \pm 0.034$~fm. The ANC agrees with the results by
\textcite{gagliardi1999}, \textcite{artemov2009}, and
\textcite{huang2010}, who found values of ($80.6\pm 4.2$)~fm$^{-1/2}$,
($75.5\pm 1.5$)~fm$^{-1/2}$, and 77.2~fm$^{-1/2}$, respectively.  Our
effective range parameters are $a_0=4080 \pm 430$~fm, and $r_0= 1.17
\pm 0.01$~fm. Within their uncertainties, these values agree with
those of Refs.~\cite{kamouni2007,yarmukhamedov2011}.

Let us also compare to Coulomb halo EFT. For the excited $1/2^+$
state, \textcite{ryberg2014} employed one parameter at leading order
and found that the relative distance $\langle r^2\rangle=
(0.59~\mbox{fm})^2$ between the proton and the core and the ANC
$C_0=21.4$~fm$^{-1/2}$ are too small.  At next-to-leading order,
effective range contributions enter, and the charge radius is
increased by a factor 3.6--3.8~\cite{ryberg2016}.

\begin{figure}[tb]
  \centering
  \includegraphics[width=\columnwidth]{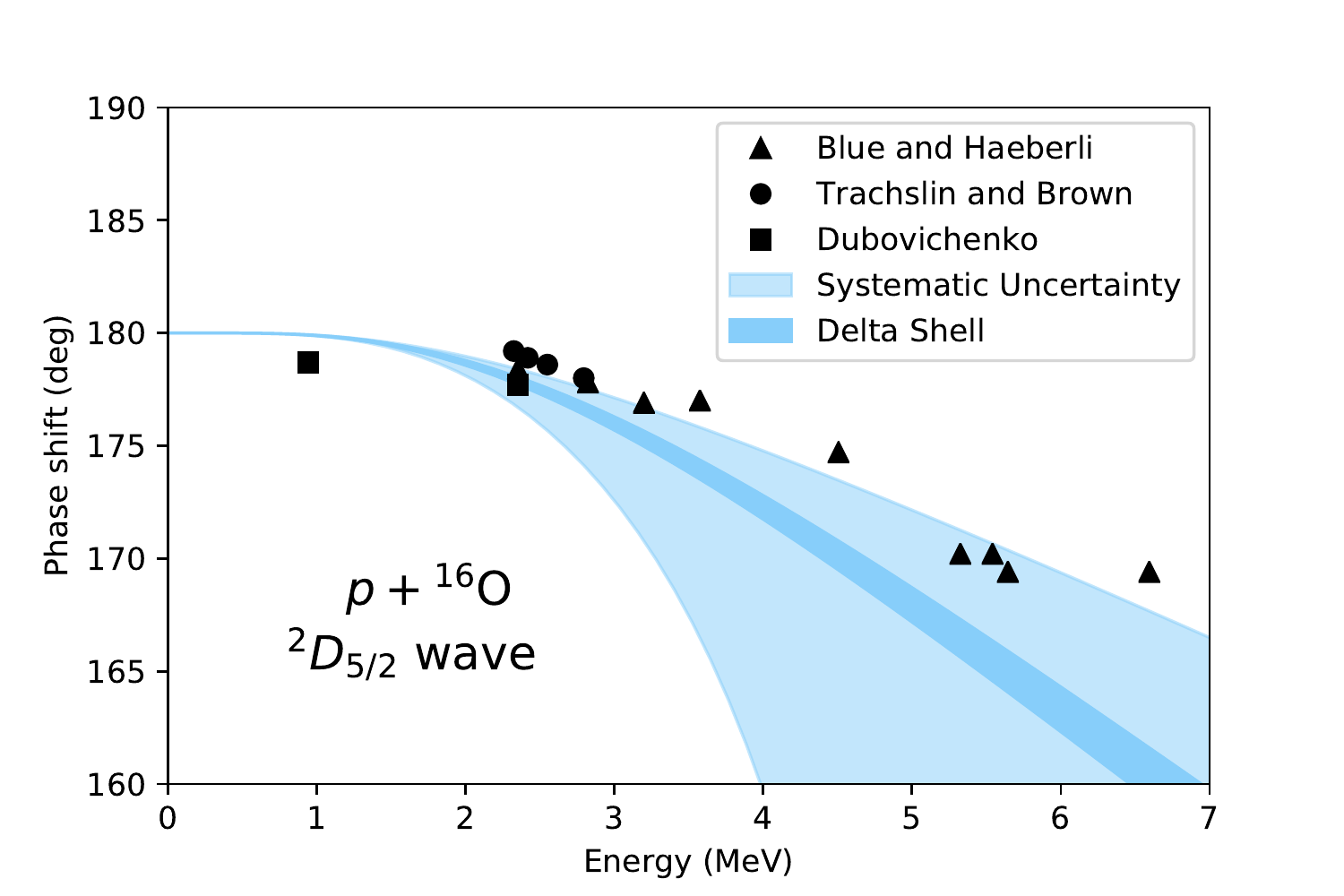}
  \caption{(Color online) Phase shifts of $p+^{16}\mbox{O}$ scattering in the
    $^2D_{5/2}$ partial wave, as a function of the energy in the
    center-of-mass frame. Data taken from Refs.~\cite{blue1965,trachslin1967,dubovichenko2017b}.}
  \label{17F}
\end{figure}

Let us turn to the $^{17}$F ground state.  Its charge radius is not
yet known but its measurement is currently an active experiment at
CERN Isolde~\cite{garciaruiz2016c}.  We want to make a prediction for
this observable.  To put things into perspective we note that the
charge radius of $^{19}$F is $R_c=2.8976(25)$~fm~\cite{angeli2013};
the ground-state of that nucleus has spin/parity ${1/2}^+$. We adjust
our model parameters to the binding energy and the ANC. The
ground-state ANC extracted from transfer reaction data via potential
models is $1.04\pm
0.05$~fm$^{-1/2}$~\cite{gagliardi1999,artemov2009}. The resulting
phase shifts are shown in Fig.~\ref{17F}. Unfortunately, the phase
shift analysis lacks uncertainties, but we see a systematic
deviation. We compute a scattering length of $a_2 = 1.15(11) \times
10^3$~fm$^5$ and an effective range of $r_2 = -0.068(7)$~fm$^{-3}$, in
agreement with results by \textcite{yarmukhamedov2011} (who were also
informed by the ANCs we used). We compute a charge radius of $R_c =
2.88(1)$~fm.  This is a large radius for a $d$-wave state and
practically as large as the charge radius of the $1/2^+$ ground state
of $^{19}$F.

To estimate the reliability of our computations, we alternatively fit
to the potential parameters to the phase shifts and the binding energy
and find $R \approx 2.957$~fm and $\lambda_0 \approx
-1.924$~fm$^{-1}$. We note the the resulting $\chi^2$ per degree of
freedom is about 11, hinting at phase-shift uncertainties of about
three degrees (assuming them to be of statistical nature).  In this
case, we compute an ANC of $C_2 = 0.7286$~fm$^{-1/2}$, and a charge
radius $R_c = 2.80(2)$~fm.  These values are significant smaller than
those given in the previous paragraph, and the uncertainties do not
overlap. It seems to us that the phase shift
data~\cite{blue1965,trachslin1967,dubovichenko2017b} and the transfer
reaction data~\cite{gagliardi1999} are probably not compatible. We
note, however, that the accurate determination of $d$-wave phase
shifts from low-energy scattering is complicated because $s$ and $p$
waves dominate. We also note that somewhat smaller ANCs of 0.91 and
0.88~fm$^{-1/2}$ have been computed by \textcite{huang2010} and
\textcite{blokhintsev2018}, respectively. As the extraction of the ANC
by \textcite{gagliardi1999} is more recent than the phase shift
analysis (and includes uncertainties), we base our computation on the
ANC and predict a charge radius of 2.88(1)~fm for $^{17}$F. The
measurement \cite{garciaruiz2016c} will certainly be useful to yield
insight into the low-energy properties of the $p+\mbox{$^{16}$O}$
system. We also note that this nucleus is in reach of ab initio
computations~\cite{hagen2010a}, but its charge radius and ANC have not
been computed, yet.

\subsection{$^6$Li as a $\alpha+d$ bound state}
The ground state of $^6$Li is only bound by about $E=1.47$~MeV with
respect to the $d +\alpha$ threshold. This corresponds to a
bound-state momentum of $\gamma\approx 0.31$~fm$^{-1}$. Its
spin/parity is identical to that of the deuteron, and the
estimate~(\ref{magmom}) for its magnetic moment yields $0.86$ nuclear
magnetons, which is close to the observed value of $0.822$~\cite{stone2014a}.  These
basic properties suggest that the $^6$Li ground state exhibits a
dominant $s$-wave halo structure, and we will we neglect any $d$ wave
component in what follows.

Let us assess the breakdown scale.  The three-body breakup of $^6$Li
into $\alpha +n +p$ requires the breakup of the deuteron and is thus
about 2.2~MeV above threshold. This inelastic process is without
concern to us.  The first excited state with the same spin and parity
as the ground state is at 5.65~MeV, and this is the empirical
breakdown energy.  The sum of charge radii is $D\approx 3.8$~fm,
setting the theoretical breakdown momentum at $\pi/D\approx
0.82$~fm$^{-1}$, which corresponds to a high energy of 10.6~MeV. Thus,
the breakdown scale is set by the empirical properties. The binding
energy of the deuteron to the $\alpha$ core is a factor of about four
smaller than the breakdown energy, and this provides us with a
separation of scale. The Coulomb momentum of the $d+\alpha$ system is
$k_c\approx 0.09$~fm$^{-1}$, and model differences are start to get
resolved above the momentum $\Lambda_m\approx 0.22$~fm$^{-1}$,
corresponding to an energy $E_m\approx 0.75$~MeV. As this energy is
smaller than the binding energy of the $\alpha +d$ system, model
dependencies could be relevant. However, below we will see that the
$\delta$-shell model yields an accurate description of existing
low-energy data.

\begin{figure}[tb]
  \centering
  \includegraphics[width=\columnwidth]{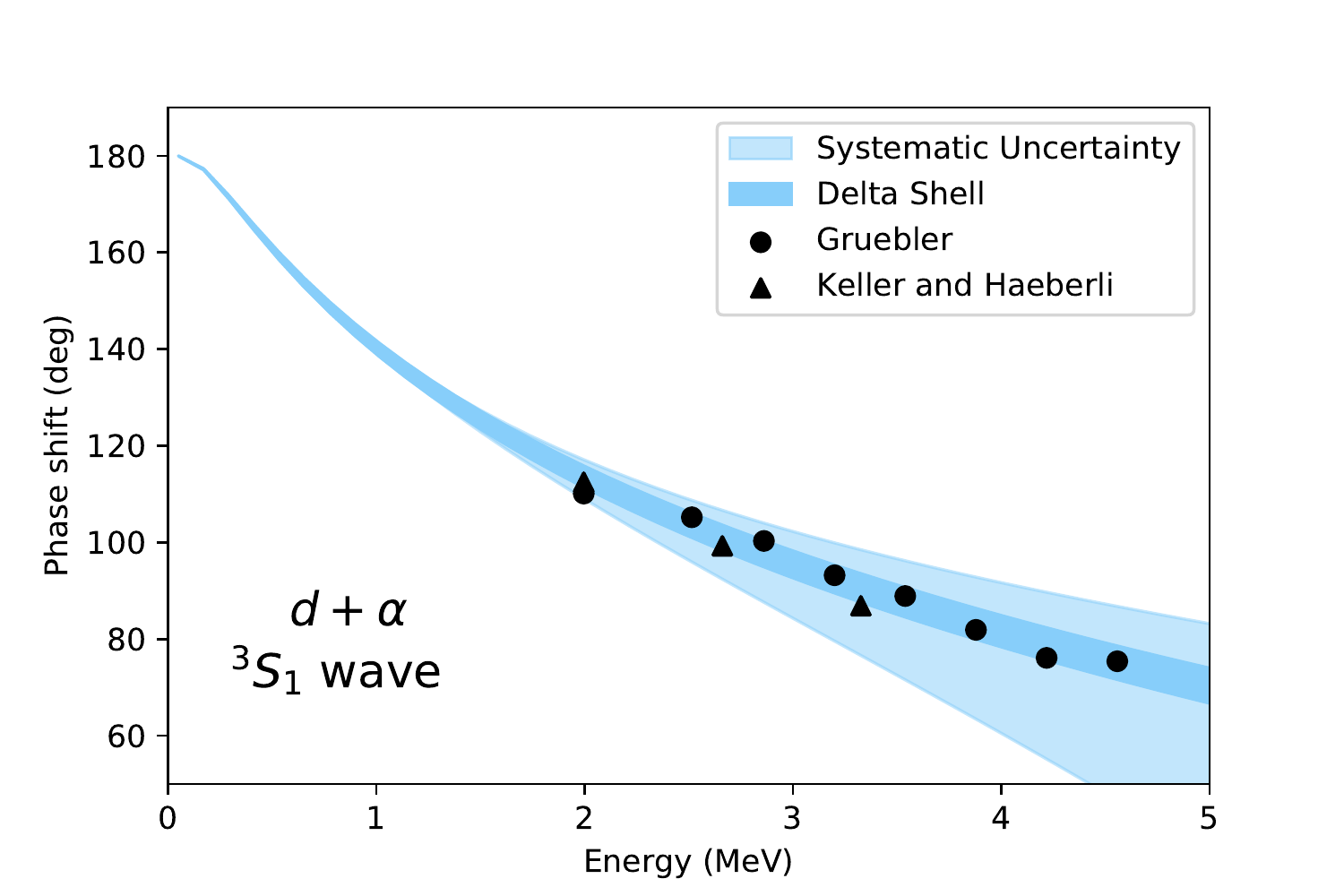}
  \caption{(Color online) Phase shifts of $d+\alpha$ scattering in the
    $^3S_1$ partial wave, as a function of the energy in the
    center-of-mass frame. Data taken from
    Refs.~\cite{keller1970,gruebler1975}.}
  \label{6Ligs}
\end{figure}

We model the $^6$Li ground state using the $\delta$-shell potential in
the $s$ partial wave. \textcite{ryberg2014b} pointed out that charge
radii can be used to constrain low-energy observables that are
relevant in astrophysics. Together with the binding energy, these are
the most precise data available at low energies. We therefore adjust
the two parameters of our potential to the binding energy and the
charge radius. The charge radius of $^6$Li is $2.589 \pm
0.039$~fm~\cite{angeli2013} and we perform a total of three
calculations, adjusting to its central, lower, and upper values. For
the relevant $s$ wave we compute an ANC of $C_0 =
2.23(11)$~fm$^{-1/2}$, a scattering length $a_0= 29.1 \pm 1.7$~fm, and
an effective range $r_0=1.85(5)$~fm.  The uncertainties reflects the
uncertainty in the charge radius. The central value of the inter-ion
distance is $\sqrt{\langle r^2\rangle}=3.86$~fm, and this marginally
exceeds the sum of charge radii of its constituents, 3.82~fm. The
resulting phase shifts are shown in Fig.~\ref{6Ligs}, and they agree
with data~\cite{keller1970,gruebler1975}. This gives us confidence in
the accuracy of our results.

It is interesting to compare our prediction for the ANC and the
effective range parameters with the literature. The effective range
parameters agree with Ref.~\cite{krasnopolsky1991}, which states
$a_0=30.8$~fm and $r_0=1.88$~fm; the ANC agrees with the values of
Refs.~\cite{mukhamedzhanov2011,blokhintsev2014,grassi2017}. However,
we note that ANCs have clearly evolved (and decreased) over time, as
the
papers~\cite{blokhintsev1993,blokhintsev2006,mukhamedzhanov2011,tursunov2015}
show.  We note that the ab initio computation by
\textcite{nollett2001} reports an ANC of $2.28\pm 0.02$~fm$^{-1/2}$
(in agreement with recent cluster models and our result), while
\textcite{hupin2015} found a larger ANC of about
$2.7$~fm$^{-1/2}$. While the calculation of Ref.~\cite{nollett2001} is
informed by charge radii through its variational wave function, the
paper~\cite{hupin2015} did not present results for charge radii.  We
believe our calculations, through their consistency for all low-energy
observables, add further weight to an ANC around 2.2~fm$^{-1/2}$.

\subsection{$^7$Be as $\alpha+^3\mbox{He}$ bound state}
The $^7$Li ground state has quantum numbers $J^\pi=3/2^-$ and is bound
by 1.6~MeV with respect to the $\alpha+\mbox{$^3$He}$ threshold. The
only other bound state is at about 0.4~MeV of excitation energy and
has quantum numbers $J^\pi=1/2^-$. Both states are thus weakly bound
and can be viewed as $p$ waves of the $\alpha+\mbox{$^3$He}$
system. We note that the estimate~(\ref{magmom}) for the ground
state's magnetic moment, $-1.556$ nuclear magnetons, is close to the
experimental value of $-1.398$~\cite{stone2014a}. This all suggests
that we can describe $^7$Be as an $\alpha+^3\mbox{He}$ system.

The empirical breakdown energy is set by the energy of excited states
9.9~MeV for quantum numbers $J^\pi=3/2^-$; it is about twice as high
for the numbers $J^\pi=1/2^-$ state. Of course, the $^3$He nucleus
breaks up at an excitation energy of about 6~MeV, but this inelastic
channel is of no concern for us. The sum of the two charge radii is
$D\approx 3.6$~fm, setting the theoretical breakdown momentum to
$\pi/D\approx 0.86$~fm$^{-1}$, corresponding to an energy of
9~MeV. Thus the breakdown energy is about 9~MeV. Model dependencies
become visible above the momentum scale $\Lambda_m\approx
0.36$~fm$^{-1}$, corresponding to an energy of 1.6~MeV. We note that
this energy is similar to the ground-state energy.

\begin{figure}[tb]
  \centering
  \includegraphics[width=\columnwidth]{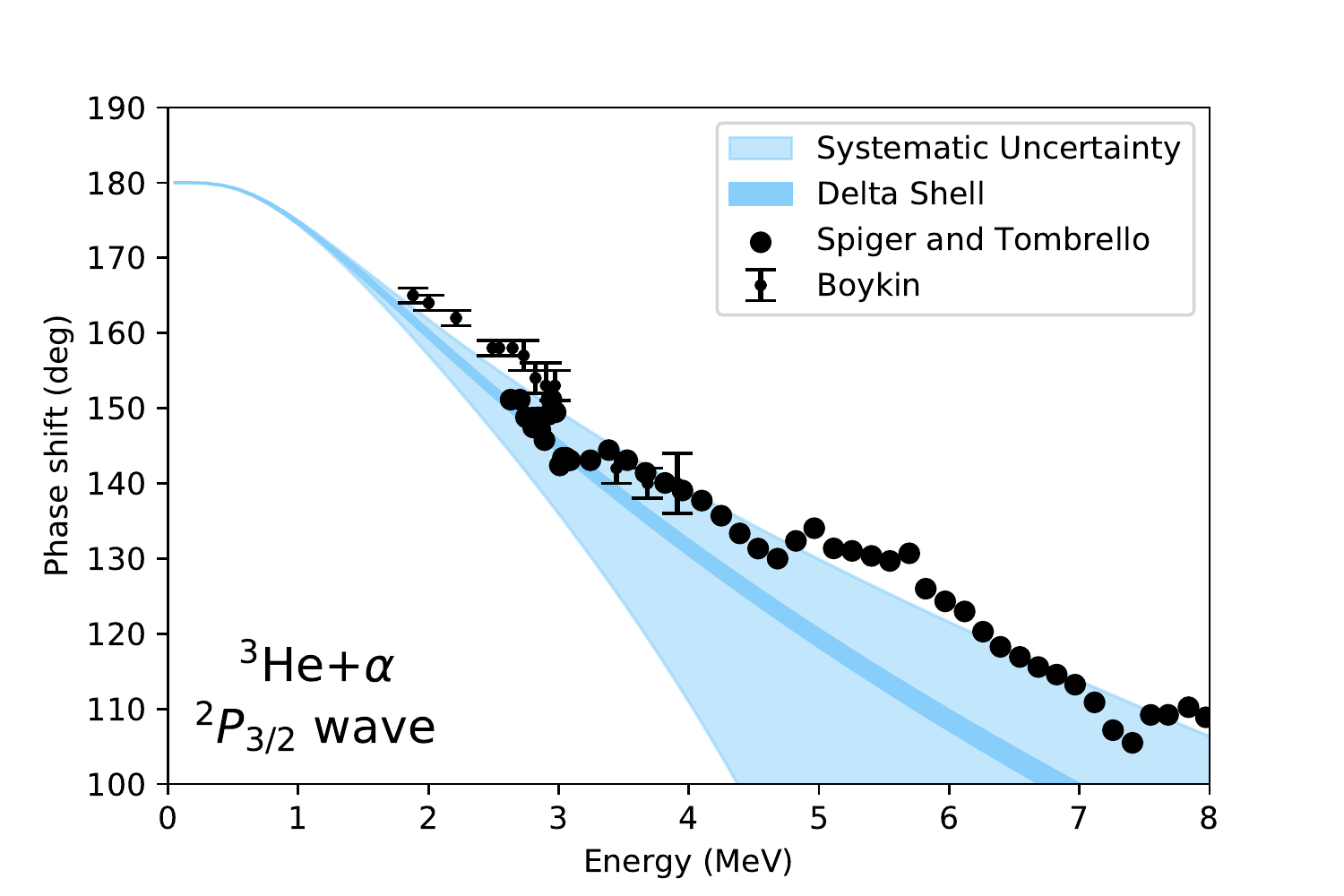}
  \caption{(Color online) Phase shifts of $^3\mbox{He}+\alpha$ scattering in the
    $^2P_{3/2}$ partial wave, as a function of the energy in the
    center-of-mass frame. Data taken from Refs.~\cite{spiger1967,boykin1972}}
  \label{7Begs}
\end{figure}

For the $^2P_{3/2}$ partial wave, we adjust the two parameters of the
$\delta$-shell potential to the binding energy of the ground state and
its charge radius of $2.646 \pm 0.016$~fm~\cite{angeli2013}. As
before, we propagate the uncertainty of the charge radius to
low-energy observables. Then, the ground-state ANC is $C_1=3.6\pm
0.1$~fm$^{-1/2}$, and the effective-range parameters are $a_1= 207\pm
8$~fm$^3$ and $r_1=-0.041\pm 0.004$~fm$^{-1}$.  The predicted phase
shifts are shown in Fig.~\ref{7Begs} and compared to
data~~\cite{spiger1967,boykin1972}. The agreement is
fair. Unfortunately, the older data by \textcite{spiger1967} lacks
uncertainties.

Let us compare with other approaches for the $3/2^-$ partial
wave. \textcite{descouvemont2004} found an ANC of
$C_1=3.79$~fm$^{-1/2}$ (close to our value) from an R matrix analysis,
while \textcite{tursunmahatov2012} found an ANC of
$C_1=4.83^{+0.1}_{-0.25}$ from evaluations of capture reactions. We
refer to the latter paper for a review of literature values. The ab
initio computation by \textcite{doheteraly2016} found a scattering
volume of $a_1=210.4$~fm$^3$ (close to our result), while the
effective range expansion techniques~\cite{yarmukhamedov2011} found
effective range parameters $a_1=301 \pm 6$~fm$^3$ and $r_1=−0.0170 \pm
0.0026$~fm$^{-1}$ (and a squared ANC of $C_1^2=23.3$~fm$^{-1}$). We
note that the ab initio computation~\cite{doheteraly2016} yields a
charge radius that is close to data.

We note that Coulomb halo EFT was very recently applied to the
$\alpha+\mbox{$^3$He}$ system~\cite{higa2018,zhang2018} for a
computation of the astrophysical S factor.  \textcite{zhang2018}
pursued a Bayesian approach based on data from capture reactions,
avoiding the need to adjust parameters to phse shifts.
\textcite{higa2018} employed the ANC from
Ref.~\cite{tursunmahatov2012} for their computation of the
astrophysical S factor. At leading order (a one-parameter or a
three-parameter theory, depending on the power counting), the
resulting phase shifts are visibly above the data~\cite{boykin1972}.

It seems to us that this $\alpha+\mbox{$^3$He}$ system is still not
sufficiently well understood. Existing theoretical results are in
conflict with each other, and no calculation seems to be able to
reproduce charge radii, phase shifts, and capture data.

\subsection{$^7$Li as $\alpha+\mbox{$^3$H}$ bound state}

The $3/2^-$ ground state of $^7$Li is bound by about 2.5~MeV with
respect to the threshold of the $\alpha+\mbox{$^3$H}$ system. Based on
a cluster assumption~(\ref{magmom}), its magnetic moment is 3.4
nuclear magnetons, which is close to the experimental datum of
3.256~\cite{stone2014a}.  This suggests that one can describe $^7$Li
as the bound state of the $\alpha+\mbox{$^3$H}$ system with orbital
angular momentum $l=1$.

The next $3/2^-$ state is at about 9.8~MeV, setting the empirical
breakdown scale. The breakup of the triton at about 6~MeV is an
inelastic channel we are not concerned with. The sum of the charge
radii of the constituent ions is $D\approx 3.4$~fm, and the
theoretical breakdown momentum is $\pi/D\approx 0.91$~fm$^{-1}$,
corresponding to an energy of about 10~MeV. Thus the breakdown energy
is at about 10~MeV. At the momentum $\Lambda_m=0.26$~fm$^{-1}$,
corresponding to an energy of 0.84~MeV, model dependencies become
visible. We note that this energy is smaller than the bound-state
energy, and model dependencies could thus be notable.

\begin{figure}[tb]
  \centering
  \includegraphics[width=\columnwidth]{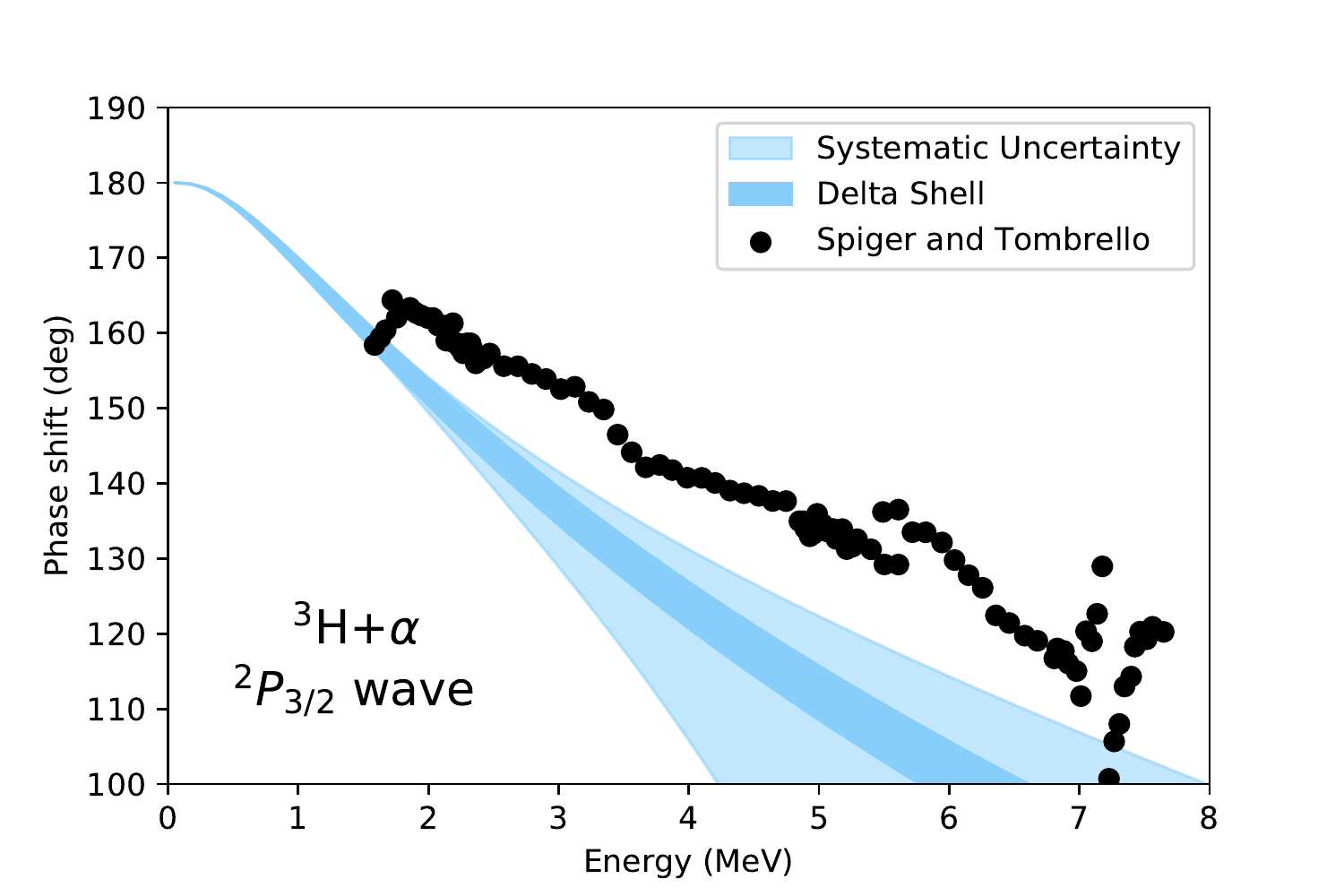}
  \caption{(Color online) Phase shifts of $\mbox{$^3$H}+\alpha$ scattering in the
    $^2P_{3/2}$ partial wave, as a function of the energy in the
    center-of-mass frame. Data taken from Ref.~\cite{spiger1967}.}
  \label{7Ligs}
\end{figure}

We adjust the $\delta$-shell parameters to the $\alpha$-separation
energy and the charge radius ($2.444 \pm 0.042$~fm~\cite{angeli2013})
of the $^7$Li. The resulting phase shifts are shown in
Fig.~\ref{7Ligs} and compared to a phase shift
analysis~\cite{spiger1967}. The agreement is poor. However, the
scatter of the points from the phase shift analysis also suggests that
the uncertainties are significant.

For the $3/2^-$ channel, we compute a scattering volume $a_1=74\pm
8$~fm$^3$, an effective range $r_1=-0.24\pm 0.02$~fm$^{-1}$, and an
ANC $C_1=3.0\pm0.2$~fm$^{-1/2}$. The ab initio computations by
\textcite{doheteraly2016} found a scattering volume of 70~fm$^3$
(which agrees with our result), and their computed charge radius is
close to data.  \textcite{kamouni2007} fit a model to phase shifts and
report effective-range parameters $a_1=72.77$~fm$^3$ and $r_1=
−0.27$~fm$^{-1}$ (which are close to our results); however, the
ground-state energy of the $^3\mbox{H}+\alpha$ system was about twice
as large as the data. However, \textcite{descouvemont2004} found an
ANC of $C_1=3.49$~fm$^{-1/2}$ from an R matrix analysis, while
\textcite{yarmukhamedov2011} computed effective-range parameters
$a_1=58.10 \pm 0.65$~fm$^3$ and $r_1=−0.346 \pm 0.005$~fm$^{-1}$ (with
an ANC of $C_1=3.57\pm 0.15$ from Ref.~\cite{igamov2007}).

We see that there is no consensus yet about low-energy observables for
the $\alpha+\mbox{$^3$H}$ system. However, the simplicity of the
$\delta$-shell potential, its economical use of only two low-energy
data, its agreement with ab initio computations, and its ability to
estimate uncertainties of models make it an attractive potential also
here.

\section{Summary}
\label{summary}

We employed a simple two-parameter model to describe a number of
nuclear light-ion systems that exhibit a separation of scale.
Whenever possible, the model parameters were constrained by the energy
and width of a low-energy resonance or by the energy and charge radius
of a weakly bound state. In those cases, we predicted phase shifts,
effective range parameters and ANCs.  Our analysis of ANCs, charge
radii and resonance widths shows that the inclusion of a finite range
is relevant for systems with strong Coulomb interactions. We also
proposed a way to account for systematic corrections and model
uncertainties. This allowed us to present uncertainty estimates for
the computed observables.  The presented approach provides us with a
constructive criticism of Coulomb halo EFT.  We predicted a charge
radius of $2.88(1)$~fm for the $^{17}$F ground state, taking its
energy and ANC to constrain the model.

The potential model employs two parameters in each partial wave. When
applied to a single partial wave, it is a minimal model whose results
compete well at low energies with traditional Woods-Saxon potential
models or $R$ matrix analyses that employ more parameters. We pointed
out that the $\delta$-shell model practically delivers
model-independent results below a momentum $\Lambda_m$ when it is
adjusted to low-energy data. We also presented simple formulas that
estimate the sizes of effective-range parameters and ANCs based on
energies of low-energy states and charge radii of the involved
ions. Such estimates are useful in the construction of EFTs, and they
seemed to be missing in the literature.

\begin{acknowledgments}
  We are indebted to Lucas Platter for many stimulating and insightful
  discussions. We also thank Hans-Werner Hammer, Sebastian K\"onig,
  Titus Morris, and Daniel Phillips for useful discussions; Chenyi Gu
  for pointing out a numerical problem with Coulomb wave functions,
  and Petr Navr{\'a}til for providing us with data. We thank the
  Institute for Nuclear Theory at the University of Washington for its
  hospitality and the U.S. Department of Energy for partial support
  during the completion of this work.  This work has been supported by
  the U.S. Department of Energy under grant Nos. DE-FG02-96ER40963,
  and DE-SC0016988, and contract DE-AC05-00OR22725 with UT-Battelle,
  LLC (Oak Ridge National Laboratory).
\end{acknowledgments}

\section{Appendix}
\label{appendix}
The Appendix presents some details that could be looked up or are
straightforward (but sometimes tedious) to derive. We present them
here briefly to make the paper self contained.

\subsection{Coulomb wave function}
For the Coulomb wave functions we followed \textcite{gaspard2018}. In
that paper, the analytical properties are emphazised. This is relevant
to us because we call the Coulomb wave functions at real and purely
imaginary arguments. We employed {\sc Mathematica} and {\sc scipy}
special functions in {\sc Python} for our numerical implementation. We
checked for a number of arguments that our implementation agrees with
the precise numerical routines by \textcite{michel2007}.

The regular Coulomb wave function is
\ba
\label{coulF}
F_l(\eta,\rho) = C_l(\eta) \rho^{l+1} e^{i\rho} M(l+1+i\eta,2l+2,-i2\rho) . \nonumber\\
\ea
Here, $\eta =k_c/k$ is the Sommerfeld parameter, and $\rho=kr$. For
bound states with energy $E=-\hbar^2\gamma^2/(2m)$ we have
$k=-i\gamma$, and the arguments $\eta=-ik_c/\gamma$, and $\rho =
i2\gamma r$ are purely imaginary.  In Eq.~(\ref{coulF}), we employed
Kummer's function $M(a,b,z)$, or the confluent hypergeometric function
${_1F_1}(a,b,z)=M(a,b,z)$. The $\eta$-dependent normalization
$C_l(\eta)$, distinct from the ANC by its argument, is
\be
C_l(\eta) = \frac{(2\eta)^l}{(2l+1)!}\sqrt{\frac{2\pi\eta w_l(\eta)}{e^{2\pi\eta}-1}} , 
\ee
with
\be
w_l(\eta) = \prod_{j=0}^l \left(1+{j^2\over \eta^2}\right) .
\ee
The incoming and outgoing Coulomb wave functions are
\ba
H^{\pm}_l(\eta,\rho) =D^{\pm}_l(\eta) \rho^{l+1} e^{\pm i\rho} U(l+1\pm i\eta,2l+2,\mp i2\rho) .\nonumber\\ 
\ea
Here, $U$ denotes Tricomi's function (or the confluent hypergeometric function of
the second kind) and the normalization is
\be
D^\pm_l(\eta) = \mp i2 (-1)^l e^{\pi\eta}\frac{(2l+1)!C_l(\eta)}{\Gamma(l+1\mp i\eta)} .
\ee
The irregular Coulomb wave function is then defined as
\be
G_l(\eta,\rho) = {1\over 2}\left[ H^+_l(\eta,\rho) + H^-_l(\eta,\rho) \right] .
\ee

We are interested in low-energy phenomena and therefor seek
approximations for Coulomb wave functions for $\eta\gg 1$. Following
\cite[Chapter 33.9]{dlmf} we expand the Coulomb wave functions into a
series of modified Bessel functions whose coefficients decrease with
inverse powers of $\eta$. Thus, ($\eta\equiv k_c/k$ and $\rho\equiv k
R$)
\ba
\label{expandBessel}
F_0(\eta,\rho) &=& {C_0(\eta)\over 2\eta} \sum_{n=1}^\infty b_n (2k_cR)^{n\over 2} I_n(2\sqrt{2k_cR}) , \nonumber \\
G_0(\eta,\rho) &=& {2\over \beta_0(\eta)C_0(\eta)} \sum_{n=1}^\infty (-1)^nb_n (2k_cR)^{n\over 2} K_n(2\sqrt{2k_cR}) \nonumber\\.
\ea
Here,
\ba
b_1 &=& 1 , \nonumber\\
b_2 &=& 0 , \nonumber\\
b_3 &=& -{1\over 4\eta^2} , \nonumber\\
b_4 &=& -{1\over 12\eta^2} , 
\ea
and all other $b_n$ are of order ${\cal O}(\eta^{-4})$ or smaller. We have
\be
\beta_0(\eta) = -1 +{\cal O}(\eta^{-4}) .
\ee
Similar expressions exist for nonzero orbital angular momentum.

We also need to know similar approximations for the Coulomb wave
functions for purely imaginary momentum $k=i\gamma$. In the
weak-binding limit $\gamma\to 0$, the regular Coulomb wave function
becomes~\cite[Eq. 13.8.12]{dlmf}
\ba
\label{Ifun}
\lefteqn{F_l\left({k_c\over i \gamma},i\gamma r\right)\approx}\nonumber\\
&& (i\gamma r)^{l+1} C_l(-ik_c/\gamma)
\frac{(2l+1)!}{(2k_cr)^{l+1/2}}I_{2l+1}(2\sqrt{2k_cr}) .
\ea
The Coulomb wave functions $G_l$ and $H^+_l$ are based on Tricomi's function. 
For $\gamma\to 0$ we use (see Ref.~\cite{abramowitz1964})
\ba
\label{Ufun}
\lim_{a\to\infty} U(a,b,z/a) \Gamma(1+a-b) = 2z^{{1-b\over 2}}K_{b-1}(2\sqrt{z}).  \nonumber\\
\ea

We see that Coulomb wave functions are approximated by modified Bessel
functions as the momentum goes to zero. Let us also consider
approximations of the latter.  We have
\ba
\label{small_z}
I_n(z) &\approx& \left({z\over 2}\right)^n\left( {1\over n!} + {z^2\over 4(n+1)!} \right) ,  \nonumber \\
K_n(z) &\approx& {1\over 2} \left({2\over z}\right)^n\left( (n-1)! + {(n-2)! z^2\over 4} \right) ,  \nonumber \\
\ea
valid for $z \ll 1$, see \cite[Chapters~10.25 and 10.31]{dlmf}.
We also have 
\ba
\label{approxBessel}
I_n(z) &\approx& {e^z\over \sqrt{2\pi z}}\left(1-{a_1(n)\over z} \right) , \nonumber\\
K_n(z) &\approx& \sqrt{\pi \over 2z} e^{-z}\left(1+{a_1(n)\over z} \right) , \nonumber\\
\ea
valid for $z\to\infty$, see \cite[Chapter~10.40]{dlmf}.
Here,
\ba
a_1(n) \equiv {4n^2-1\over 8} .
\ea

\subsection{Estimate for the asymptotic normalization coefficient and inter-ion distance}
We want to compute an estimate for the ANC. For the $\delta$-shell
potential, the bound-state wave function can be written as follows.
\ba
u_l(r) = \left\{\begin{array}{ll}
    C_l \frac{W_{-{k_c\over \gamma}, l+{1\over 2}}(2\gamma R)}{F_l\left({k_c\over i \gamma},i\gamma R\right)}F_l\left({k_c\over i \gamma},i\gamma r\right)  &  \mbox{, for $r<R$,}\\
    C_l W_{-{k_c\over \gamma}, l+{1\over 2}}(2\gamma r)  &  \mbox{, for $r>R$.}
    \end{array}\right. \nonumber
\ea
Here, we employed the Whittaker function $W$ (which is proportional to
the outgoing Coulomb wave function for bound states~\cite{dlmf}), and
$C_l$ is the ANC by definition. We have
\be
W_{\kappa,\mu}(z) = e^{-{z\over 2}}z^{\mu+{1\over 2}} U({1/2}+\mu-\kappa,1+2\mu,z)
\ee
The ANC is determined by the normalization condition
\ba
\label{nnorm}
1 &=& \int\limits_0^\infty dr \left|u_l(r)\right|^2 \nonumber\\
&=& C_l^2 \int\limits_R^\infty dr \left(W_{-{k_c\over \gamma},l+{1\over 2}}(2\gamma r)\right)^2 \nonumber\\
&&+ \left|C_l \frac{W_{-{k_c\over \gamma}, l+{1\over 2}}(2\gamma R)}{F_l\left({k_c\over i \gamma},i\gamma R\right)}\right|^2\int\limits_0^R dr
\left|F_l\left({k_c\over i \gamma},i\gamma r\right)\right|^2. 
\ea
To perform the integration, we need to make approximations.  As we are
interested in the case of weak binding, i.e. $\gamma\to 0$, we use
the approximation~(\ref{Ufun}).  Thus,
\ba
W_{-{k_c\over \gamma}, l+{1\over 2}}(2\gamma r) &\approx&
\frac{2(2k_cr)^{-l-{1\over 2}}}{\Gamma(k_c/\gamma-l)} K_{2l+1}\left(2\sqrt{2k_cr}\right) .
\ea
Here, we approximated $e^{-\gamma r}\approx 1$. 
We note that the bound-state momentum enters as the argument of the
$\Gamma$ function. To simplify matters further, we approximate
\be
\Gamma(k_c/\gamma-l)\left({k_c\over \gamma}\right)^{l+1}\approx \Gamma(k_c/\gamma+1), 
\ee
which is correct in leading order when $\gamma\ll k_c$. Then
\ba
\label{Kfun}
W_{-{k_c\over \gamma}, l+{1\over 2}}(2\gamma r) &\approx&
\frac{2\sqrt{2k_c r}}{\Gamma(1+k_c/\gamma)} K_{2l+1}\left(2\sqrt{2k_cr}\right) .
\ea
In the weak-binding limit $\gamma\to 0$, we use the
approximation~(\ref{Ifun}) for the regular Coulomb wave function. The
integral~(\ref{nnorm}) can now be evaluated exactly (e.g. via {\sc
  Mathematica}), but we did not find
the result particularly illuminating. However, for $l=0$ one can then take
the limit $R\to 0$ and finds
\be
\label{ANC-haloEFT}
C_0 \approx \sqrt{6k_c}\Gamma(1+k_c/\gamma) .
\ee
This is the result from leading-order Coulomb halo EFT~\cite{ryberg2016}. 

For further analytical insights we return to Eqs.~(\ref{Kfun}) and
(\ref{Ifun}), and assume $k_cR \gg 1$. This allows us to use the
leading terms of Eqs.~(\ref{expandBessel}).  We change the integration
variable to $z=\sqrt{2k_cr}$ and perform the
integration~(\ref{nnorm}). Keeping only the leading term in $k_cR\gg
1$ yields
\be
C_l \approx {\Gamma(1+k_c/\gamma)\over \sqrt{\pi R}}e^{2\sqrt{2k_cR}} . 
\ee
Replacing $R\to D$ yields the result presented in Table~\ref{tabnaive}.

Similar computations allow us also to give an estimate for the squared
inter-ion distance~(\ref{iid}).  Making the same approximations as in
the computation of the ANC we find (for orbital angular momentum $l=0$)
\ba
\langle r^2\rangle \approx\left\{\begin{array}{ll}
        {9\over 35} k_c^{-2} ,   & \mbox{for $R\to0$, }\\
        R^2 , & \mbox{for $k_c R\gg 1$.}\end{array}\right.
\ea
The results are strikingly different from each other because the wave
function is strongly localized and peaked around $r=R$ in for large
Coulomb momenta. We see in particular that the inter-ion distance does
not depend on the bound-state momentum, and this is in stark contrast
to the case without Coulomb, where $\langle
r^2\rangle\propto\gamma^{-2}$. Replacing $R\to D$ yields the
expressions presented in Table~\ref{tabnaive}.

\subsection{Estimate for the resonance width}
For the $\delta$-shell potential, the resonance width is given in
Eq.~(\ref{width}).

Using the approximation~(\ref{expandBessel}) the inverse width becomes
in leading order of $k_c/\kappa \gg 1$
\ba
{E\over \Gamma} &\approx& {\kappa R\over 3 C_0^2(\eta)} \frac{\sqrt{2k_cR}(I_1K_4-I_4K_1)-3(I_1K_3+I_3K_1)}{I_1^2} \nonumber\\.
\ea
Here, we have suppressed the arguments of the modified Bessel function,
i.e. $I_n \equiv I_n(2\sqrt{2k_cR})$ and $K_n \equiv
K_n(2\sqrt{2k_cR})$.

We consider two cases. For zero-range interactions, we take $R\to 0$ and obtain
\be
{\Gamma\over E}\bigg|_{R\to 0} \approx 24\pi {k_c^2\over \kappa^2} e^{-2\pi {k_c\over\kappa}} .  
\ee

Here, we used the expansions~(\ref{small_z}).  The physically relevant
case $k_cR\gg 1$ is more interesting. We use the
expansions~(\ref{approxBessel}) and find
\ba
   {\Gamma\over E}\bigg|_{k_cR \gg 1} \approx 4 {k_c\over \kappa^2R}e^{4\sqrt{2k_cR}} e^{-2\pi{k_c\over \kappa}} .
\ea
Replacing $R\to D$ yields the results presented in Subsection~\ref{eft}.

\subsection{Estimates for effective-range parameters}
We start from the effective-range parameters given in
Eq.~(\ref{ere}). These expressions contain the strength $\lambda_0$ of
the $\delta$-shell potential. For a resonance with energy
$E=\hbar^2\kappa^2/(2m)$ this parameter fulfills
Eq.~(\ref{lambda_res}). We assume $\kappa\ll k_c$ and use the
approximation~(\ref{expandBessel}), focusing on orbital angular
momentum $l=0$. This yields
\ba
\label{coupling}
(\lambda_0R)^{-1} \approx -2I_1K_1 -\frac{\kappa^2R}{8k_c\sqrt{2k_cR}} , 
\ea
and we have omitted higher-order corrections in $\kappa/k_c$. Here,
and in what follows the modified Bessel functions have arguments
$I_n\equiv I_n(2\sqrt{2k_cR})$ and similar for $K_n$.

We insert the expression~(\ref{coupling}) into the Eq.~(\ref{ere}) for
the $s$-wave scattering length and find
\ba
a_0^{-1} = -\frac{\kappa^2R}{4I_1^2\sqrt{2k_cR}} .
\ea
Again, we consider two approximations. For $2\sqrt{2k_cR}\gg 1$, we
take the leading approximation of Eqs.~(\ref{approxBessel}) and find
\ba
a_0 &=& -(\pi\kappa^2 R)^{-1} e^{4\sqrt{2kc_D}} .
\ea
For $R\to 0$, we take the approximations~(\ref{small_z}) and find
$a_0=-6k_c/\kappa^2$. Replacing $R\to D$ yields the expressions given
in Table~\ref{tabnaive}.

We turn to the effective range of Eq.~(\ref{ere}) and employ the
leading term $(\lambda_0R)^{-1} \approx -2I_1K_1$ from
Eq.~(\ref{coupling}). This yields
\ba
r_0 = {1\over 3k_c }\left( 1 +\frac{2(2kcR)^{3/2}I_2K_1-k_cR}{I_1^2} \right) .
\ea
Again, we consider two approximations. For $2\sqrt{2k_cR}\gg 1$, we
take the Eqs.~(\ref{approxBessel}) and find~\cite{mur1993}
\ba
r_0 = (3k_c)^{-1} -\pi R \, e^{-4\sqrt{2k_cR}} .  
\ea
For $R\to 0$, we employ the approximations~(\ref{small_z}) and find
$r_0= {\cal O}(R))$. Replacing $R\to D$ yields the expressions given
in Table~\ref{tabnaive} and in Eq.~(\ref{effrange}).

\subsection{Derivatives of Coulomb wave functions}
We limit the discussion to orbital angular momentum $l=0$ and positive
energies. For $k\ll k_c$ we find ($z\equiv 2\sqrt{2k_cr}$) from
Eq.~(\ref{expandBessel}) that
\ba
F_0(k_c/k,kr)&\propto& zI_1(z) , \nonumber\\
G_0(k_c/k,kr)&\propto& zK_1(z) .
\ea
Here, we neglected any constants and functions that depend on $k$ and
$k_c$, but not on $r$. We see that only the combination $2k_c r$
enters, and it is clear that a derivative with respect to $r$ will
yield a factor $k_c$ rather than $k$. Taking a derivative becomes
particularly simple for strong Coulomb interactions as $k_cr \gg 1$
practically holds for all distances exceeding 1~fm or so.  We use the
approximations~(\ref{approxBessel}) and find
\ba
F_0(k_c/k,kr)&\propto& \sqrt{z\over 2\pi}e^z , \nonumber\\
G_0(k_c/k,kr)&\propto& \sqrt{\pi z\over 2}e^{-z} .
\ea
Taking the derivative with respect to $r$, and using $z\gg 1$ yields
\ba
{d\over dr}F_0(k_c/k,kr)&\approx& +4k_c F_0(k_c/k,kr), \nonumber\\
{d\over dr}G_0(k_c/k,kr)&\approx& -4k_c G_0(k_c/k,kr).
\ea

\subsection{Square well plus Coulomb}
The potential is
\ba
\label{sq}
V(r) = \left\{
\begin{array}{ll}
  -{\hbar^2 q^2\over 2 m},  & r < R\\
  {Z_1Z_2\alpha\hbar \over r} & r > R .
\end{array}
\right.
\ea
We limit ourselves to $s$ waves. Solutions with positive energy $E=\hbar^2k^2/(2m)$ are
\ba
\lefteqn{u(r) =}\nonumber\\
&&\left\{
\begin{array}{ll}
  \frac{\cos\delta F_0\left({k_c\over k}, kR\right)+\sin\delta G_0\left({k_c\over k}, kR\right)}{\sin p_kR}
  \sin{p_k r},  & r < R\\
  \cos\delta F_0\left({k_c\over k}, kr\right)+\sin\delta G_0\left({k_c\over k}, kr\right), & r > R .
\end{array}
\right.
\ea
Here, $p_k \equiv \sqrt{k^2 +q^2}$. The phase shifts fulfill
\ba
\label{sqPhase}
\lefteqn{\cot\delta =}\nonumber\\
&&\frac{G'_0\left({k_c\over k}, kR\right)\sin{p_kR}-{p\over k}G_0\left({k_c\over k}, kR\right)\cos{p_kR}}
           {F'_0\left({k_c\over k}, kR\right)\sin{p_kR}-{p_k\over k} F_0\left({k_c\over k}, kR\right)\cos{p_kR}}
\ea
Here, we used $F'_0(\eta,z) \equiv {d\over dz} F_0(\eta,z)$ and
similar for the irregular Coulomb wave function. A resonance at energy
$E_\kappa \equiv \hbar^2\kappa^2/(2m)$ fulfills
\ba
\label{sqE}
p_\kappa\cot{p_\kappa R} = \kappa \frac{G'_0\left({k_c\over \kappa}, \kappa R\right)}
{G_0\left({k_c\over \kappa}, \kappa R\right)}
\ea
The resonance width $\Gamma$ fulfills
\ba
\label{sqGamma}
{E_\kappa\over\Gamma} =
        &&{G_0\over 4}\left({q^2 \over p_\kappa^2} G_0' +\kappa\dot{G}'_0 -\kappa {G'_0\dot{G}_0\over G_0}
   +{\kappa R G_0\over \sin^2{p_\kappa R}} \right) \nonumber\\.
\ea
Here, we used $\dot{G}_0(kc/\kappa,\kappa R) \equiv {d\over d\kappa}
G_0(k_c/\kappa,\kappa R)$, and we dropped the arguments for all
Coulomb wave function. For a given resonance energy and width, one can
solve Eqs.~(\ref{sqE}) and (\ref{sqGamma}) for the parameters $(q,R)$
of the potential. Once these are known, the phase shifts result from
Eq.~(\ref{sqPhase}). As the square well can hold an arbitrary number
of bound states, the solutions are not unique. However, low-energy
data such as the $\alpha-\alpha$ phase shifts exhibit sensitivity to
such details only at energies above about 1.7~MeV.


\begin{thebibliography}{77}%
\makeatletter
\providecommand \@ifxundefined [1]{%
 \@ifx{#1\undefined}
}%
\providecommand \@ifnum [1]{%
 \ifnum #1\expandafter \@firstoftwo
 \else \expandafter \@secondoftwo
 \fi
}%
\providecommand \@ifx [1]{%
 \ifx #1\expandafter \@firstoftwo
 \else \expandafter \@secondoftwo
 \fi
}%
\providecommand \natexlab [1]{#1}%
\providecommand \enquote  [1]{``#1''}%
\providecommand \bibnamefont  [1]{#1}%
\providecommand \bibfnamefont [1]{#1}%
\providecommand \citenamefont [1]{#1}%
\providecommand \href@noop [0]{\@secondoftwo}%
\providecommand \href [0]{\begingroup \@sanitize@url \@href}%
\providecommand \@href[1]{\@@startlink{#1}\@@href}%
\providecommand \@@href[1]{\endgroup#1\@@endlink}%
\providecommand \@sanitize@url [0]{\catcode `\\12\catcode `\$12\catcode
  `\&12\catcode `\#12\catcode `\^12\catcode `\_12\catcode `\%12\relax}%
\providecommand \@@startlink[1]{}%
\providecommand \@@endlink[0]{}%
\providecommand \url  [0]{\begingroup\@sanitize@url \@url }%
\providecommand \@url [1]{\endgroup\@href {#1}{\urlprefix }}%
\providecommand \urlprefix  [0]{URL }%
\providecommand \Eprint [0]{\href }%
\providecommand \doibase [0]{http://dx.doi.org/}%
\providecommand \selectlanguage [0]{\@gobble}%
\providecommand \bibinfo  [0]{\@secondoftwo}%
\providecommand \bibfield  [0]{\@secondoftwo}%
\providecommand \translation [1]{[#1]}%
\providecommand \BibitemOpen [0]{}%
\providecommand \bibitemStop [0]{}%
\providecommand \bibitemNoStop [0]{.\EOS\space}%
\providecommand \EOS [0]{\spacefactor3000\relax}%
\providecommand \BibitemShut  [1]{\csname bibitem#1\endcsname}%
\let\auto@bib@innerbib\@empty
\bibitem [{\citenamefont {Adelberger}\ \emph {et~al.}(2011)\citenamefont
  {Adelberger}, \citenamefont {Garc\'{\i}a}, \citenamefont {Robertson},
  \citenamefont {Snover}, \citenamefont {Balantekin}, \citenamefont {Heeger},
  \citenamefont {Ramsey-Musolf}, \citenamefont {Bemmerer}, \citenamefont
  {Junghans}, \citenamefont {Bertulani}, \citenamefont {Chen}, \citenamefont
  {Costantini}, \citenamefont {Prati}, \citenamefont {Couder}, \citenamefont
  {Uberseder}, \citenamefont {Wiescher}, \citenamefont {Cyburt}, \citenamefont
  {Davids}, \citenamefont {Freedman}, \citenamefont {Gai}, \citenamefont
  {Gazit}, \citenamefont {Gialanella}, \citenamefont {Imbriani}, \citenamefont
  {Greife}, \citenamefont {Hass}, \citenamefont {Haxton}, \citenamefont
  {Itahashi}, \citenamefont {Kubodera}, \citenamefont {Langanke}, \citenamefont
  {Leitner}, \citenamefont {Leitner}, \citenamefont {Vetter}, \citenamefont
  {Winslow}, \citenamefont {Marcucci}, \citenamefont {Motobayashi},
  \citenamefont {Mukhamedzhanov}, \citenamefont {Tribble}, \citenamefont
  {Nollett}, \citenamefont {Nunes}, \citenamefont {Park}, \citenamefont
  {Parker}, \citenamefont {Schiavilla}, \citenamefont {Simpson}, \citenamefont
  {Spitaleri}, \citenamefont {Strieder}, \citenamefont {Trautvetter},
  \citenamefont {Suemmerer},\ and\ \citenamefont {Typel}}]{adelberger2011}%
  \BibitemOpen
  \bibfield  {author} {\bibinfo {author} {\bibfnamefont {E.~G.}\ \bibnamefont
  {Adelberger}}, \bibinfo {author} {\bibfnamefont {A.}~\bibnamefont
  {Garc\'{\i}a}}, \bibinfo {author} {\bibfnamefont {R.~G.~Hamish}\ \bibnamefont
  {Robertson}}, \bibinfo {author} {\bibfnamefont {K.~A.}\ \bibnamefont
  {Snover}}, \bibinfo {author} {\bibfnamefont {A.~B.}\ \bibnamefont
  {Balantekin}}, \bibinfo {author} {\bibfnamefont {K.}~\bibnamefont {Heeger}},
  \bibinfo {author} {\bibfnamefont {M.~J.}\ \bibnamefont {Ramsey-Musolf}},
  \bibinfo {author} {\bibfnamefont {D.}~\bibnamefont {Bemmerer}}, \bibinfo
  {author} {\bibfnamefont {A.}~\bibnamefont {Junghans}}, \bibinfo {author}
  {\bibfnamefont {C.~A.}\ \bibnamefont {Bertulani}}, \bibinfo {author}
  {\bibfnamefont {J.-W.}\ \bibnamefont {Chen}}, \bibinfo {author}
  {\bibfnamefont {H.}~\bibnamefont {Costantini}}, \bibinfo {author}
  {\bibfnamefont {P.}~\bibnamefont {Prati}}, \bibinfo {author} {\bibfnamefont
  {M.}~\bibnamefont {Couder}}, \bibinfo {author} {\bibfnamefont
  {E.}~\bibnamefont {Uberseder}}, \bibinfo {author} {\bibfnamefont
  {M.}~\bibnamefont {Wiescher}}, \bibinfo {author} {\bibfnamefont
  {R.}~\bibnamefont {Cyburt}}, \bibinfo {author} {\bibfnamefont
  {B.}~\bibnamefont {Davids}}, \bibinfo {author} {\bibfnamefont {S.~J.}\
  \bibnamefont {Freedman}}, \bibinfo {author} {\bibfnamefont {M.}~\bibnamefont
  {Gai}}, \bibinfo {author} {\bibfnamefont {D.}~\bibnamefont {Gazit}}, \bibinfo
  {author} {\bibfnamefont {L.}~\bibnamefont {Gialanella}}, \bibinfo {author}
  {\bibfnamefont {G.}~\bibnamefont {Imbriani}}, \bibinfo {author}
  {\bibfnamefont {U.}~\bibnamefont {Greife}}, \bibinfo {author} {\bibfnamefont
  {M.}~\bibnamefont {Hass}}, \bibinfo {author} {\bibfnamefont {W.~C.}\
  \bibnamefont {Haxton}}, \bibinfo {author} {\bibfnamefont {T.}~\bibnamefont
  {Itahashi}}, \bibinfo {author} {\bibfnamefont {K.}~\bibnamefont {Kubodera}},
  \bibinfo {author} {\bibfnamefont {K.}~\bibnamefont {Langanke}}, \bibinfo
  {author} {\bibfnamefont {D.}~\bibnamefont {Leitner}}, \bibinfo {author}
  {\bibfnamefont {M.}~\bibnamefont {Leitner}}, \bibinfo {author} {\bibfnamefont
  {P.}~\bibnamefont {Vetter}}, \bibinfo {author} {\bibfnamefont
  {L.}~\bibnamefont {Winslow}}, \bibinfo {author} {\bibfnamefont {L.~E.}\
  \bibnamefont {Marcucci}}, \bibinfo {author} {\bibfnamefont {T.}~\bibnamefont
  {Motobayashi}}, \bibinfo {author} {\bibfnamefont {A.}~\bibnamefont
  {Mukhamedzhanov}}, \bibinfo {author} {\bibfnamefont {R.~E.}\ \bibnamefont
  {Tribble}}, \bibinfo {author} {\bibfnamefont {Kenneth~M.}\ \bibnamefont
  {Nollett}}, \bibinfo {author} {\bibfnamefont {F.~M.}\ \bibnamefont {Nunes}},
  \bibinfo {author} {\bibfnamefont {T.-S.}\ \bibnamefont {Park}}, \bibinfo
  {author} {\bibfnamefont {P.~D.}\ \bibnamefont {Parker}}, \bibinfo {author}
  {\bibfnamefont {R.}~\bibnamefont {Schiavilla}}, \bibinfo {author}
  {\bibfnamefont {E.~C.}\ \bibnamefont {Simpson}}, \bibinfo {author}
  {\bibfnamefont {C.}~\bibnamefont {Spitaleri}}, \bibinfo {author}
  {\bibfnamefont {F.}~\bibnamefont {Strieder}}, \bibinfo {author}
  {\bibfnamefont {H.-P.}\ \bibnamefont {Trautvetter}}, \bibinfo {author}
  {\bibfnamefont {K.}~\bibnamefont {Suemmerer}}, \ and\ \bibinfo {author}
  {\bibfnamefont {S.}~\bibnamefont {Typel}},\ }\bibfield  {title} {\enquote
  {\bibinfo {title} {Solar fusion cross sections. ii. the $pp$ chain and cno
  cycles},}\ }\href {\doibase 10.1103/RevModPhys.83.195} {\bibfield  {journal}
  {\bibinfo  {journal} {Rev. Mod. Phys.}\ }\textbf {\bibinfo {volume} {83}},\
  \bibinfo {pages} {195--245} (\bibinfo {year} {2011})}\BibitemShut {NoStop}%
\bibitem [{\citenamefont {Buck}\ \emph {et~al.}(1975)\citenamefont {Buck},
  \citenamefont {Dover},\ and\ \citenamefont {Vary}}]{buck1975}%
  \BibitemOpen
  \bibfield  {author} {\bibinfo {author} {\bibfnamefont {B.}~\bibnamefont
  {Buck}}, \bibinfo {author} {\bibfnamefont {C.~B.}\ \bibnamefont {Dover}}, \
  and\ \bibinfo {author} {\bibfnamefont {J.~P.}\ \bibnamefont {Vary}},\
  }\bibfield  {title} {\enquote {\bibinfo {title} {Simple potential model for
  cluster states in light nuclei},}\ }\href {\doibase 10.1103/PhysRevC.11.1803}
  {\bibfield  {journal} {\bibinfo  {journal} {Phys. Rev. C}\ }\textbf {\bibinfo
  {volume} {11}},\ \bibinfo {pages} {1803--1821} (\bibinfo {year}
  {1975})}\BibitemShut {NoStop}%
\bibitem [{\citenamefont {Kulik}\ and\ \citenamefont {Mur}(2003)}]{kulik2003}%
  \BibitemOpen
  \bibfield  {author} {\bibinfo {author} {\bibfnamefont {A.~V.}\ \bibnamefont
  {Kulik}}\ and\ \bibinfo {author} {\bibfnamefont {V.~D.}\ \bibnamefont
  {Mur}},\ }\bibfield  {title} {\enquote {\bibinfo {title} {$dt$, $d$$^3$he,
  and $p\alpha$ scattering in the vicinity of the 5he$^*$ and 5li$^*$
  resonances},}\ }\href {\doibase 10.1134/1.1540661} {\bibfield  {journal}
  {\bibinfo  {journal} {Phys. Atom. Nuclei}\ }\textbf {\bibinfo {volume}
  {66}},\ \bibinfo {pages} {87} (\bibinfo {year} {2003})}\BibitemShut {NoStop}%
\bibitem [{\citenamefont {Typel}\ and\ \citenamefont {Baur}(2005)}]{typel2005}%
  \BibitemOpen
  \bibfield  {author} {\bibinfo {author} {\bibfnamefont {S.}~\bibnamefont
  {Typel}}\ and\ \bibinfo {author} {\bibfnamefont {G.}~\bibnamefont {Baur}},\
  }\bibfield  {title} {\enquote {\bibinfo {title} {Electromagnetic strength of
  neutron and proton single-particle halo nuclei},}\ }\href {\doibase
  10.1016/j.nuclphysa.2005.05.145} {\bibfield  {journal} {\bibinfo  {journal}
  {Nuclear Physics A}\ }\textbf {\bibinfo {volume} {759}},\ \bibinfo {pages}
  {247 -- 308} (\bibinfo {year} {2005})}\BibitemShut {NoStop}%
\bibitem [{\citenamefont {Grassi}\ \emph {et~al.}(2017)\citenamefont {Grassi},
  \citenamefont {Mangano}, \citenamefont {Marcucci},\ and\ \citenamefont
  {Pisanti}}]{grassi2017}%
  \BibitemOpen
  \bibfield  {author} {\bibinfo {author} {\bibfnamefont {A.}~\bibnamefont
  {Grassi}}, \bibinfo {author} {\bibfnamefont {G.}~\bibnamefont {Mangano}},
  \bibinfo {author} {\bibfnamefont {L.~E.}\ \bibnamefont {Marcucci}}, \ and\
  \bibinfo {author} {\bibfnamefont {O.}~\bibnamefont {Pisanti}},\ }\bibfield
  {title} {\enquote {\bibinfo {title}
  {$\ensuremath{\alpha}+d\ensuremath{\rightarrow}^{6}\mathrm{Li}+\ensuremath{\gamma}$
  astrophysical $s$ factor and its implications for big bang
  nucleosynthesis},}\ }\href {\doibase 10.1103/PhysRevC.96.045807} {\bibfield
  {journal} {\bibinfo  {journal} {Phys. Rev. C}\ }\textbf {\bibinfo {volume}
  {96}},\ \bibinfo {pages} {045807} (\bibinfo {year} {2017})}\BibitemShut
  {NoStop}%
\bibitem [{\citenamefont {Hamilton}\ \emph {et~al.}(1973)\citenamefont
  {Hamilton}, \citenamefont {{\O}verb{\"o}},\ and\ \citenamefont
  {Tromborg}}]{hamilton1973}%
  \BibitemOpen
  \bibfield  {author} {\bibinfo {author} {\bibfnamefont {J.}~\bibnamefont
  {Hamilton}}, \bibinfo {author} {\bibfnamefont {I.}~\bibnamefont
  {{\O}verb{\"o}}}, \ and\ \bibinfo {author} {\bibfnamefont {B.}~\bibnamefont
  {Tromborg}},\ }\bibfield  {title} {\enquote {\bibinfo {title} {Coulomb
  corrections in non-relativistic scattering},}\ }\href {\doibase
  10.1016/0550-3213(73)90193-4} {\bibfield  {journal} {\bibinfo  {journal}
  {Nuclear Physics B}\ }\textbf {\bibinfo {volume} {60}},\ \bibinfo {pages}
  {443 -- 477} (\bibinfo {year} {1973})}\BibitemShut {NoStop}%
\bibitem [{\citenamefont {Mur}\ and\ \citenamefont {Popov}(1985)}]{mur1985}%
  \BibitemOpen
  \bibfield  {author} {\bibinfo {author} {\bibfnamefont {V.~D.}\ \bibnamefont
  {Mur}}\ and\ \bibinfo {author} {\bibfnamefont {V.~S.}\ \bibnamefont
  {Popov}},\ }\bibfield  {title} {\enquote {\bibinfo {title} {Coulomb problem
  with short-range interaction: Exactly solvable model},}\ }\href {\doibase
  10.1007/BF01017937} {\bibfield  {journal} {\bibinfo  {journal} {Theoretical
  and Mathematical Physics}\ }\textbf {\bibinfo {volume} {65}},\ \bibinfo
  {pages} {1132--1140} (\bibinfo {year} {1985})}\BibitemShut {NoStop}%
\bibitem [{\citenamefont {{Mur}}\ \emph {et~al.}(1993)\citenamefont {{Mur}},
  \citenamefont {{Karnakov}}, \citenamefont {{Pozdnyakov}},\ and\ \citenamefont
  {{Popov}}}]{mur1993}%
  \BibitemOpen
  \bibfield  {author} {\bibinfo {author} {\bibfnamefont {V.~D.}\ \bibnamefont
  {{Mur}}}, \bibinfo {author} {\bibfnamefont {B.~M.}\ \bibnamefont
  {{Karnakov}}}, \bibinfo {author} {\bibfnamefont {S.~G.}\ \bibnamefont
  {{Pozdnyakov}}}, \ and\ \bibinfo {author} {\bibfnamefont {V.~S.}\
  \bibnamefont {{Popov}}},\ }\bibfield  {title} {\enquote {\bibinfo {title}
  {{Low-energy parameters of the dt and d$^{3}$He systems}},}\ }\href
  {https://ui.adsabs.harvard.edu/abs/1993PAN....56..217M} {\bibfield  {journal}
  {\bibinfo  {journal} {Physics of Atomic Nuclei}\ }\textbf {\bibinfo {volume}
  {56}},\ \bibinfo {pages} {217--226} (\bibinfo {year} {1993})}\BibitemShut
  {NoStop}%
\bibitem [{\citenamefont {Sparenberg}\ \emph {et~al.}(2010)\citenamefont
  {Sparenberg}, \citenamefont {Capel},\ and\ \citenamefont
  {Baye}}]{sparenberg2010}%
  \BibitemOpen
  \bibfield  {author} {\bibinfo {author} {\bibfnamefont {Jean-Marc}\
  \bibnamefont {Sparenberg}}, \bibinfo {author} {\bibfnamefont {Pierre}\
  \bibnamefont {Capel}}, \ and\ \bibinfo {author} {\bibfnamefont {Daniel}\
  \bibnamefont {Baye}},\ }\bibfield  {title} {\enquote {\bibinfo {title}
  {Influence of low-energy scattering on loosely bound states},}\ }\href
  {\doibase 10.1103/PhysRevC.81.011601} {\bibfield  {journal} {\bibinfo
  {journal} {Phys. Rev. C}\ }\textbf {\bibinfo {volume} {81}},\ \bibinfo
  {pages} {011601} (\bibinfo {year} {2010})}\BibitemShut {NoStop}%
\bibitem [{\citenamefont {Yarmukhamedov}\ and\ \citenamefont
  {Baye}(2011)}]{yarmukhamedov2011}%
  \BibitemOpen
  \bibfield  {author} {\bibinfo {author} {\bibfnamefont {R.}~\bibnamefont
  {Yarmukhamedov}}\ and\ \bibinfo {author} {\bibfnamefont {D.}~\bibnamefont
  {Baye}},\ }\bibfield  {title} {\enquote {\bibinfo {title} {Connection between
  effective-range expansion and nuclear vertex constant or asymptotic
  normalization coefficient},}\ }\href {\doibase 10.1103/PhysRevC.84.024603}
  {\bibfield  {journal} {\bibinfo  {journal} {Phys. Rev. C}\ }\textbf {\bibinfo
  {volume} {84}},\ \bibinfo {pages} {024603} (\bibinfo {year}
  {2011})}\BibitemShut {NoStop}%
\bibitem [{\citenamefont {Ram\'{\i}rez~Su\'arez}\ and\ \citenamefont
  {Sparenberg}(2017)}]{suarez2017}%
  \BibitemOpen
  \bibfield  {author} {\bibinfo {author} {\bibfnamefont {O.~L.}\ \bibnamefont
  {Ram\'{\i}rez~Su\'arez}}\ and\ \bibinfo {author} {\bibfnamefont {J.-M.}\
  \bibnamefont {Sparenberg}},\ }\bibfield  {title} {\enquote {\bibinfo {title}
  {Phase-shift parametrization and extraction of asymptotic normalization
  constants from elastic-scattering data},}\ }\href {\doibase
  10.1103/PhysRevC.96.034601} {\bibfield  {journal} {\bibinfo  {journal} {Phys.
  Rev. C}\ }\textbf {\bibinfo {volume} {96}},\ \bibinfo {pages} {034601}
  (\bibinfo {year} {2017})}\BibitemShut {NoStop}%
\bibitem [{\citenamefont {Blokhintsev}\ \emph {et~al.}(2018)\citenamefont
  {Blokhintsev}, \citenamefont {Kadyrov}, \citenamefont {Mukhamedzhanov},\ and\
  \citenamefont {Savin}}]{blokhintsev2018}%
  \BibitemOpen
  \bibfield  {author} {\bibinfo {author} {\bibfnamefont {L.~D.}\ \bibnamefont
  {Blokhintsev}}, \bibinfo {author} {\bibfnamefont {A.~S.}\ \bibnamefont
  {Kadyrov}}, \bibinfo {author} {\bibfnamefont {A.~M.}\ \bibnamefont
  {Mukhamedzhanov}}, \ and\ \bibinfo {author} {\bibfnamefont {D.~A.}\
  \bibnamefont {Savin}},\ }\bibfield  {title} {\enquote {\bibinfo {title}
  {Extrapolation of scattering data to the negative-energy region. iii.
  application to the $p\ensuremath{-}^{16}\mathrm{O}$ system},}\ }\href
  {\doibase 10.1103/PhysRevC.98.064610} {\bibfield  {journal} {\bibinfo
  {journal} {Phys. Rev. C}\ }\textbf {\bibinfo {volume} {98}},\ \bibinfo
  {pages} {064610} (\bibinfo {year} {2018})}\BibitemShut {NoStop}%
\bibitem [{\citenamefont {Higa}\ \emph {et~al.}(2008)\citenamefont {Higa},
  \citenamefont {Hammer},\ and\ \citenamefont {van Kolck}}]{higa2008}%
  \BibitemOpen
  \bibfield  {author} {\bibinfo {author} {\bibfnamefont {R.}~\bibnamefont
  {Higa}}, \bibinfo {author} {\bibfnamefont {H.-W.}\ \bibnamefont {Hammer}}, \
  and\ \bibinfo {author} {\bibfnamefont {U.}~\bibnamefont {van Kolck}},\
  }\bibfield  {title} {\enquote {\bibinfo {title} {$\alpha\alpha$ scattering in
  halo effective field theory},}\ }\href {\doibase
  10.1016/j.nuclphysa.2008.06.003} {\bibfield  {journal} {\bibinfo  {journal}
  {Nuclear Physics A}\ }\textbf {\bibinfo {volume} {809}},\ \bibinfo {pages}
  {171 -- 188} (\bibinfo {year} {2008})}\BibitemShut {NoStop}%
\bibitem [{\citenamefont {Ryberg}\ \emph
  {et~al.}(2014{\natexlab{a}})\citenamefont {Ryberg}, \citenamefont
  {Forss\'en}, \citenamefont {Hammer},\ and\ \citenamefont
  {Platter}}]{ryberg2014}%
  \BibitemOpen
  \bibfield  {author} {\bibinfo {author} {\bibfnamefont {Emil}\ \bibnamefont
  {Ryberg}}, \bibinfo {author} {\bibfnamefont {Christian}\ \bibnamefont
  {Forss\'en}}, \bibinfo {author} {\bibfnamefont {H.-W.}\ \bibnamefont
  {Hammer}}, \ and\ \bibinfo {author} {\bibfnamefont {Lucas}\ \bibnamefont
  {Platter}},\ }\bibfield  {title} {\enquote {\bibinfo {title} {Effective field
  theory for proton halo nuclei},}\ }\href {\doibase
  10.1103/PhysRevC.89.014325} {\bibfield  {journal} {\bibinfo  {journal} {Phys.
  Rev. C}\ }\textbf {\bibinfo {volume} {89}},\ \bibinfo {pages} {014325}
  (\bibinfo {year} {2014}{\natexlab{a}})}\BibitemShut {NoStop}%
\bibitem [{\citenamefont {Zhang}\ \emph {et~al.}(2014)\citenamefont {Zhang},
  \citenamefont {Smith}, \citenamefont {Kang},\ and\ \citenamefont
  {Zhao}}]{zhang2014}%
  \BibitemOpen
  \bibfield  {author} {\bibinfo {author} {\bibfnamefont {Shi-Sheng}\
  \bibnamefont {Zhang}}, \bibinfo {author} {\bibfnamefont {M.~S.}\ \bibnamefont
  {Smith}}, \bibinfo {author} {\bibfnamefont {Zhong-Shu}\ \bibnamefont {Kang}},
  \ and\ \bibinfo {author} {\bibfnamefont {Jie}\ \bibnamefont {Zhao}},\
  }\bibfield  {title} {\enquote {\bibinfo {title} {Microscopic self-consistent
  study of neon halos with resonant contributions},}\ }\href {\doibase
  10.1016/j.physletb.2014.01.023} {\bibfield  {journal} {\bibinfo  {journal}
  {Phys. Lett. B}\ }\textbf {\bibinfo {volume} {730}},\ \bibinfo {pages} {30 --
  35} (\bibinfo {year} {2014})}\BibitemShut {NoStop}%
\bibitem [{\citenamefont {Higa}\ \emph {et~al.}(2018)\citenamefont {Higa},
  \citenamefont {Rupak},\ and\ \citenamefont {Vaghani}}]{higa2018}%
  \BibitemOpen
  \bibfield  {author} {\bibinfo {author} {\bibfnamefont {Renato}\ \bibnamefont
  {Higa}}, \bibinfo {author} {\bibfnamefont {Gautam}\ \bibnamefont {Rupak}}, \
  and\ \bibinfo {author} {\bibfnamefont {Akshay}\ \bibnamefont {Vaghani}},\
  }\bibfield  {title} {\enquote {\bibinfo {title} {Radiative
  3he($\alpha$,$\gamma$)7be reaction in halo effective field theory},}\ }\href
  {\doibase 10.1140/epja/i2018-12486-5} {\bibfield  {journal} {\bibinfo
  {journal} {The European Physical Journal A}\ }\textbf {\bibinfo {volume}
  {54}},\ \bibinfo {pages} {89} (\bibinfo {year} {2018})}\BibitemShut {NoStop}%
\bibitem [{\citenamefont {Capel}\ \emph {et~al.}(2018)\citenamefont {Capel},
  \citenamefont {Phillips},\ and\ \citenamefont {Hammer}}]{capel2018}%
  \BibitemOpen
  \bibfield  {author} {\bibinfo {author} {\bibfnamefont {P.}~\bibnamefont
  {Capel}}, \bibinfo {author} {\bibfnamefont {D.~R.}\ \bibnamefont {Phillips}},
  \ and\ \bibinfo {author} {\bibfnamefont {H.-W.}\ \bibnamefont {Hammer}},\
  }\bibfield  {title} {\enquote {\bibinfo {title} {Dissecting reaction
  calculations using halo effective field theory and ab initio input},}\ }\href
  {\doibase 10.1103/PhysRevC.98.034610} {\bibfield  {journal} {\bibinfo
  {journal} {Phys. Rev. C}\ }\textbf {\bibinfo {volume} {98}},\ \bibinfo
  {pages} {034610} (\bibinfo {year} {2018})}\BibitemShut {NoStop}%
\bibitem [{\citenamefont {{Zhang}}\ \emph {et~al.}(2018)\citenamefont
  {{Zhang}}, \citenamefont {{Nollett}},\ and\ \citenamefont
  {{Phillips}}}]{zhang2018}%
  \BibitemOpen
  \bibfield  {author} {\bibinfo {author} {\bibfnamefont {Xilin}\ \bibnamefont
  {{Zhang}}}, \bibinfo {author} {\bibfnamefont {Kenneth~M.}\ \bibnamefont
  {{Nollett}}}, \ and\ \bibinfo {author} {\bibfnamefont {Daniel~R.}\
  \bibnamefont {{Phillips}}},\ }\bibfield  {title} {\enquote {\bibinfo {title}
  {{$S$-factor and scattering parameters from ${}^3$He + ${}^4$He $\rightarrow
  {}^7$Be + $\gamma$ data}},}\ }\href
  {https://ui.adsabs.harvard.edu/abs/2018arXiv181107611Z} {\bibfield  {journal}
  {\bibinfo  {journal} {arXiv e-prints}\ ,\ \bibinfo {eid} {arXiv:1811.07611}}
  (\bibinfo {year} {2018})},\ \Eprint {http://arxiv.org/abs/1811.07611}
  {arXiv:1811.07611 [nucl-th]} \BibitemShut {NoStop}%
\bibitem [{\citenamefont {Neff}(2011)}]{neff2011}%
  \BibitemOpen
  \bibfield  {author} {\bibinfo {author} {\bibfnamefont {Thomas}\ \bibnamefont
  {Neff}},\ }\bibfield  {title} {\enquote {\bibinfo {title} {Microscopic
  calculation of the
  $^{3}\mathrm{He}(\ensuremath{\alpha},\ensuremath{\gamma})^{7}\mathrm{Be}$ and
  $^{3}\mathbf{H}(\ensuremath{\alpha},\ensuremath{\gamma})^{7}\mathrm{Li}$
  capture cross sections using realistic interactions},}\ }\href {\doibase
  10.1103/PhysRevLett.106.042502} {\bibfield  {journal} {\bibinfo  {journal}
  {Phys. Rev. Lett.}\ }\textbf {\bibinfo {volume} {106}},\ \bibinfo {pages}
  {042502} (\bibinfo {year} {2011})}\BibitemShut {NoStop}%
\bibitem [{\citenamefont {Nollett}\ \emph {et~al.}(2001)\citenamefont
  {Nollett}, \citenamefont {Wiringa},\ and\ \citenamefont
  {Schiavilla}}]{nollett2001}%
  \BibitemOpen
  \bibfield  {author} {\bibinfo {author} {\bibfnamefont {K.~M.}\ \bibnamefont
  {Nollett}}, \bibinfo {author} {\bibfnamefont {R.~B.}\ \bibnamefont
  {Wiringa}}, \ and\ \bibinfo {author} {\bibfnamefont {R.}~\bibnamefont
  {Schiavilla}},\ }\bibfield  {title} {\enquote {\bibinfo {title} {Six-body
  calculation of the $\ensuremath{\alpha}$-deuteron radiative capture cross
  section},}\ }\href {\doibase 10.1103/PhysRevC.63.024003} {\bibfield
  {journal} {\bibinfo  {journal} {Phys. Rev. C}\ }\textbf {\bibinfo {volume}
  {63}},\ \bibinfo {pages} {024003} (\bibinfo {year} {2001})}\BibitemShut
  {NoStop}%
\bibitem [{\citenamefont {Quaglioni}\ and\ \citenamefont
  {Navr\'atil}(2008)}]{quaglioni2008}%
  \BibitemOpen
  \bibfield  {author} {\bibinfo {author} {\bibfnamefont {Sofia}\ \bibnamefont
  {Quaglioni}}\ and\ \bibinfo {author} {\bibfnamefont {Petr}\ \bibnamefont
  {Navr\'atil}},\ }\bibfield  {title} {\enquote {\bibinfo {title}
  {\textit{Ab~Initio} many-body calculations of $n\mathrm{-}^{3}\mathrm{H}$,
  $n\mathrm{-}^{4}\mathrm{He}$, $p\mathrm{-}^{3,4}\mathrm{He}$, and
  $n\mathrm{-}^{10}\mathrm{Be}$ scattering},}\ }\href {\doibase
  10.1103/PhysRevLett.101.092501} {\bibfield  {journal} {\bibinfo  {journal}
  {Phys. Rev. Lett.}\ }\textbf {\bibinfo {volume} {101}},\ \bibinfo {pages}
  {092501} (\bibinfo {year} {2008})}\BibitemShut {NoStop}%
\bibitem [{\citenamefont {Hupin}\ \emph {et~al.}(2015)\citenamefont {Hupin},
  \citenamefont {Quaglioni},\ and\ \citenamefont {Navr\'atil}}]{hupin2015}%
  \BibitemOpen
  \bibfield  {author} {\bibinfo {author} {\bibfnamefont {Guillaume}\
  \bibnamefont {Hupin}}, \bibinfo {author} {\bibfnamefont {Sofia}\ \bibnamefont
  {Quaglioni}}, \ and\ \bibinfo {author} {\bibfnamefont {Petr}\ \bibnamefont
  {Navr\'atil}},\ }\bibfield  {title} {\enquote {\bibinfo {title} {Unified
  description of $^{6}\mathrm{Li}$ structure and deuterium-$^{4}\mathrm{He}$
  dynamics with chiral two- and three-nucleon forces},}\ }\href {\doibase
  10.1103/PhysRevLett.114.212502} {\bibfield  {journal} {\bibinfo  {journal}
  {Phys. Rev. Lett.}\ }\textbf {\bibinfo {volume} {114}},\ \bibinfo {pages}
  {212502} (\bibinfo {year} {2015})}\BibitemShut {NoStop}%
\bibitem [{\citenamefont {Dohet-Eraly}\ \emph {et~al.}(2016)\citenamefont
  {Dohet-Eraly}, \citenamefont {Navr{\'a}til}, \citenamefont {Quaglioni},
  \citenamefont {Horiuchi}, \citenamefont {Hupin},\ and\ \citenamefont
  {Raimondi}}]{doheteraly2016}%
  \BibitemOpen
  \bibfield  {author} {\bibinfo {author} {\bibfnamefont {J{\'e}r{\'e}my}\
  \bibnamefont {Dohet-Eraly}}, \bibinfo {author} {\bibfnamefont {Petr}\
  \bibnamefont {Navr{\'a}til}}, \bibinfo {author} {\bibfnamefont {Sofia}\
  \bibnamefont {Quaglioni}}, \bibinfo {author} {\bibfnamefont {Wataru}\
  \bibnamefont {Horiuchi}}, \bibinfo {author} {\bibfnamefont {Guillaume}\
  \bibnamefont {Hupin}}, \ and\ \bibinfo {author} {\bibfnamefont {Francesco}\
  \bibnamefont {Raimondi}},\ }\bibfield  {title} {\enquote {\bibinfo {title}
  {He3$(\alpha,\gamma)$be7 and h3$(\alpha,\gamma)$li7 astrophysical s factors
  from the no-core shell model with continuum},}\ }\href {\doibase
  10.1016/j.physletb.2016.04.021} {\bibfield  {journal} {\bibinfo  {journal}
  {Physics Letters B}\ }\textbf {\bibinfo {volume} {757}},\ \bibinfo {pages}
  {430 -- 436} (\bibinfo {year} {2016})}\BibitemShut {NoStop}%
\bibitem [{\citenamefont {Dubovichenko}\ and\ \citenamefont
  {Dzhazairov-Kakhramanov}(2017)}]{dubovichenko2017}%
  \BibitemOpen
  \bibfield  {author} {\bibinfo {author} {\bibfnamefont {Sergey}\ \bibnamefont
  {Dubovichenko}}\ and\ \bibinfo {author} {\bibfnamefont {Albert}\ \bibnamefont
  {Dzhazairov-Kakhramanov}},\ }\bibfield  {title} {\enquote {\bibinfo {title}
  {Study of the nucleon radiative captures 8li($n,\gamma$)9li,
  9be($p,\gamma$)10b, 10be($n,\gamma$)11be, 10b($p,\gamma$)11c, and
  16o($p,\gamma$)17f at thermal and astrophysical energies},}\ }\href {\doibase
  10.1142/S0218301316300095} {\bibfield  {journal} {\bibinfo  {journal}
  {International Journal of Modern Physics E}\ }\textbf {\bibinfo {volume}
  {26}},\ \bibinfo {pages} {1630009} (\bibinfo {year} {2017})}\BibitemShut
  {NoStop}%
\bibitem [{\citenamefont {Descouvemont}\ \emph {et~al.}(2004)\citenamefont
  {Descouvemont}, \citenamefont {Adahchour}, \citenamefont {Angulo},
  \citenamefont {Coc},\ and\ \citenamefont {Vangioni-Flam}}]{descouvemont2004}%
  \BibitemOpen
  \bibfield  {author} {\bibinfo {author} {\bibfnamefont {Pierre}\ \bibnamefont
  {Descouvemont}}, \bibinfo {author} {\bibfnamefont {Abderrahim}\ \bibnamefont
  {Adahchour}}, \bibinfo {author} {\bibfnamefont {Carmen}\ \bibnamefont
  {Angulo}}, \bibinfo {author} {\bibfnamefont {Alain}\ \bibnamefont {Coc}}, \
  and\ \bibinfo {author} {\bibfnamefont {Elisabeth}\ \bibnamefont
  {Vangioni-Flam}},\ }\bibfield  {title} {\enquote {\bibinfo {title}
  {Compilation and r-matrix analysis of big bang nuclear reaction rates},}\
  }\href {\doibase 10.1016/j.adt.2004.08.001} {\bibfield  {journal} {\bibinfo
  {journal} {Atomic Data and Nuclear Data Tables}\ }\textbf {\bibinfo {volume}
  {88}},\ \bibinfo {pages} {203 -- 236} (\bibinfo {year} {2004})}\BibitemShut
  {NoStop}%
\bibitem [{\citenamefont {Huang}\ \emph {et~al.}(2010)\citenamefont {Huang},
  \citenamefont {Bertulani},\ and\ \citenamefont {Guimar{\~a}es}}]{huang2010}%
  \BibitemOpen
  \bibfield  {author} {\bibinfo {author} {\bibfnamefont {J.~T.}\ \bibnamefont
  {Huang}}, \bibinfo {author} {\bibfnamefont {C.~A.}\ \bibnamefont
  {Bertulani}}, \ and\ \bibinfo {author} {\bibfnamefont {V.}~\bibnamefont
  {Guimar{\~a}es}},\ }\bibfield  {title} {\enquote {\bibinfo {title} {Radiative
  capture of nucleons at astrophysical energies with single-particle states},}\
  }\href {\doibase 10.1016/j.adt.2010.06.004} {\bibfield  {journal} {\bibinfo
  {journal} {Atomic Data and Nuclear Data Tables}\ }\textbf {\bibinfo {volume}
  {96}},\ \bibinfo {pages} {824 -- 847} (\bibinfo {year} {2010})}\BibitemShut
  {NoStop}%
\bibitem [{\citenamefont {Dobaczewski}\ \emph {et~al.}(2014)\citenamefont
  {Dobaczewski}, \citenamefont {Nazarewicz},\ and\ \citenamefont
  {Reinhard}}]{dobaczewski2014}%
  \BibitemOpen
  \bibfield  {author} {\bibinfo {author} {\bibfnamefont {J.}~\bibnamefont
  {Dobaczewski}}, \bibinfo {author} {\bibfnamefont {W.}~\bibnamefont
  {Nazarewicz}}, \ and\ \bibinfo {author} {\bibfnamefont {P.-G.}\ \bibnamefont
  {Reinhard}},\ }\bibfield  {title} {\enquote {\bibinfo {title} {Error
  estimates of theoretical models: a guide},}\ }\href
  {http://stacks.iop.org/0954-3899/41/i=7/a=074001} {\bibfield  {journal}
  {\bibinfo  {journal} {Journal of Physics G: Nuclear and Particle Physics}\
  }\textbf {\bibinfo {volume} {41}},\ \bibinfo {pages} {074001} (\bibinfo
  {year} {2014})}\BibitemShut {NoStop}%
\bibitem [{\citenamefont {{Furnstahl}}\ \emph {et~al.}(2015)\citenamefont
  {{Furnstahl}}, \citenamefont {{Phillips}},\ and\ \citenamefont
  {{Wesolowski}}}]{furnstahl2014c}%
  \BibitemOpen
  \bibfield  {author} {\bibinfo {author} {\bibfnamefont {R.~J.}\ \bibnamefont
  {{Furnstahl}}}, \bibinfo {author} {\bibfnamefont {D.~R.}\ \bibnamefont
  {{Phillips}}}, \ and\ \bibinfo {author} {\bibfnamefont {S.}~\bibnamefont
  {{Wesolowski}}},\ }\bibfield  {title} {\enquote {\bibinfo {title} {A recipe
  for eft uncertainty quantification in nuclear physics},}\ }\href
  {http://stacks.iop.org/0954-3899/42/i=3/a=034028} {\bibfield  {journal}
  {\bibinfo  {journal} {Journal of Physics G: Nuclear and Particle Physics}\
  }\textbf {\bibinfo {volume} {42}},\ \bibinfo {pages} {034028} (\bibinfo
  {year} {2015})}\BibitemShut {NoStop}%
\bibitem [{\citenamefont {Schindler}\ and\ \citenamefont
  {Phillips}(2009)}]{schindler2009}%
  \BibitemOpen
  \bibfield  {author} {\bibinfo {author} {\bibfnamefont {M.~R.}\ \bibnamefont
  {Schindler}}\ and\ \bibinfo {author} {\bibfnamefont {D.~R.}\ \bibnamefont
  {Phillips}},\ }\bibfield  {title} {\enquote {\bibinfo {title} {Bayesian
  methods for parameter estimation in effective field theories},}\ }\href
  {\doibase 10.1016/j.aop.2008.09.003} {\bibfield  {journal} {\bibinfo
  {journal} {Ann. Phys.}\ }\textbf {\bibinfo {volume} {324}},\ \bibinfo {pages}
  {682 -- 708} (\bibinfo {year} {2009})}\BibitemShut {NoStop}%
\bibitem [{\citenamefont {Furnstahl}\ \emph {et~al.}(2015)\citenamefont
  {Furnstahl}, \citenamefont {Hagen}, \citenamefont {Papenbrock},\ and\
  \citenamefont {Wendt}}]{furnstahl2015}%
  \BibitemOpen
  \bibfield  {author} {\bibinfo {author} {\bibfnamefont {R.~J.}\ \bibnamefont
  {Furnstahl}}, \bibinfo {author} {\bibfnamefont {G.}~\bibnamefont {Hagen}},
  \bibinfo {author} {\bibfnamefont {T.}~\bibnamefont {Papenbrock}}, \ and\
  \bibinfo {author} {\bibfnamefont {K.~A.}\ \bibnamefont {Wendt}},\ }\bibfield
  {title} {\enquote {\bibinfo {title} {Infrared extrapolations for atomic
  nuclei},}\ }\href {\doibase 10.1088/0954-3899/42/3/034032} {\bibfield
  {journal} {\bibinfo  {journal} {Journal of Physics G: Nuclear and Particle
  Physics}\ }\textbf {\bibinfo {volume} {42}},\ \bibinfo {pages} {034032}
  (\bibinfo {year} {2015})}\BibitemShut {NoStop}%
\bibitem [{\citenamefont {Coello~P\'erez}\ and\ \citenamefont
  {Papenbrock}(2015)}]{coelloperez2015b}%
  \BibitemOpen
  \bibfield  {author} {\bibinfo {author} {\bibfnamefont {E.~A.}\ \bibnamefont
  {Coello~P\'erez}}\ and\ \bibinfo {author} {\bibfnamefont {T.}~\bibnamefont
  {Papenbrock}},\ }\bibfield  {title} {\enquote {\bibinfo {title} {Effective
  field theory for nuclear vibrations with quantified uncertainties},}\ }\href
  {\doibase 10.1103/PhysRevC.92.064309} {\bibfield  {journal} {\bibinfo
  {journal} {Phys. Rev. C}\ }\textbf {\bibinfo {volume} {92}},\ \bibinfo
  {pages} {064309} (\bibinfo {year} {2015})}\BibitemShut {NoStop}%
\bibitem [{\citenamefont {Carlsson}\ \emph {et~al.}(2016)\citenamefont
  {Carlsson}, \citenamefont {Ekstr\"om}, \citenamefont {Forss\'en},
  \citenamefont {Str\"omberg}, \citenamefont {Jansen}, \citenamefont {Lilja},
  \citenamefont {Lindby}, \citenamefont {Mattsson},\ and\ \citenamefont
  {Wendt}}]{carlsson2016}%
  \BibitemOpen
  \bibfield  {author} {\bibinfo {author} {\bibfnamefont {B.~D.}\ \bibnamefont
  {Carlsson}}, \bibinfo {author} {\bibfnamefont {A.}~\bibnamefont {Ekstr\"om}},
  \bibinfo {author} {\bibfnamefont {C.}~\bibnamefont {Forss\'en}}, \bibinfo
  {author} {\bibfnamefont {D.~Fahlin}\ \bibnamefont {Str\"omberg}}, \bibinfo
  {author} {\bibfnamefont {G.~R.}\ \bibnamefont {Jansen}}, \bibinfo {author}
  {\bibfnamefont {O.}~\bibnamefont {Lilja}}, \bibinfo {author} {\bibfnamefont
  {M.}~\bibnamefont {Lindby}}, \bibinfo {author} {\bibfnamefont {B.~A.}\
  \bibnamefont {Mattsson}}, \ and\ \bibinfo {author} {\bibfnamefont {K.~A.}\
  \bibnamefont {Wendt}},\ }\bibfield  {title} {\enquote {\bibinfo {title}
  {Uncertainty analysis and order-by-order optimization of chiral nuclear
  interactions},}\ }\href {\doibase 10.1103/PhysRevX.6.011019} {\bibfield
  {journal} {\bibinfo  {journal} {Phys. Rev. X}\ }\textbf {\bibinfo {volume}
  {6}},\ \bibinfo {pages} {011019} (\bibinfo {year} {2016})}\BibitemShut
  {NoStop}%
\bibitem [{\citenamefont {{Bedaque}}\ and\ \citenamefont {{van
  Kolck}}(2002)}]{bedaque2002}%
  \BibitemOpen
  \bibfield  {author} {\bibinfo {author} {\bibfnamefont {P.~F.}\ \bibnamefont
  {{Bedaque}}}\ and\ \bibinfo {author} {\bibfnamefont {U.}~\bibnamefont {{van
  Kolck}}},\ }\bibfield  {title} {\enquote {\bibinfo {title} {{Effective field
  theory for few-nucleon systems}},}\ }\href {\doibase
  10.1146/annurev.nucl.52.050102.090637} {\bibfield  {journal} {\bibinfo
  {journal} {Annual Review of Nuclear and Particle Science}\ }\textbf {\bibinfo
  {volume} {52}},\ \bibinfo {pages} {339--396} (\bibinfo {year} {2002})},\
  \Eprint {http://arxiv.org/abs/nucl-th/0203055} {nucl-th/0203055} \BibitemShut
  {NoStop}%
\bibitem [{\citenamefont {Bertulani}\ \emph {et~al.}(2002)\citenamefont
  {Bertulani}, \citenamefont {Hammer},\ and\ \citenamefont {van
  Kolck}}]{bertulani2002}%
  \BibitemOpen
  \bibfield  {author} {\bibinfo {author} {\bibfnamefont {C.~A.}\ \bibnamefont
  {Bertulani}}, \bibinfo {author} {\bibfnamefont {H.-W.}\ \bibnamefont
  {Hammer}}, \ and\ \bibinfo {author} {\bibfnamefont {U.}~\bibnamefont {van
  Kolck}},\ }\bibfield  {title} {\enquote {\bibinfo {title} {Effective field
  theory for halo nuclei: shallow p-wave states},}\ }\href {\doibase
  10.1016/S0375-9474(02)01270-8} {\bibfield  {journal} {\bibinfo  {journal}
  {Nucl. Phys. A}\ }\textbf {\bibinfo {volume} {712}},\ \bibinfo {pages} {37 --
  58} (\bibinfo {year} {2002})}\BibitemShut {NoStop}%
\bibitem [{\citenamefont {Hammer}\ \emph {et~al.}(2017)\citenamefont {Hammer},
  \citenamefont {Ji},\ and\ \citenamefont {Phillips}}]{hammer2017}%
  \BibitemOpen
  \bibfield  {author} {\bibinfo {author} {\bibfnamefont {H.-W.}\ \bibnamefont
  {Hammer}}, \bibinfo {author} {\bibfnamefont {C.}~\bibnamefont {Ji}}, \ and\
  \bibinfo {author} {\bibfnamefont {D.~R.}\ \bibnamefont {Phillips}},\
  }\bibfield  {title} {\enquote {\bibinfo {title} {Effective field theory
  description of halo nuclei},}\ }\href {\doibase 10.1088/1361-6471/aa83db}
  {\bibfield  {journal} {\bibinfo  {journal} {Journal of Physics G: Nuclear and
  Particle Physics}\ }\textbf {\bibinfo {volume} {44}},\ \bibinfo {pages}
  {103002} (\bibinfo {year} {2017})}\BibitemShut {NoStop}%
\bibitem [{\citenamefont {Papenbrock}()}]{papenbrock2019}%
  \BibitemOpen
  \bibfield  {author} {\bibinfo {author} {\bibfnamefont {T.}~\bibnamefont
  {Papenbrock}},\ }\href@noop {} {}\bibinfo {howpublished} {Talk at the
  workshop {\it Progress in Ab Initio Techniques in Nuclear Physics}, TRIUMF,
  Vancouver, B.C. (February 2019)}\BibitemShut {NoStop}%
\bibitem [{\citenamefont {{Schmickler}}\ \emph {et~al.}(2019)\citenamefont
  {{Schmickler}}, \citenamefont {{Hammer}},\ and\ \citenamefont
  {{Volosniev}}}]{schmickler2019}%
  \BibitemOpen
  \bibfield  {author} {\bibinfo {author} {\bibfnamefont {C.~H.}\ \bibnamefont
  {{Schmickler}}}, \bibinfo {author} {\bibfnamefont {H.~W.}\ \bibnamefont
  {{Hammer}}}, \ and\ \bibinfo {author} {\bibfnamefont {A.~G.}\ \bibnamefont
  {{Volosniev}}},\ }\bibfield  {title} {\enquote {\bibinfo {title} {{Universal
  physics of bound states of a few charged particles}},}\ }\href@noop {}
  {\bibfield  {journal} {\bibinfo  {journal} {arXiv e-prints}\ ,\ \bibinfo
  {eid} {arXiv:1904.00913}} (\bibinfo {year} {2019})},\ \Eprint
  {http://arxiv.org/abs/1904.00913} {arXiv:1904.00913 [nucl-th]} \BibitemShut
  {NoStop}%
\bibitem [{\citenamefont {Buck}\ and\ \citenamefont {Pilt}(1977)}]{buck1977}%
  \BibitemOpen
  \bibfield  {author} {\bibinfo {author} {\bibfnamefont {B.}~\bibnamefont
  {Buck}}\ and\ \bibinfo {author} {\bibfnamefont {A.~A.}\ \bibnamefont
  {Pilt}},\ }\bibfield  {title} {\enquote {\bibinfo {title} {Alpha-particle and
  triton cluster states in 19f},}\ }\href {\doibase
  10.1016/0375-9474(77)90300-1} {\bibfield  {journal} {\bibinfo  {journal}
  {Nuclear Physics A}\ }\textbf {\bibinfo {volume} {280}},\ \bibinfo {pages}
  {133 -- 160} (\bibinfo {year} {1977})}\BibitemShut {NoStop}%
\bibitem [{\citenamefont {Ryberg}\ \emph {et~al.}(2019)\citenamefont {Ryberg},
  \citenamefont {Forss{\'e}n}, \citenamefont {Phillips},\ and\ \citenamefont
  {van Kolck}}]{ryberg2019}%
  \BibitemOpen
  \bibfield  {author} {\bibinfo {author} {\bibfnamefont {E.}~\bibnamefont
  {Ryberg}}, \bibinfo {author} {\bibfnamefont {C.}~\bibnamefont {Forss{\'e}n}},
  \bibinfo {author} {\bibfnamefont {D.~R.}\ \bibnamefont {Phillips}}, \ and\
  \bibinfo {author} {\bibfnamefont {U.}~\bibnamefont {van Kolck}},\ }\bibfield
  {title} {\enquote {\bibinfo {title} {Finite-size effects in heavy halo nuclei
  from effective field theory},}\ }\href
  {https://ui.adsabs.harvard.edu/abs/2019arXiv190501107R} {\bibfield  {journal}
  {\bibinfo  {journal} {arXiv e-prints}\ ,\ \bibinfo {eid} {arXiv:1905.01107}}
  (\bibinfo {year} {2019})},\ \Eprint {http://arxiv.org/abs/1905.01107}
  {arXiv:1905.01107 [nucl-th]} \BibitemShut {NoStop}%
\bibitem [{\citenamefont {K{\"o}nig}\ \emph {et~al.}(2013)\citenamefont
  {K{\"o}nig}, \citenamefont {Lee},\ and\ \citenamefont {Hammer}}]{konig2013}%
  \BibitemOpen
  \bibfield  {author} {\bibinfo {author} {\bibfnamefont {Sebastian}\
  \bibnamefont {K{\"o}nig}}, \bibinfo {author} {\bibfnamefont {Dean}\
  \bibnamefont {Lee}}, \ and\ \bibinfo {author} {\bibfnamefont {H.-W.}\
  \bibnamefont {Hammer}},\ }\bibfield  {title} {\enquote {\bibinfo {title}
  {Causality constraints for charged particles},}\ }\href {\doibase
  10.1088/0954-3899/40/4/045106} {\bibfield  {journal} {\bibinfo  {journal}
  {Journal of Physics G: Nuclear and Particle Physics}\ }\textbf {\bibinfo
  {volume} {40}},\ \bibinfo {pages} {045106} (\bibinfo {year}
  {2013})}\BibitemShut {NoStop}%
\bibitem [{\citenamefont {Gagliardi}\ \emph {et~al.}(1999)\citenamefont
  {Gagliardi}, \citenamefont {Tribble}, \citenamefont {Azhari}, \citenamefont
  {Clark}, \citenamefont {Lui}, \citenamefont {Mukhamedzhanov}, \citenamefont
  {Sattarov}, \citenamefont {Trache}, \citenamefont {Burjan}, \citenamefont
  {Cejpek}, \citenamefont {Kroha}, \citenamefont {Pisko\ifmmode~\check{r}\else
  \v{r}\fi{}},\ and\ \citenamefont {Vincour}}]{gagliardi1999}%
  \BibitemOpen
  \bibfield  {author} {\bibinfo {author} {\bibfnamefont {C.~A.}\ \bibnamefont
  {Gagliardi}}, \bibinfo {author} {\bibfnamefont {R.~E.}\ \bibnamefont
  {Tribble}}, \bibinfo {author} {\bibfnamefont {A.}~\bibnamefont {Azhari}},
  \bibinfo {author} {\bibfnamefont {H.~L.}\ \bibnamefont {Clark}}, \bibinfo
  {author} {\bibfnamefont {Y.-W.}\ \bibnamefont {Lui}}, \bibinfo {author}
  {\bibfnamefont {A.~M.}\ \bibnamefont {Mukhamedzhanov}}, \bibinfo {author}
  {\bibfnamefont {A.}~\bibnamefont {Sattarov}}, \bibinfo {author}
  {\bibfnamefont {L.}~\bibnamefont {Trache}}, \bibinfo {author} {\bibfnamefont
  {V.}~\bibnamefont {Burjan}}, \bibinfo {author} {\bibfnamefont
  {J.}~\bibnamefont {Cejpek}}, \bibinfo {author} {\bibfnamefont
  {V.}~\bibnamefont {Kroha}}, \bibinfo {author} {\bibfnamefont {\ifmmode
  \check{S}\else~\v{S}\fi{}.}\ \bibnamefont {Pisko\ifmmode~\check{r}\else
  \v{r}\fi{}}}, \ and\ \bibinfo {author} {\bibfnamefont {J.}~\bibnamefont
  {Vincour}},\ }\bibfield  {title} {\enquote {\bibinfo {title} {Tests of
  transfer reaction determinations of astrophysical s factors},}\ }\href
  {\doibase 10.1103/PhysRevC.59.1149} {\bibfield  {journal} {\bibinfo
  {journal} {Phys. Rev. C}\ }\textbf {\bibinfo {volume} {59}},\ \bibinfo
  {pages} {1149--1153} (\bibinfo {year} {1999})}\BibitemShut {NoStop}%
\bibitem [{\citenamefont {Artemov}\ \emph {et~al.}(2009)\citenamefont
  {Artemov}, \citenamefont {Igamov}, \citenamefont {Tursunmakhatov},\ and\
  \citenamefont {Yarmukhamedov}}]{artemov2009}%
  \BibitemOpen
  \bibfield  {author} {\bibinfo {author} {\bibfnamefont {S.~V.}\ \bibnamefont
  {Artemov}}, \bibinfo {author} {\bibfnamefont {S.~B.}\ \bibnamefont {Igamov}},
  \bibinfo {author} {\bibfnamefont {K.~I.}\ \bibnamefont {Tursunmakhatov}}, \
  and\ \bibinfo {author} {\bibfnamefont {R.}~\bibnamefont {Yarmukhamedov}},\
  }\bibfield  {title} {\enquote {\bibinfo {title} {Determination of nuclear
  vertex constants (asymptotic normalization coefficients) for the virtual
  decays 3he $\rightarrow$ d + p and 17f $\rightarrow$ 16o + p and their use
  for extrapolating astrophysical s-factors of the radiative proton capture by
  the deuteron and the 16o nucleus at very low energies},}\ }\href {\doibase
  10.3103/S1062873809020075} {\bibfield  {journal} {\bibinfo  {journal}
  {Bulletin of the Russian Academy of Sciences: Physics}\ }\textbf {\bibinfo
  {volume} {73}},\ \bibinfo {pages} {165--170} (\bibinfo {year}
  {2009})}\BibitemShut {NoStop}%
\bibitem [{\citenamefont {Breit}\ and\ \citenamefont
  {Bouricius}(1948)}]{breit1948}%
  \BibitemOpen
  \bibfield  {author} {\bibinfo {author} {\bibfnamefont {G.}~\bibnamefont
  {Breit}}\ and\ \bibinfo {author} {\bibfnamefont {W.~G.}\ \bibnamefont
  {Bouricius}},\ }\bibfield  {title} {\enquote {\bibinfo {title} {A boundary
  value condition for proton-proton scattering},}\ }\href {\doibase
  10.1103/PhysRev.74.1546} {\bibfield  {journal} {\bibinfo  {journal} {Phys.
  Rev.}\ }\textbf {\bibinfo {volume} {74}},\ \bibinfo {pages} {1546--1547}
  (\bibinfo {year} {1948})}\BibitemShut {NoStop}%
\bibitem [{\citenamefont {Kok}\ \emph {et~al.}(1982)\citenamefont {Kok},
  \citenamefont {de~Maag}, \citenamefont {Brouwer},\ and\ \citenamefont {van
  Haeringen}}]{kok1982}%
  \BibitemOpen
  \bibfield  {author} {\bibinfo {author} {\bibfnamefont {L.~P.}\ \bibnamefont
  {Kok}}, \bibinfo {author} {\bibfnamefont {J.~W.}\ \bibnamefont {de~Maag}},
  \bibinfo {author} {\bibfnamefont {H.~H.}\ \bibnamefont {Brouwer}}, \ and\
  \bibinfo {author} {\bibfnamefont {H.}~\bibnamefont {van Haeringen}},\
  }\bibfield  {title} {\enquote {\bibinfo {title} {Formulas for the
  $\ensuremath{\delta}$-shell-plus-coulomb potential for all partial waves},}\
  }\href {\doibase 10.1103/PhysRevC.26.2381} {\bibfield  {journal} {\bibinfo
  {journal} {Phys. Rev. C}\ }\textbf {\bibinfo {volume} {26}},\ \bibinfo
  {pages} {2381--2396} (\bibinfo {year} {1982})}\BibitemShut {NoStop}%
\bibitem [{\citenamefont {Gaspard}\ and\ \citenamefont
  {Sparenberg}(2018)}]{gaspard2018}%
  \BibitemOpen
  \bibfield  {author} {\bibinfo {author} {\bibfnamefont {David}\ \bibnamefont
  {Gaspard}}\ and\ \bibinfo {author} {\bibfnamefont {Jean-Marc}\ \bibnamefont
  {Sparenberg}},\ }\bibfield  {title} {\enquote {\bibinfo {title}
  {Effective-range function methods for charged particle collisions},}\ }\href
  {\doibase 10.1103/PhysRevC.97.044003} {\bibfield  {journal} {\bibinfo
  {journal} {Phys. Rev. C}\ }\textbf {\bibinfo {volume} {97}},\ \bibinfo
  {pages} {044003} (\bibinfo {year} {2018})}\BibitemShut {NoStop}%
\bibitem [{\citenamefont {Wigner}(1955)}]{wigner1955}%
  \BibitemOpen
  \bibfield  {author} {\bibinfo {author} {\bibfnamefont {Eugene~P.}\
  \bibnamefont {Wigner}},\ }\bibfield  {title} {\enquote {\bibinfo {title}
  {Lower limit for the energy derivative of the scattering phase shift},}\
  }\href {\doibase 10.1103/PhysRev.98.145} {\bibfield  {journal} {\bibinfo
  {journal} {Phys. Rev.}\ }\textbf {\bibinfo {volume} {98}},\ \bibinfo {pages}
  {145--147} (\bibinfo {year} {1955})}\BibitemShut {NoStop}%
\bibitem [{\citenamefont {{Garcia Ruiz et. al}}(2016)}]{garciaruiz2016c}%
  \BibitemOpen
  \bibfield  {author} {\bibinfo {author} {\bibfnamefont {R.~F.}\ \bibnamefont
  {{Garcia Ruiz et. al}}},\ }\href@noop {} {\emph {\bibinfo {title} {Towards
  laser spectroscopy of exotic fluorine isotopes}}},\ \bibinfo {type} {Proposal
  INTC-I-171}\ \bibinfo {number} {CERN-INTC-2016-037}\ (\bibinfo  {institution}
  {CERN},\ \bibinfo {year} {2016})\BibitemShut {NoStop}%
\bibitem [{\citenamefont {Sonzogni}(2019)}]{nndc}%
  \BibitemOpen
  \bibfield  {author} {\bibinfo {author} {\bibfnamefont {Alejandro}\
  \bibnamefont {Sonzogni}},\ }\href {https://www.nndc.bnl.gov/} {\emph
  {\bibinfo {title} {NuDat2.7}}},\ \bibinfo {type} {Tech. Rep.}\ (\bibinfo
  {institution} {National Nuclear Data Center (NNDC)},\ \bibinfo {address}
  {Brookhaven National Laboratory},\ \bibinfo {year} {2019})\BibitemShut
  {NoStop}%
\bibitem [{\citenamefont {Heydenburg}\ and\ \citenamefont
  {Temmer}(1956)}]{heydenburg1956}%
  \BibitemOpen
  \bibfield  {author} {\bibinfo {author} {\bibfnamefont {N.~P.}\ \bibnamefont
  {Heydenburg}}\ and\ \bibinfo {author} {\bibfnamefont {G.~M.}\ \bibnamefont
  {Temmer}},\ }\bibfield  {title} {\enquote {\bibinfo {title} {Alpha-alpha
  scattering at low energies},}\ }\href {\doibase 10.1103/PhysRev.104.123}
  {\bibfield  {journal} {\bibinfo  {journal} {Phys. Rev.}\ }\textbf {\bibinfo
  {volume} {104}},\ \bibinfo {pages} {123--134} (\bibinfo {year}
  {1956})}\BibitemShut {NoStop}%
\bibitem [{\citenamefont {Afzal}\ \emph {et~al.}(1969)\citenamefont {Afzal},
  \citenamefont {Ahmad},\ and\ \citenamefont {Ali}}]{afzal1969}%
  \BibitemOpen
  \bibfield  {author} {\bibinfo {author} {\bibfnamefont {S.~A.}\ \bibnamefont
  {Afzal}}, \bibinfo {author} {\bibfnamefont {A.~A.~Z.}\ \bibnamefont {Ahmad}},
  \ and\ \bibinfo {author} {\bibfnamefont {S.}~\bibnamefont {Ali}},\ }\bibfield
   {title} {\enquote {\bibinfo {title} {Systematic survey of the $\alpha -
  \alpha$ interaction},}\ }\href {\doibase 10.1103/RevModPhys.41.247}
  {\bibfield  {journal} {\bibinfo  {journal} {Rev. Mod. Phys.}\ }\textbf
  {\bibinfo {volume} {41}},\ \bibinfo {pages} {247--273} (\bibinfo {year}
  {1969})}\BibitemShut {NoStop}%
\bibitem [{\citenamefont {Rasche}(1967)}]{rasche1967}%
  \BibitemOpen
  \bibfield  {author} {\bibinfo {author} {\bibfnamefont {G.}~\bibnamefont
  {Rasche}},\ }\bibfield  {title} {\enquote {\bibinfo {title} {Effective range
  analysis of s- and d-wave $\alpha-\alpha$ scattering},}\ }\href {\doibase
  10.1016/0375-9474(67)90005-X} {\bibfield  {journal} {\bibinfo  {journal}
  {Nuclear Physics A}\ }\textbf {\bibinfo {volume} {94}},\ \bibinfo {pages}
  {301 -- 312} (\bibinfo {year} {1967})}\BibitemShut {NoStop}%
\bibitem [{\citenamefont {Kamouni}\ and\ \citenamefont
  {Baye}(2007)}]{kamouni2007}%
  \BibitemOpen
  \bibfield  {author} {\bibinfo {author} {\bibfnamefont {R.}~\bibnamefont
  {Kamouni}}\ and\ \bibinfo {author} {\bibfnamefont {D.}~\bibnamefont {Baye}},\
  }\bibfield  {title} {\enquote {\bibinfo {title} {Scattering length and
  effective range for collisions between light ions within a microscopic
  model},}\ }\href {\doibase 10.1016/j.nuclphysa.2007.04.009} {\bibfield
  {journal} {\bibinfo  {journal} {Nuclear Physics A}\ }\textbf {\bibinfo
  {volume} {791}},\ \bibinfo {pages} {68 -- 83} (\bibinfo {year}
  {2007})}\BibitemShut {NoStop}%
\bibitem [{\citenamefont {Elhatisari}\ \emph {et~al.}(2015)\citenamefont
  {Elhatisari}, \citenamefont {Lee}, \citenamefont {Rupak}, \citenamefont
  {Epelbaum}, \citenamefont {Krebs}, \citenamefont {L{\"a}hde}, \citenamefont
  {Luu},\ and\ \citenamefont {Mei{\ss}ner}}]{elhatisari2015}%
  \BibitemOpen
  \bibfield  {author} {\bibinfo {author} {\bibfnamefont {S.}~\bibnamefont
  {Elhatisari}}, \bibinfo {author} {\bibfnamefont {D.}~\bibnamefont {Lee}},
  \bibinfo {author} {\bibfnamefont {G.}~\bibnamefont {Rupak}}, \bibinfo
  {author} {\bibfnamefont {E.}~\bibnamefont {Epelbaum}}, \bibinfo {author}
  {\bibfnamefont {H.}~\bibnamefont {Krebs}}, \bibinfo {author} {\bibfnamefont
  {T.~A.}\ \bibnamefont {L{\"a}hde}}, \bibinfo {author} {\bibfnamefont
  {T.}~\bibnamefont {Luu}}, \ and\ \bibinfo {author} {\bibfnamefont {U.-G.}\
  \bibnamefont {Mei{\ss}ner}},\ }\bibfield  {title} {\enquote {\bibinfo {title}
  {Ab initio alpha--alpha scattering},}\ }\href {\doibase 10.1038/nature16067}
  {\bibfield  {journal} {\bibinfo  {journal} {Nature}\ }\textbf {\bibinfo
  {volume} {528}},\ \bibinfo {pages} {111--114} (\bibinfo {year}
  {2015})}\BibitemShut {NoStop}%
\bibitem [{\citenamefont {Morlock}\ \emph {et~al.}(1997)\citenamefont
  {Morlock}, \citenamefont {Kunz}, \citenamefont {Mayer}, \citenamefont
  {Jaeger}, \citenamefont {M\"uller}, \citenamefont {Hammer}, \citenamefont
  {Mohr}, \citenamefont {Oberhummer}, \citenamefont {Staudt},\ and\
  \citenamefont {K\"olle}}]{morlock1997}%
  \BibitemOpen
  \bibfield  {author} {\bibinfo {author} {\bibfnamefont {R.}~\bibnamefont
  {Morlock}}, \bibinfo {author} {\bibfnamefont {R.}~\bibnamefont {Kunz}},
  \bibinfo {author} {\bibfnamefont {A.}~\bibnamefont {Mayer}}, \bibinfo
  {author} {\bibfnamefont {M.}~\bibnamefont {Jaeger}}, \bibinfo {author}
  {\bibfnamefont {A.}~\bibnamefont {M\"uller}}, \bibinfo {author}
  {\bibfnamefont {J.~W.}\ \bibnamefont {Hammer}}, \bibinfo {author}
  {\bibfnamefont {P.}~\bibnamefont {Mohr}}, \bibinfo {author} {\bibfnamefont
  {H.}~\bibnamefont {Oberhummer}}, \bibinfo {author} {\bibfnamefont
  {G.}~\bibnamefont {Staudt}}, \ and\ \bibinfo {author} {\bibfnamefont
  {V.}~\bibnamefont {K\"olle}},\ }\bibfield  {title} {\enquote {\bibinfo
  {title} {Halo properties of the first $1/{2}^{+}$ state in $^{17}f$ from the
  $^{16}o(\mathit{p},\ensuremath{\gamma})^{17}f$ reaction},}\ }\href {\doibase
  10.1103/PhysRevLett.79.3837} {\bibfield  {journal} {\bibinfo  {journal}
  {Phys. Rev. Lett.}\ }\textbf {\bibinfo {volume} {79}},\ \bibinfo {pages}
  {3837--3840} (\bibinfo {year} {1997})}\BibitemShut {NoStop}%
\bibitem [{\citenamefont {Dubovichenko}\ \emph {et~al.}(2017)\citenamefont
  {Dubovichenko}, \citenamefont {Burtebayev}, \citenamefont
  {Dzhazairov-Kakhramanov}, \citenamefont {Zazulin}, \citenamefont
  {Kerimkulov}, \citenamefont {Nassurlla}, \citenamefont {Omarov},
  \citenamefont {Tkachenko}, \citenamefont {Shmygaleva}, \citenamefont
  {Kliczewski},\ and\ \citenamefont {Sadykov}}]{dubovichenko2017b}%
  \BibitemOpen
  \bibfield  {author} {\bibinfo {author} {\bibfnamefont {Sergey}\ \bibnamefont
  {Dubovichenko}}, \bibinfo {author} {\bibfnamefont {Nassurlla}\ \bibnamefont
  {Burtebayev}}, \bibinfo {author} {\bibfnamefont {Albert}\ \bibnamefont
  {Dzhazairov-Kakhramanov}}, \bibinfo {author} {\bibfnamefont {Denis}\
  \bibnamefont {Zazulin}}, \bibinfo {author} {\bibfnamefont {Zhambul}\
  \bibnamefont {Kerimkulov}}, \bibinfo {author} {\bibfnamefont {Marzhan}\
  \bibnamefont {Nassurlla}}, \bibinfo {author} {\bibfnamefont {Chingis}\
  \bibnamefont {Omarov}}, \bibinfo {author} {\bibfnamefont {Alesya}\
  \bibnamefont {Tkachenko}}, \bibinfo {author} {\bibfnamefont {Tatyana}\
  \bibnamefont {Shmygaleva}}, \bibinfo {author} {\bibfnamefont {Stanislaw}\
  \bibnamefont {Kliczewski}}, \ and\ \bibinfo {author} {\bibfnamefont {Turlan}\
  \bibnamefont {Sadykov}},\ }\bibfield  {title} {\enquote {\bibinfo {title}
  {New measurements and phase shift analysis of p 16o elastic scattering at
  astrophysical energies},}\ }\href {\doibase 10.1088/1674-1137/41/1/014001}
  {\bibfield  {journal} {\bibinfo  {journal} {Chinese Physics C}\ }\textbf
  {\bibinfo {volume} {41}},\ \bibinfo {pages} {014001} (\bibinfo {year}
  {2017})}\BibitemShut {NoStop}%
\bibitem [{\citenamefont {Blue}\ and\ \citenamefont
  {Haeberli}(1965)}]{blue1965}%
  \BibitemOpen
  \bibfield  {author} {\bibinfo {author} {\bibfnamefont {R.~A.}\ \bibnamefont
  {Blue}}\ and\ \bibinfo {author} {\bibfnamefont {W.}~\bibnamefont
  {Haeberli}},\ }\bibfield  {title} {\enquote {\bibinfo {title} {Polarization
  of protons elastically scattered by oxygen},}\ }\href {\doibase
  10.1103/PhysRev.137.B284} {\bibfield  {journal} {\bibinfo  {journal} {Phys.
  Rev.}\ }\textbf {\bibinfo {volume} {137}},\ \bibinfo {pages} {B284--B293}
  (\bibinfo {year} {1965})}\BibitemShut {NoStop}%
\bibitem [{\citenamefont {Tr{\"a}chslin}\ and\ \citenamefont
  {Brown}(1967)}]{trachslin1967}%
  \BibitemOpen
  \bibfield  {author} {\bibinfo {author} {\bibfnamefont {Walter}\ \bibnamefont
  {Tr{\"a}chslin}}\ and\ \bibinfo {author} {\bibfnamefont {Louis}\ \bibnamefont
  {Brown}},\ }\bibfield  {title} {\enquote {\bibinfo {title} {Polarization and
  phase shifts in 12c(p,p)12c and 16o(p,p)16o from 1.5 to 3 mev},}\ }\href
  {\doibase 10.1016/0375-9474(67)90187-X} {\bibfield  {journal} {\bibinfo
  {journal} {Nuclear Physics A}\ }\textbf {\bibinfo {volume} {101}},\ \bibinfo
  {pages} {273 -- 287} (\bibinfo {year} {1967})}\BibitemShut {NoStop}%
\bibitem [{\citenamefont {Ryberg}\ \emph {et~al.}(2016)\citenamefont {Ryberg},
  \citenamefont {Forss{\'e}n}, \citenamefont {Hammer},\ and\ \citenamefont
  {Platter}}]{ryberg2016}%
  \BibitemOpen
  \bibfield  {author} {\bibinfo {author} {\bibfnamefont {Emil}\ \bibnamefont
  {Ryberg}}, \bibinfo {author} {\bibfnamefont {Christian}\ \bibnamefont
  {Forss{\'e}n}}, \bibinfo {author} {\bibfnamefont {H.-W.}\ \bibnamefont
  {Hammer}}, \ and\ \bibinfo {author} {\bibfnamefont {Lucas}\ \bibnamefont
  {Platter}},\ }\bibfield  {title} {\enquote {\bibinfo {title} {Range
  corrections in proton halo nuclei},}\ }\href {\doibase
  10.1016/j.aop.2016.01.008} {\bibfield  {journal} {\bibinfo  {journal} {Annals
  of Physics}\ }\textbf {\bibinfo {volume} {367}},\ \bibinfo {pages} {13 -- 32}
  (\bibinfo {year} {2016})}\BibitemShut {NoStop}%
\bibitem [{\citenamefont {Angeli}\ and\ \citenamefont
  {Marinova}(2013)}]{angeli2013}%
  \BibitemOpen
  \bibfield  {author} {\bibinfo {author} {\bibfnamefont {I.}~\bibnamefont
  {Angeli}}\ and\ \bibinfo {author} {\bibfnamefont {K.P.}\ \bibnamefont
  {Marinova}},\ }\bibfield  {title} {\enquote {\bibinfo {title} {Table of
  experimental nuclear ground state charge radii: An update},}\ }\href
  {\doibase 10.1016/j.adt.2011.12.006} {\bibfield  {journal} {\bibinfo
  {journal} {At. Data Nucl. Data Tables}\ }\textbf {\bibinfo {volume} {99}},\
  \bibinfo {pages} {69 -- 95} (\bibinfo {year} {2013})}\BibitemShut {NoStop}%
\bibitem [{\citenamefont {Hagen}\ \emph {et~al.}(2010)\citenamefont {Hagen},
  \citenamefont {Papenbrock},\ and\ \citenamefont
  {Hjorth-Jensen}}]{hagen2010a}%
  \BibitemOpen
  \bibfield  {author} {\bibinfo {author} {\bibfnamefont {G.}~\bibnamefont
  {Hagen}}, \bibinfo {author} {\bibfnamefont {T.}~\bibnamefont {Papenbrock}}, \
  and\ \bibinfo {author} {\bibfnamefont {M.}~\bibnamefont {Hjorth-Jensen}},\
  }\bibfield  {title} {\enquote {\bibinfo {title} {{\textit{Ab~Initio}
  Computation of the $^{17}\mathbf{F}$ Proton Halo State and Resonances in $A =
  17$ Nuclei}},}\ }\href {\doibase 10.1103/PhysRevLett.104.182501} {\bibfield
  {journal} {\bibinfo  {journal} {Phys. Rev. Lett.}\ }\textbf {\bibinfo
  {volume} {104}},\ \bibinfo {pages} {182501} (\bibinfo {year}
  {2010})}\BibitemShut {NoStop}%
\bibitem [{\citenamefont {Stone}(2014)}]{stone2014a}%
  \BibitemOpen
  \bibfield  {author} {\bibinfo {author} {\bibfnamefont {N.~J.}\ \bibnamefont
  {Stone}},\ }\href
  {https://www-nds.iaea.org/publications/indc/indc-nds-0658.pdf} {\emph
  {\bibinfo {title} {Table of Nuclear Magnetic Dipole and Electric Quadrupole
  Moments}}},\ \bibinfo {type} {Tech. Rep.}\ \bibinfo {number}
  {INDC(NDS)--0658}\ (\bibinfo  {institution} {International Atomic Energy
  Agency (IAEA)},\ \bibinfo {year} {2014})\BibitemShut {NoStop}%
\bibitem [{\citenamefont {Ryberg}\ \emph
  {et~al.}(2014{\natexlab{b}})\citenamefont {Ryberg}, \citenamefont
  {Forss{\'e}n}, \citenamefont {Hammer},\ and\ \citenamefont
  {Platter}}]{ryberg2014b}%
  \BibitemOpen
  \bibfield  {author} {\bibinfo {author} {\bibfnamefont {Emil}\ \bibnamefont
  {Ryberg}}, \bibinfo {author} {\bibfnamefont {Christian}\ \bibnamefont
  {Forss{\'e}n}}, \bibinfo {author} {\bibfnamefont {H.-W.}\ \bibnamefont
  {Hammer}}, \ and\ \bibinfo {author} {\bibfnamefont {Lucas}\ \bibnamefont
  {Platter}},\ }\bibfield  {title} {\enquote {\bibinfo {title} {Constraining
  low-energy proton capture on beryllium-7 through charge radius
  measurements},}\ }\href {\doibase 10.1140/epja/i2014-14170-2} {\bibfield
  {journal} {\bibinfo  {journal} {The European Physical Journal A}\ }\textbf
  {\bibinfo {volume} {50}},\ \bibinfo {pages} {170} (\bibinfo {year}
  {2014}{\natexlab{b}})}\BibitemShut {NoStop}%
\bibitem [{\citenamefont {Keller}\ and\ \citenamefont
  {Haeberli}(1970)}]{keller1970}%
  \BibitemOpen
  \bibfield  {author} {\bibinfo {author} {\bibfnamefont {L.~G.}\ \bibnamefont
  {Keller}}\ and\ \bibinfo {author} {\bibfnamefont {W.}~\bibnamefont
  {Haeberli}},\ }\bibfield  {title} {\enquote {\bibinfo {title}
  {Vector-polarization measurements and phase-shift analysis for $d-\alpha$
  scattering between 3 and 11 mev},}\ }\href {\doibase
  10.1016/0375-9474(70)90244-7} {\bibfield  {journal} {\bibinfo  {journal}
  {Nuclear Physics A}\ }\textbf {\bibinfo {volume} {156}},\ \bibinfo {pages}
  {465 -- 476} (\bibinfo {year} {1970})}\BibitemShut {NoStop}%
\bibitem [{\citenamefont {Gr{\"u}ebler}\ \emph {et~al.}(1975)\citenamefont
  {Gr{\"u}ebler}, \citenamefont {Schmelzbach}, \citenamefont {K{\"o}nig},
  \citenamefont {Risler},\ and\ \citenamefont {Boerma}}]{gruebler1975}%
  \BibitemOpen
  \bibfield  {author} {\bibinfo {author} {\bibfnamefont {W.}~\bibnamefont
  {Gr{\"u}ebler}}, \bibinfo {author} {\bibfnamefont {P.~A.}\ \bibnamefont
  {Schmelzbach}}, \bibinfo {author} {\bibfnamefont {V.}~\bibnamefont
  {K{\"o}nig}}, \bibinfo {author} {\bibfnamefont {R.}~\bibnamefont {Risler}}, \
  and\ \bibinfo {author} {\bibfnamefont {D.}~\bibnamefont {Boerma}},\
  }\bibfield  {title} {\enquote {\bibinfo {title} {Phase-shift analysis of
  $d-\alpha$ elastic scattering between 3 and 17 mev},}\ }\href {\doibase
  10.1016/0375-9474(75)90048-2} {\bibfield  {journal} {\bibinfo  {journal}
  {Nuclear Physics A}\ }\textbf {\bibinfo {volume} {242}},\ \bibinfo {pages}
  {265 -- 284} (\bibinfo {year} {1975})}\BibitemShut {NoStop}%
\bibitem [{\citenamefont {Krasnopol'sky}\ \emph {et~al.}(1991)\citenamefont
  {Krasnopol'sky}, \citenamefont {Kukulin}, \citenamefont {Kuznetsova},
  \citenamefont {Hor\'aek},\ and\ \citenamefont {Queen}}]{krasnopolsky1991}%
  \BibitemOpen
  \bibfield  {author} {\bibinfo {author} {\bibfnamefont {V.~M.}\ \bibnamefont
  {Krasnopol'sky}}, \bibinfo {author} {\bibfnamefont {V.~I.}\ \bibnamefont
  {Kukulin}}, \bibinfo {author} {\bibfnamefont {E.~V.}\ \bibnamefont
  {Kuznetsova}}, \bibinfo {author} {\bibfnamefont {J.}~\bibnamefont
  {Hor\'aek}}, \ and\ \bibinfo {author} {\bibfnamefont {N.~M.}\ \bibnamefont
  {Queen}},\ }\bibfield  {title} {\enquote {\bibinfo {title} {Energy-dependent
  phase-shift analysis of $^{2}\mathrm{H}$${+}^{4}$he scattering in the energy
  range $0.87 < e_d < 5.24$~mev},}\ }\href {\doibase 10.1103/PhysRevC.43.822}
  {\bibfield  {journal} {\bibinfo  {journal} {Phys. Rev. C}\ }\textbf {\bibinfo
  {volume} {43}},\ \bibinfo {pages} {822--834} (\bibinfo {year}
  {1991})}\BibitemShut {NoStop}%
\bibitem [{\citenamefont {Mukhamedzhanov}\ \emph {et~al.}(2011)\citenamefont
  {Mukhamedzhanov}, \citenamefont {Blokhintsev},\ and\ \citenamefont
  {Irgaziev}}]{mukhamedzhanov2011}%
  \BibitemOpen
  \bibfield  {author} {\bibinfo {author} {\bibfnamefont {A.~M.}\ \bibnamefont
  {Mukhamedzhanov}}, \bibinfo {author} {\bibfnamefont {L.~D.}\ \bibnamefont
  {Blokhintsev}}, \ and\ \bibinfo {author} {\bibfnamefont {B.~F.}\ \bibnamefont
  {Irgaziev}},\ }\bibfield  {title} {\enquote {\bibinfo {title} {Reexamination
  of the astrophysical $s$ factor for the
  $\ensuremath{\alpha}$+$d\ensuremath{\rightarrow}{}^{6}$li+$\ensuremath{\gamma}$
  reaction},}\ }\href {\doibase 10.1103/PhysRevC.83.055805} {\bibfield
  {journal} {\bibinfo  {journal} {Phys. Rev. C}\ }\textbf {\bibinfo {volume}
  {83}},\ \bibinfo {pages} {055805} (\bibinfo {year} {2011})}\BibitemShut
  {NoStop}%
\bibitem [{\citenamefont {Blokhintsev}\ and\ \citenamefont
  {Savin}(2014)}]{blokhintsev2014}%
  \BibitemOpen
  \bibfield  {author} {\bibinfo {author} {\bibfnamefont {L.~D.}\ \bibnamefont
  {Blokhintsev}}\ and\ \bibinfo {author} {\bibfnamefont {D.~A.}\ \bibnamefont
  {Savin}},\ }\bibfield  {title} {\enquote {\bibinfo {title} {Analytic
  continuation of the effective-range expansion as a method for determining the
  features of bound states: Application to the 6li nucleus},}\ }\href {\doibase
  10.1134/S1063778814030041} {\bibfield  {journal} {\bibinfo  {journal}
  {Physics of Atomic Nuclei}\ }\textbf {\bibinfo {volume} {77}},\ \bibinfo
  {pages} {351--361} (\bibinfo {year} {2014})}\BibitemShut {NoStop}%
\bibitem [{\citenamefont {Blokhintsev}\ \emph {et~al.}(1993)\citenamefont
  {Blokhintsev}, \citenamefont {Kukulin}, \citenamefont {Sakharuk},
  \citenamefont {Savin},\ and\ \citenamefont {Kuznetsova}}]{blokhintsev1993}%
  \BibitemOpen
  \bibfield  {author} {\bibinfo {author} {\bibfnamefont {L.~D.}\ \bibnamefont
  {Blokhintsev}}, \bibinfo {author} {\bibfnamefont {V.~I.}\ \bibnamefont
  {Kukulin}}, \bibinfo {author} {\bibfnamefont {A.~A.}\ \bibnamefont
  {Sakharuk}}, \bibinfo {author} {\bibfnamefont {D.~A.}\ \bibnamefont {Savin}},
  \ and\ \bibinfo {author} {\bibfnamefont {E.~V.}\ \bibnamefont {Kuznetsova}},\
  }\bibfield  {title} {\enquote {\bibinfo {title} {Determination of the
  $^{6}\mathrm{\ensuremath{\alpha}}$+d vertex constant (asymptotic coefficient)
  from the $^{4}\mathrm{He}$+d phase-shift analysis},}\ }\href {\doibase
  10.1103/PhysRevC.48.2390} {\bibfield  {journal} {\bibinfo  {journal} {Phys.
  Rev. C}\ }\textbf {\bibinfo {volume} {48}},\ \bibinfo {pages} {2390--2394}
  (\bibinfo {year} {1993})}\BibitemShut {NoStop}%
\bibitem [{\citenamefont {Blokhintsev}\ \emph {et~al.}(2006)\citenamefont
  {Blokhintsev}, \citenamefont {Igamov}, \citenamefont {Nishonov},\ and\
  \citenamefont {Yarmukhamedov}}]{blokhintsev2006}%
  \BibitemOpen
  \bibfield  {author} {\bibinfo {author} {\bibfnamefont {L.~D.}\ \bibnamefont
  {Blokhintsev}}, \bibinfo {author} {\bibfnamefont {S.~B.}\ \bibnamefont
  {Igamov}}, \bibinfo {author} {\bibfnamefont {M.~M.}\ \bibnamefont
  {Nishonov}}, \ and\ \bibinfo {author} {\bibfnamefont {R.}~\bibnamefont
  {Yarmukhamedov}},\ }\bibfield  {title} {\enquote {\bibinfo {title}
  {Calculation of the nuclear vertex constant (asymptotic normalization
  coefficient) for the virtual decay 6li $\to\alpha$ + d on the basis of the
  three-body model and application of the result in describing the
  astrophysical nuclear reaction d($\alpha$, $\gamma$)6li at ultralow
  energies},}\ }\href {\doibase 10.1134/S1063778806030069} {\bibfield
  {journal} {\bibinfo  {journal} {Physics of Atomic Nuclei}\ }\textbf {\bibinfo
  {volume} {69}},\ \bibinfo {pages} {433--444} (\bibinfo {year}
  {2006})}\BibitemShut {NoStop}%
\bibitem [{\citenamefont {Tursunov}\ \emph {et~al.}(2015)\citenamefont
  {Tursunov}, \citenamefont {Turakulov},\ and\ \citenamefont
  {Descouvemont}}]{tursunov2015}%
  \BibitemOpen
  \bibfield  {author} {\bibinfo {author} {\bibfnamefont {E.~M.}\ \bibnamefont
  {Tursunov}}, \bibinfo {author} {\bibfnamefont {S.~A.}\ \bibnamefont
  {Turakulov}}, \ and\ \bibinfo {author} {\bibfnamefont {P.}~\bibnamefont
  {Descouvemont}},\ }\bibfield  {title} {\enquote {\bibinfo {title}
  {Theoretical analysis of the astrophysical s-factor for the capture reaction
  $\alpha + d \to ^6\mbox{Li} + \gamma$ in the two-body model},}\ }\href
  {\doibase 10.1134/S1063778815010196} {\bibfield  {journal} {\bibinfo
  {journal} {Physics of Atomic Nuclei}\ }\textbf {\bibinfo {volume} {78}},\
  \bibinfo {pages} {193--200} (\bibinfo {year} {2015})}\BibitemShut {NoStop}%
\bibitem [{\citenamefont {Spiger}\ and\ \citenamefont
  {Tombrello}(1967)}]{spiger1967}%
  \BibitemOpen
  \bibfield  {author} {\bibinfo {author} {\bibfnamefont {R.~J.}\ \bibnamefont
  {Spiger}}\ and\ \bibinfo {author} {\bibfnamefont {T.~A.}\ \bibnamefont
  {Tombrello}},\ }\bibfield  {title} {\enquote {\bibinfo {title} {Scattering of
  ${\mathrm{he}}^{3}$ by ${\mathrm{he}}^{4}$ and of ${\mathrm{he}}^{4}$ by
  tritium},}\ }\href {\doibase 10.1103/PhysRev.163.964} {\bibfield  {journal}
  {\bibinfo  {journal} {Phys. Rev.}\ }\textbf {\bibinfo {volume} {163}},\
  \bibinfo {pages} {964--984} (\bibinfo {year} {1967})}\BibitemShut {NoStop}%
\bibitem [{\citenamefont {Boykin}\ \emph {et~al.}(1972)\citenamefont {Boykin},
  \citenamefont {Baker},\ and\ \citenamefont {Hardy}}]{boykin1972}%
  \BibitemOpen
  \bibfield  {author} {\bibinfo {author} {\bibfnamefont {W.~R.}\ \bibnamefont
  {Boykin}}, \bibinfo {author} {\bibfnamefont {S.~D.}\ \bibnamefont {Baker}}, \
  and\ \bibinfo {author} {\bibfnamefont {D.~M.}\ \bibnamefont {Hardy}},\
  }\bibfield  {title} {\enquote {\bibinfo {title} {Scattering of 3he and 4he
  from polarized 3he between 4 and 10 mev},}\ }\href {\doibase
  10.1016/0375-9474(72)90732-4} {\bibfield  {journal} {\bibinfo  {journal}
  {Nuclear Physics A}\ }\textbf {\bibinfo {volume} {195}},\ \bibinfo {pages}
  {241 -- 249} (\bibinfo {year} {1972})}\BibitemShut {NoStop}%
\bibitem [{\citenamefont {Tursunmahatov}\ and\ \citenamefont
  {Yarmukhamedov}(2012)}]{tursunmahatov2012}%
  \BibitemOpen
  \bibfield  {author} {\bibinfo {author} {\bibfnamefont {Q.~I.}\ \bibnamefont
  {Tursunmahatov}}\ and\ \bibinfo {author} {\bibfnamefont {R.}~\bibnamefont
  {Yarmukhamedov}},\ }\bibfield  {title} {\enquote {\bibinfo {title}
  {Determination of the $^3$he $+\alpha\to ^7$be asymptotic normalization
  coefficients, the nuclear vertex constants, and their application for the
  extrapolation of the $^3$he($\alpha,\gamma$)$^7$be astrophysical $s$ factors
  to the solar energy region},}\ }\href {\doibase 10.1103/PhysRevC.85.045807}
  {\bibfield  {journal} {\bibinfo  {journal} {Phys. Rev. C}\ }\textbf {\bibinfo
  {volume} {85}},\ \bibinfo {pages} {045807} (\bibinfo {year}
  {2012})}\BibitemShut {NoStop}%
\bibitem [{\citenamefont {Igamov}\ and\ \citenamefont
  {Yarmukhamedov}(2007)}]{igamov2007}%
  \BibitemOpen
  \bibfield  {author} {\bibinfo {author} {\bibfnamefont {S.~B.}\ \bibnamefont
  {Igamov}}\ and\ \bibinfo {author} {\bibfnamefont {R.}~\bibnamefont
  {Yarmukhamedov}},\ }\bibfield  {title} {\enquote {\bibinfo {title} {Modified
  two-body potential approach to the peripheral direct capture astrophysical
  $a+a\to b+\gamma$ reaction and asymptotic normalization coefficients},}\
  }\href {\doibase 10.1016/j.nuclphysa.2006.10.041} {\bibfield  {journal}
  {\bibinfo  {journal} {Nuclear Physics A}\ }\textbf {\bibinfo {volume}
  {781}},\ \bibinfo {pages} {247 -- 276} (\bibinfo {year} {2007})}\BibitemShut
  {NoStop}%
\bibitem [{\citenamefont {Michel}(2007)}]{michel2007}%
  \BibitemOpen
  \bibfield  {author} {\bibinfo {author} {\bibfnamefont {N.}~\bibnamefont
  {Michel}},\ }\bibfield  {title} {\enquote {\bibinfo {title} {Precise coulomb
  wave functions for a wide range of complex $l$, $\eta$, and $z$},}\ }\href
  {\doibase 10.1016/j.cpc.2006.10.004} {\bibfield  {journal} {\bibinfo
  {journal} {Computer Physics Communications}\ }\textbf {\bibinfo {volume}
  {176}},\ \bibinfo {pages} {232 -- 249} (\bibinfo {year} {2007})}\BibitemShut
  {NoStop}%
\bibitem [{{\relax DLMF}()}]{dlmf}%
  \BibitemOpen
  {\relax DLMF},\ \href {http://dlmf.nist.gov/} {\enquote {\bibinfo {title}
  {{NIST Digital Library of Mathematical Functions}},}\ }\bibinfo
  {howpublished} {http://dlmf.nist.gov/, Release 1.0.23 of 2019-06-15},\
  \bibinfo {note} {{F.~W.~J. Olver, A.~B. {Olde Daalhuis}, D.~W. Lozier, B.~I.
  Schneider, R.~F. Boisvert, C.~W. Clark, B.~R. Miller and B.~V. Saunders,
  eds.}}\BibitemShut {Stop}%
\bibitem [{\citenamefont {{Abramowitz}}\ and\ \citenamefont
  {{Stegun}}(1964)}]{abramowitz1964}%
  \BibitemOpen
  \bibfield  {author} {\bibinfo {author} {\bibfnamefont {Milton}\ \bibnamefont
  {{Abramowitz}}}\ and\ \bibinfo {author} {\bibfnamefont {Irene~A.}\
  \bibnamefont {{Stegun}}},\ }\href@noop {} {\emph {\bibinfo {title} {Handbook
  of Mathematical Functions with Formulas, Graphs, and Mathematical Tables}}},\
  \bibinfo {edition} {ninth dover printing, tenth gpo printing}\ ed.\ (\bibinfo
   {publisher} {Dover},\ \bibinfo {address} {New York City},\ \bibinfo {year}
  {1964})\BibitemShut {NoStop}%
\end{thebibliography}
\end{document}